\documentclass{aa}
\usepackage{multirow}

\usepackage{multirow}
\usepackage{comment}
\usepackage{amsmath}
\usepackage{yhmath}
\usepackage{amssymb}
\usepackage{bm}
\usepackage[colorlinks,linkcolor=blue,anchorcolor=blue,citecolor=blue]{hyperref}
\usepackage{graphicx,subfig} 
\usepackage{lscape}
\usepackage{longtable}
\usepackage{txfonts}
\usepackage{natbib}
\usepackage{color}
\usepackage{booktabs}
\usepackage[dvipsnames,table,xcdraw]{xcolor}

\makeatletter
\renewcommand*\aa@pageof{, page \thepage{} of \pageref*{LastPage}}
\usepackage{hyperref}
\hypersetup{colorlinks=true,allcolors=[rgb]{0,0,0.8}}

\newcommand{\erosita}{eROSITA\xspace}
\newcommand{\planck}{\emph{Planck}\xspace}

\newcommand{\esass}{\texttt{eSASS}\xspace}

\newcommand{\erass}[1][1]{eRASS1\xspace}
\newcommand{\extlike}{$\mathcal{L}_{ext}$\xspace}

\newcommand{\mbproj}{MBProj2D\xspace}

\usepackage[normalem]{ulem} 

\newcommand{\sig}{\sigma_\textrm{8}}
\newcommand{\Sh}{S_\textrm{8}}
\newcommand{\Om}{\Omega_\textrm{m}}
\newcommand{\Ode}{\Omega_\Lambda}

\usepackage{natbib,twoopt}
\defcitealias{Ghirardini2023}{G24}
\defcitealias{PlanckCollaboration2020}{Planck20}
\usepackage{xcolor}

\makeatletter
\newcommandtwoopt{\citeads}[3][][]{\href{http://adsabs.harvard.edu/abs/#3}%
    {\def\hyper@linkstart##1##2{}%
     \let\hyper@linkend\@empty\citealp[#1][#2]{#3}}}
  \newcommandtwoopt{\citepads}[3][][]{\href{http://adsabs.harvard.edu/abs/#3}%
    {\def\hyper@linkstart##1##2{}%
     \let\hyper@linkend\@empty\citep[#1][#2]{#3}}}
  \newcommandtwoopt{\citetads}[3][][]{\href{http://adsabs.harvard.edu/abs/#3}%
    {\def\hyper@linkstart##1##2{}%
     \let\hyper@linkend\@empty\citet[#1][#2]{#3}}}
  \newcommandtwoopt{\citeyearads}[3][][]%
    {\href{http://adsabs.harvard.edu/abs/#3}
    {\def\hyper@linkstart##1##2{}%
     \let\hyper@linkend\@empty\citeyear[#1][#2]{#3}}}
\makeatother
\usepackage{graphicx}
\usepackage{multirow}
\usepackage{siunitx}  %unit
\usepackage{txfonts}
\title{The SRG/eROSITA All-Sky Survey}
\subtitle{\Large Constraints on the structure growth  from cluster number counts} %and the $S_\mathrm{8}$ tension

\author{E. Artis\inst{1},
E. Bulbul\inst{1}, 
S.~Grandis\inst{2,5},
V. Ghirardini\inst{1,17}, 
N.~Clerc\inst{3},  
R.~Seppi\inst{1, 7},
J. Comparat\inst{1},
M. Cataneo\inst{4,16},
A. von der Linden\inst{15,1},
Y.~E.~Bahar\inst{1}, 
F.~Balzer\inst{1}, 
I.~Chiu\inst{8}, 
D.~Gruen\inst{5},
F.~Kleinebreil\inst{2, 4}, 
M.~Kluge\inst{1}, 
S.~Krippendorf\inst{5, 6}, 
X.~Li\inst{10},
A.~Liu\inst{1},
N.~Malavasi\inst{1},
A.~Merloni\inst{1},
H.~Miyatake\inst{12, 13, 14},
S.~Miyazaki\inst{11},
K.~Nandra\inst{1},
N.~Okabe\inst{9},
F.~Pacaud\inst{4}, 
P.~Predehl\inst{1},
M.~E.~Ramos-Ceja\inst{1}, 
T.~H.~Reiprich\inst{4}, 
J.~S.~Sanders\inst{1}, 
T.~Schrabback\inst{2, 4}, 
S.~Zelmer\inst{1}, and
X.~Zhang\inst{1}}
\institute{
Max Planck Institute for Extraterrestrial Physics, Giessenbachstrasse 1, 85748 Garching, Germany
\and
Universit\"at Innsbruck,  Institut f\"ur Astro- und Teilchenphysik, Technikerstr. 25/8, 6020 Innsbruck, Austria
\and
IRAP, Université de Toulouse, CNRS, UPS, CNES, F-31028 Toulouse, France
\and
Argelander-Institut f\"ur Astronomie (AIfA), Universit\"at Bonn, Auf dem H\"ugel 71, 53121 Bonn, Germany
\and
Universit\"ats-Sternwarte, LMU Munich, Scheinerstr. 1, 81679 M\"unchen, Germany
\and
Arnold Sommerfeld Center for Theoretical Physics, LMU Munich, Theresienstr. 37, 80333 M\"unchen, Germany
\and
Department of Astronomy, University of Geneva, Ch. d’Ecogia 16, CH-1290 Versoix, Switzerland
\and
Department of Physics, National Cheng Kung University, 70101 Tainan, Taiwan
\and
Department of Physical Science, Hiroshima University, 1-3-1 Kagamiyama,
Higashi-Hiroshima, Hiroshima 739-8526, Japan
\and 
McWilliams Center for Cosmology, Department of Physics, Carnegie Mellon University, Pittsburgh, PA 15213, USA
\and
Subaru Telescope, National Astronomical Observatory of Japan, 650 N Aohoku Place Hilo HI 96720 USA
\and
Kobayashi-Maskawa Institute for the Origin of Particles and the Universe (KMI), Nagoya University, Nagoya, 464-8602, Japan
\and
Institute for Advanced Research, Nagoya University, Nagoya 464-8601, Japan
\and
Kavli Institute for the Physics and Mathematics of the Universe (WPI), The University of Tokyo Institutes for Advanced Study (UTIAS), The University of Tokyo, Chiba 277-8583, Japan
\and
Department of Physics and Astronomy, Stony Brook University, Stony Brook, NY 11794, USA
\and
Ruhr University Bochum, Faculty of Physics and Astronomy, Astronomical Institute (AIRUB), German Centre for Cosmological Lensing, 44780 Bochum, Germany
\and
INAF, Osservatorio di Astrofisica e Scienza dello Spazio, via Piero Gobetti 93/3, 40129 Bologna, Italy
}
\date{\today}

\titlerunning{Constraints on the Growth of Structures}
\authorrunning{Artis et al.}

\begin{document}

\abstract{Recent advancements in methods utilized in wide-area surveys have demonstrated the reliability of the number density of galaxy clusters as a viable tool for precision cosmology. Beyond testing the current cosmological paradigm, cluster number counts can also be utilized to investigate the discrepancies currently affecting current cosmological measurements.
In particular, cosmological studies based on cosmic shear and other large-scale structure probes routinely find a value of the amplitude of the fluctuations in the universe $\Sh = \sigma_\mathrm{8}(\Omega_\mathrm{m}/0.3)^{0.5}$ smaller than the one inferred from the primary cosmic microwave background. In this work, we investigate this tension by measuring structure evolution across cosmic time as probed by the number counts of the massive halos with the first SRG/\erosita\ All-Sky Survey cluster catalog in the Western Galactic Hemisphere complemented with the overlapping Dark Energy Survey Year-3, KiloDegree Survey, and Hyper Suprime-Cam data for weak lensing mass calibration, by implementing two different parameterizations and a model-agnostic method. In the first model, we measure the cosmic linear growth index as $\gamma = 1.19\pm 0.21$, in tension with the standard value of $\gamma = 0.55$, but in good statistical agreement with other large-scale structures probes. The second model is a phenomenological scenario in which we rescale the linear matter power spectrum at low redshift to investigate a potential reduction of structure formation, providing similar results. Finally, in a third strategy, we consider a standard $\Lambda\mathrm{CDM}$ cosmology, but we separate the cluster catalog into five redshift bins, measuring the cosmological parameters in each and inferring the evolution of the structure formation, finding hints of a reduction.  Interestingly, the $\Sh$ value inferred from eRASS1 cluster number counts, when we add a degree of freedom to the matter power spectrum, recover the value inferred by cosmic shear studies. The observed reduction in the growth rate or systematic uncertainties associated with various measurements may account for the discrepancy in the $\Sh$ values suggested between cosmic shear probes and \erosita\ cluster number counts, \planck\ CMB measurements.
 }

\maketitle

\section{Introduction}
\label{sec:intro}

In the standard cosmological formalism, most of the energy content of the universe is formed by a non-relativistic component called Cold Dark Matter \citep[$\mathrm{CDM}$ hereafter]{Rubin1970} and an unknown Dark Energy  \citep[DE]{Riess1998} responsible for the accelerated expansion of the universe, and frequently represented by the cosmological constant $\Lambda$. The third ingredient is an early phase of exponential expansion called Inflation \citep{Brout1978}. Under the influence of these elements, quantum fluctuations evolved in the early universe to form the large-scale structure (LSS) that we observe today. This standard formalism, the $\Lambda$CDM model, has been successful at describing a broad range of observations, from the statistical distribution of the fluctuations of the Cosmic Microwave Background (CMB) to the spatial correlations of galaxies in the late universe \citep{PlanckCollaboration2020, Amon2022, Desi2024}. Despite its success, the standard concordance model, $\Lambda$CDM, is being challenged by several tensions, such as discrepancies reported in the measurements of the Hubble constant ($H_\mathrm{0}$).
The Hubble constant measured by early time probes is in 4 to 6$\sigma$ statistical tension with the one directly measured by late-time distance ladders \citep{DiValentino2021}. Whether these differences in the expansion rate measurements originate from fundamental physics or systematic effects is still debated \citep{Freedman2024}. This tension is supplemented by numerous but less significant discordant observations, such as anomalies in the CMB anisotropies like a higher lensing amplitude than the one predicted by $\mathrm{\Lambda CDM}$\,\citep{PlanckCollaboration2020}, discrepancies in the abundance of galaxy satellites \citep{Kanehisa2024, Bullock2010}, potential anisotropies in the large-scale distribution of galaxy clusters \citep{Migkas2020}, or the unlikely presence of high velocity colliding clusters \citep{Ascencio2021} (see \cite{Peebles2022} and \cite{Perivolaropoulos2022} for a review of the most common reported anomalies). Recently, \cite{Desi2024} reported a significant preference for an evolving dark energy equation of state when measuring the cosmological parameters from the baryonic acoustic oscillations. The tension is reported as high as $3.9\sigma$ when BAO are combined with the CMB and supernovae data. Interestingly, this preference is consistent with earlier cluster observations like \cite{Mantz2015} for which the best-fit values suggested that $w_\mathrm{a} < 0$, however still in excellent agreement with $\Lambda\mathrm{CDM}$. The same was observed for galaxy clustering and weak lensing data from the Dark Energy Survey collaboration \cite{Abbott2023c}.

Another reported tension is found in the measurement of the growth of structures. The values reported by the CMB are higher than the ones measured by a number of late-time cosmological probes \citep[see][and references therein]{DiValentino2021b}. This tension is best seen in the $\Om-\sig$ plane, where the CMB constraints lie above the degeneracy of the LSS probes, and is usually quantified in terms of the $\Sh = \sig(\Om/0.3)^{0.5}$ parameter. 
Although statistically less significant than the $H_\mathrm{0}$ tension, various probes report the $\Sh$ tension as a motivation to investigate a potential physical origin. Among the most prominent early time probes, \cite{PlanckCollaboration2020} (\citetalias{PlanckCollaboration2020} hereafter), measured $\Sh = 0.832\pm 0.013$ (TT+TE+EE+lensing). Similar "early-time" measurements are reported in the results of the Wilkinson Microwave Anisotropy Probe \citep[][WMAP]{Hinshaw2013}. This value is confirmed by the results of the South Pole Telescope (SPT) data through the lensing of the CMB combined with baryon acoustic oscillation \citep{Bianchini2020}, who found $\Sh = 0.862\pm0.040$, and the recent result of the sixth data release of the Atacama Cosmology Telescope (ACT) using the same method and who measured $\Sh = 0.840\pm 0.028$ \citep{Madhavacheril2024}. Since CMB lensing probes linear scales with most of the signal coming from a redshift range of $0.5 < z < 5$, the good agreement between the primary CMB and the CMB lensing compared to other late-time probes might suggest a change in the non-linear regime at low redshift.
Conversely, cosmic shear surveys tend to find a lower amplitude of structure formation. For example, 
the Dark Energy Survey \citep{Abbott2022} finds $\Sh = 0.776\pm 0.017$.

\begin{figure}
    \centering
    \includegraphics[width=0.49\textwidth]{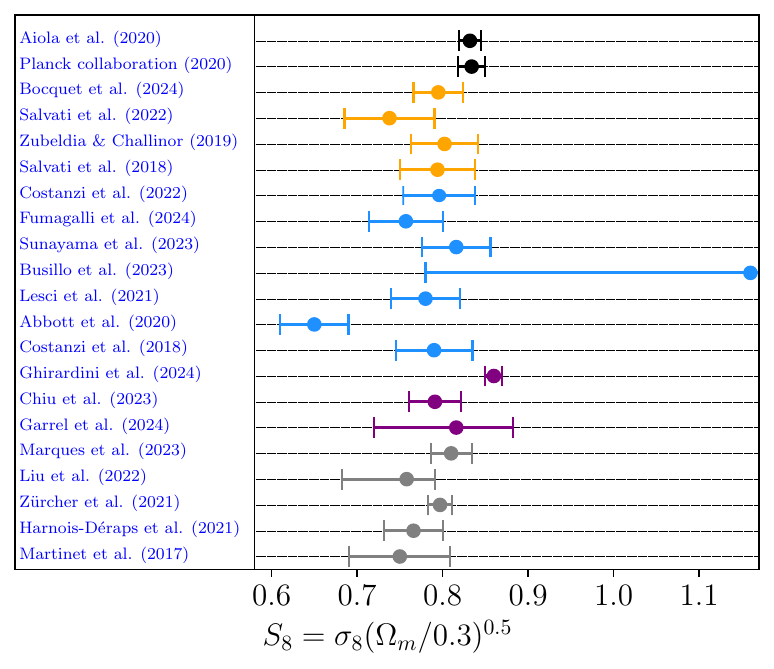}
    \caption{$\Sh$ values obtained from different surveys using the peaks of the matter density field (including cluster counts and weak lensing peaks) as a cosmological probe, compared with the values obtained by the primary CMB anisotropies shown in black (ACT DR4 data \citep{Aiola2020} is first shown followed by \citetalias{PlanckCollaboration2020}). Constraints from cluster number counts are sorted by detection wavelength; each color is assigned to the same type of experiment. The millimeter wavelength is shown in yellow. We show the constraints from SPT with weak lensing mass calibration \citep{Bocquet2024}, Planck with weak hydrostatic mass calibration  \citep{Salvati2022}, and Planck with a mass calibration based on CMB lensing \citep{Zubeldia2019}. We also show the combined Planck clusters and SZ power spectrum \citep{Salvati2018}. In blue, we show the constraints obtained from optically selected clusters. We present SDSS constraints \cite{Costanzi2019, Fumagalli2024}, HSC constraints \citep{Sunayama2023}, DES constraints \citep{DesCollaboration2020}, and KIDS constraints \citep{Lesci2022}.  We also represent constraints obtained from the extreme value statistics of KIDS clusters \cite{Busillo2023}. Those constraints are larger as extreme events are intrinsically rare. In the X-ray band, the results from eFEDS \citep{Chiu2023}, the XMM/XXL  \cite{Garrel2022}, and the eROSITA/eRASS1 results are shown in magenta \citep{Ghirardini2023}. The grey markers represent studies of the abundance of the peaks in the cosmic shear maps that strongly correlate with the presence of a galaxy cluster and probe cosmology at the nodes of the cosmic web: HSC \citep{Marques2024, Liu2023}, DES  \citep{Zurcher2022, Harnois2021}, and KIDS constraints \citep{Martinet2018}. We show that cosmological probes using the abundance of the nodes of the cosmic web find, on average, little hints for the existence of a growth tension.
    }
    \label{fig:cluster_s8}
\end{figure}

Cluster counts have also been utilized in the $\Sh$ measurements in recent years. In the millimeter wavelength, \cite{Zubeldia2019} used \planck-detected galaxy clusters with CMB lensing mass calibration to find $\Sh = 0.803 \pm 0.039$. Recently, the SPT collaboration used a sample of 1,005 galaxy clusters with weak gravitational lensing data from DES and the Hubble Space Telescope (HST) and found $\Sh = 0.795 \pm 0.029$ \citep{Bocquet2024}. Similarly, \cite{Salvati2022} combined the \planck and SPT samples and reported $\Sh = 0.74\pm0.05$. \cite{Aymerich2024} reanalyzed the same \textit{Planck} cluster sample with external \textit{Chandra} data and obtained $\Sh = 0.81 \pm 0.02$. In the optical wavelength, a preliminary study on the clusters detected in the Sloan Digital Sky Survey DR8 produced an $\Sh$ value of $0.78\pm 0.04$ \citep{Costanzi2019}, while the latest DES results combined with SPT find $\Sh=0.736\pm 0.049$ \citep{Costanzi2021}.  
Furthermore, \cite{Sunayama2023} analyzed a sample of 8,379 clusters detected by the redMaPPer algorithm with HSC weak lensing data and found $\Sh = 0.816^{+0.041}_{-0.039}$. Additionally, an analysis of a sample of 3652 galaxy clusters detected in KIDS produced $\Sh = 0.78 \pm 0.04$. 
The recent results for the first eROSITA All-Sky-Survey \citep[eRASS1,][]{Ghirardini2023, Artis2024} established cluster number counts as a precision cosmological probe, placing tight constraints on $\Sh = 0.86 \pm 0.01$. Interestingly, the eROSITA, as a later-time probe analysis, has no tension with the \citetalias{PlanckCollaboration2020} measurements. These measurements in the literature are compared in Figure~\ref{fig:cluster_s8}.

As the cluster mass function is established as a reliable cosmological probe, and the samples have reached a sizable number of clusters to ensure statistical precision with the launch of SRG/eROSITA All-Sky Survey, we take our primary cosmology analysis one step further and constrain the structure growth over cosmic time.
The overarching goal of this paper is to measure the parameterization of the growth factor using the cosmic linear growth index ($\gamma$), a late-time reduction of the power spectrum, and a model agnostic method where the sample is divided into five different redshift bins \ref{sec:direct_strucutre_growth}. 
The first approach aims at modifying the growth rate of structure formation $f(z)$ and defines it as a power-law of the matter density parameter, i.e., $\Omega_\mathrm{m}^{\gamma}$. This method has been widely discussed in the literature in the context of testing general relativity (GR) at large scales. For instance, \cite{Wang1998} shows that the non-linear theory of structure formation assuming general relativity yields a growth rate well modeled by this function when $\gamma \sim 0.55$. Thus, deviations from this value would indicate that structures do not evolve as predicted and, more generally, suggest a potential departure from GR-$\Lambda\mathrm{CDM}$. A higher $\gamma$ is equivalent to a reduction of structure formation.
Previous galaxy cluster surveys have been utilized in GR studies.  \cite{Rapetti2009} used a sample of 78 X-ray selected clusters to constrain the luminosity-mass scaling relation in combination with CMB, SNe~Ia, and X-ray cluster gas-mass fractions ($f_\mathrm{gas}$) and found consistent results with GR. \cite{Mantz2015} used a sample of 224 X-ray clusters detected by ROSAT \citep{Truemper1993}, with weak lensing (WL hereafter) mass measurements obtained from the Subaru telescope and the Canada-France Hawaii Telescope \citep{vonderLinden2014}. This cosmological study combines cluster abundance with the cluster gas fraction, primordial nucleosynthesis, and $H_\mathrm{0}$ priors to constrain the background expansion and break the $\Omega_\mathrm{m}-\sigma_\mathrm{8}$ degeneracy and found no evidence for physics beyond GR. \cite{Bocquet2015} used the same approach on a sample of 100 clusters detected by their millimeter-wave signature, the so-called Sunyaev Zel'dovich (SZ) effect, through the South Pole Telescope (SPT), agreeing with the previous studies when clusters are combined with the CMB measurement from the \cite{PlanckCollaboration2014}, as well as BAO and supernoave type Ia (SNe~Ia). 
However, recent cosmological studies based on the LSS find a higher value of the cosmic linear growth index for the GR prediction ($\gamma > 0.55$). Studies based on the galaxy clustering of the Sloan Digital Sky Survey (SDSS) data \citep{Samushia2013, Beutler2014, Gil-Marin2017}  found a lower structure growth rate than the one inferred from the CMB. This was later confirmed by DES data \citep{Basilakos2020}, once again using galaxy clustering in combination with galaxy-galaxy lensing, implying a damping of structure formation at low redshifts. This recurrent finding is intriguing and may indicate that the theory of structure formation does not describe observations accurately. If so, it might hint at physics beyond GR and $\Lambda\mathrm{CDM}$, or an unknown systematic effect related, for example, to baryonic feedback. 

Apart from the parametrization using the cosmic linear growth index $\gamma$, another method used in this work is to directly measure the evolution of the parameter $\sigma_\mathrm{8}$ across cosmic time. 
Surveys using galaxy clustering provide robust constraints on $f\sigma_\mathrm{8}(z)$ parameter \citep[][among others]{Zarrouk2018, deMattia2021}. The measurements of the evolution of the parameter, $ f\sig (z) = f(z)\times\sig (z)$, is probed by redshift space distortions (RSD) where $f$ is the linear growth rate defined as $f = \mathrm{d}\ln D/ \mathrm{d}\ln a$. Recent surveys have used weak lensing and galaxy clustering to probe this evolution \citep{Jullo2019, GarciaGarcia2021, White2022, Abbott2023}. In the context of cosmology with cluster number counts, \cite{Bocquet2024} probed the growth of structures by using a non-parametric approach on their SPT-SZ sample of 1,005 clusters combined with sound horizon priors from \citetalias{PlanckCollaboration2020}. They divided their sample into five redshift bins, allowing them to obtain a quasi-direct measurement of $\sigma_\mathrm{8}(z)$, and finding good agreement with predictions from the primary CMB probes. 

The paper is organized as follows: Section~\ref{sec:erass1} describes the survey data employed, while Section~\ref{sec:mg_framwork} discusses modifying the structure growth framework. Section~\ref{sec:res} shows our result on the cosmic linear growth index. Section~\ref{sec:power_suppression} and Section~\ref{sec:direct_strucutre_growth} present the results on the power spectrum reduction and the model agnostic method. Section~\ref{sec:robust} demonstrates the robustness of the analysis. Discussions and relation to current cosmological tensions are provided in Section~\ref{sec:discussion}.
Throughout this paper, we use the notation $\log \equiv \log_\mathrm{10}$, and $\ln \equiv \log_\mathrm{e}$. Reported uncertainties correspond to a 68\% confidence level unless noted otherwise.

\section{Survey Data}
\label{sec:erass1}

The data used in this work is identical to the catalogs presented in \citet{Bulbul2023, Kluge2023}. The weak lensing follow-up for mass calibration uses the data utilized in \citep{Grandis2023, Kleinebreil2024}. The cosmological pipeline and analysis framework is the same in \citep{Ghirardini2023, Artis2024}. We summarize the multi-wavelength data, including X-ray, optical, and weak lensing observations, utilized in this work in the following section. 

\subsection{\erosita All-Sky Survey}

The X-ray telescope, \erosita, on board the Spectrum Roentgen Gamma Mission (SRG), completed its first All-Sky Survey program on June 11, 2020, 184 days after the start of the survey. In this work, we use the data of the Western Galactic half of the \erosita\ All-Sky survey (359.9442~deg~$> l >$~179.9442~deg), where the data rights belong to the German \erosita\ consortium. Two catalogs of clusters detected are compiled in this region by \citet{Bulbul2023} and \citet{Kluge2023} based on the catalog of the X-ray sources in the soft band (0.2--2.3~keV) provided in \citet{Merloni2023}. The primary galaxy group and cluster catalog adopts a lower extent likelihood threshold to maximize the discovery space than the cosmological subsample, which maximizes the purity of the sample for scientific applications. In this work, we utilize the \erass\ cosmological subsample due to its high fidelity confirmation rate, well-defined and -tested selection function, and low contamination fraction, which adopts a selection cut of extent likelihood, \extlike~$>6$ \citep[see][for further details]{Bulbul2023, Clerc2023b}. The optical identification of the clusters selected in the Legacy Survey Data Release 10 in the Southern Hemisphere (LS DR10-South) common footprint of 12791~deg$^{2}$ yields a final cosmology sample of 5259 clusters of galaxies reaching purity levels of 95\%, an ideal sample for the studies of growth of structure \citep{Kluge2023}. Unlike the primary cluster catalog, the redshifts in the cosmological subsample are purely photometric measurements, allowing a consistent assessment of the systematics in our analysis. The count rates are extracted with the 2D-image fitting tool \mbproj\ as described in \cite{Bulbul2023} and are employed as a mass proxy in the weak lensing mass calibration likelihood (\cite{Ghirardini2023}, \citetalias{Ghirardini2023} hereafter).

\subsection{Weak Lensing Survey Data}
\label{subsec:weak_lensing_survey}

To perform the mass calibration in a minimally biased way, we utilize the deep and wide-area optical surveys for lensing measurements with the overlapping footprint with \erosita\ in the Western Galactic Hemisphere. The three major wide-area surveys used in this work for mass calibration include the Dark Energy Survey (DES), Kilo Degree Survey (KiDS), and the  Hyper Suprime-Cam Survey (HSC). We briefly describe the weak lensing survey data here.

The three-year weak-lensing data (S19A) covered by the HSC Subaru Strategic Program \citep{Aihara2018, Li2022} weak lensing measurements are made around the location of \erass clusters using in the $g$, $r$, $i$, $z$, and $Y$ bands in the optical. The total area coverage of HSC is $\approx500$~deg$^2$. The shear profile $g_{t}\left(\theta\right)$, the lensing covariance matrix that serves the measurement uncertainty, and the photometric redshift distribution of the selected source sample are obtained as the HSC weak-lensing data products for 96 \erass clusters, with a total signal-to-noise of 40.

We utilize data from the first three years of observations of the Dark Energy Survey (DES~Y3). The DES~Y3 shape catalog \citep{Gatti2021} is built from the $r, i, z$-bands using the \textsc{Metacalibration} pipeline \citep{Huff2017, Sheldon2017}. Considering the overlap between the DES~Y3 footprint and the \erass footprint, we produce tangential shear data for $2201$ \erass galaxy clusters, with a total signal-to-noise of 65 in the tangential shear profile. The analysis and shear profile extraction details are presented in \citet{Grandis2023}.

We use the gold sample of weak lensing and photometric redshift measurements from the fourth data release of the Kilo-Degree Survey \citep{Kuijken2019, Wright2020, Hildebrandt2021, Giblin2021}, hereafter referred to as KiDS-1000. We extract individual reduced tangential shear profiles for a total of 236 \erass galaxy clusters in both the KiDS-North field (101 clusters) and the KiDS-South field (136 clusters), as both have overlap with the \erass footprint with a total signal-to-noise of 19.

The weak lensing data in the mass calibration process is detailed and published in \citep{Ghirardini2023, Grandis2023}. We do not re-process the data here to obtain new shear profiles; instead, we adopt the framework and derived products described in \citetalias{Ghirardini2023}.

\section{Framework and parameterization with the cosmic linear growth index $\gamma$}
\label{sec:mg_framwork}

 Galaxy cluster number counts efficiently place constraints on the low-redshift and large-scale regime, which cannot be explored by every cosmological probe, such as the anisotropies of the CMB. In this section, we describe the first framework we explore in this paper, namely the linear growth parametrization scenario ($\gamma$-model hereafter). The power spectrum reduction model and the model agnostic method are described respectively in section \ref{sec:power_suppression} and section \ref{sec:direct_strucutre_growth}.

 \subsection{Cosmology with \erass cluster count}
 \label{subsec:cluster_counts}

The cosmology analysis presented in this paper is similar to the one introduced in \citetalias{Ghirardini2023}, and we refer the reader to this work for a detailed review. For readability, this section provides the main characteristics and fundamental assumptions of this analysis. The abundance of dark matter halos per units of mass $M$, redshift $z$, and solid angle (sky positions are noted $\mathcal{H}$) follows
\begin{equation}
    \label{eq:halo_mass_function}
    \frac{\mathrm{d}\,n}{\mathrm{d}\ln M\mathrm{d}\,z\mathrm{d}\,\mathcal{H}} = \frac{\rho_\mathrm{m,0}}{M} \frac{\mathrm{d}\ln \sigma^{-1}}{\mathrm{d}\ln M}f(\sigma)\frac{\mathrm{d}\,V}{\mathrm{d}\,z\mathrm{d}\,\mathcal{H}}, 
\end{equation}
where $\rho_\mathrm{m,0}$ is the matter density at present, $\sigma(R,z)$ is the root mean square density fluctuation defined by Equation~\eqref{eq:sigma2_definition}, $\mathrm{d}\,V/\mathrm{d}\,z\mathrm{d}\,\mathcal{H}$ is the differential comoving volume per redshift per steradian and for $f(\sigma)$ we adopt the multiplicity function introduced by \citet{Tinker2008}, with its parameters assumed to be fixed and universal in the considered models. In the case of the X-ray cluster analysis developed in \citetalias{Ghirardini2023}, the astrophysical quantities used in this study are the observed X-ray count-rates $\widehat{C_\mathrm{R}}$, the observed cluster richness $\widehat{\lambda}$ computed from the DESI Legacy Survey Data Release 10 \citep{Dey2019} and detailed in \cite{Kluge2023}, and the weak lensing shear profiles $\widehat{g_\mathrm{t}}$ of clusters that belong in the overlapping KIDS, DES, and HSC regions (see Section~\ref{sec:erass1}).
The observed number density of galaxy clusters is obtained through
\begin{equation}
    \label{eq:observed_number_density}
    \frac{\mathrm{d}\,\hat n}{\mathrm{d}\widehat{C_\mathrm{R}}\mathrm{d}\widehat{\lambda}\mathrm{d}\widehat{g_\mathrm{t}}\mathrm{d}\widehat{z}\mathrm{d}\mathcal{H}} = \int\frac{\mathrm{d}\,n}{\mathrm{d}\mathrm{ln} M\mathrm{d}z\mathrm{d}\mathcal{H}}\mathcal{P}(\widehat X|M,z)\mathcal{P}(I|\widehat X)\, \mathrm{d}\mathrm{ln} M \mathrm{d}z,
\end{equation}
where $\widehat X = (\widehat{C_\mathrm{R}},\widehat{\lambda},\widehat{z},\widehat{g_\mathrm{t}}, \mathcal{H})$ is the vector of the astrophysical observable quantities. $\mathcal{P}(\widehat X|M,z)$ is the probability distribution function related to the observables at a given mass and redshift. Following \citetalias{Ghirardini2023}, the observed cluster count (combined with a mixture model for AGN contaminants and background fluctuations) is modeled through a Poisson likelihood.

This work explores a class of effective and redshift-dependent modifications of the linear matter power spectrum. We, therefore, assume that the multiplicity function function (Equation~\ref{eq:halo_mass_function}) parameterization remains valid in the considered models. This is justified as the applied modification is only redshift-dependent. In practice, we are changing the redshift dynamic of the collapse while keeping the critical overdensity for collapse $\delta_c$ to the $\Lambda\mathrm{CDM}$ value. This method thus impacts the modeling of the non-linear scales.
However, quantities entering the computation of the halo number density like $\sigma(R,z)$ are computed from the linear matter power spectrum in the case of cluster count. Thus, we cannot consider scale-dependant changes of the matter power spectrum in the non-linear regime, like the one presented in \cite{Amon2022} and \cite{Preston2023}, without risking our parameterization of $f(\sigma)$ to be invalid. However, both methods are related, impacting the physics of non-linear structure formation.

\subsection{Growth rate parameterization}
\label{subsec:growth_rate_parameterization}
\begin{figure*}
   \resizebox{\hsize}{!}
   {\begin{minipage}{.35\textwidth}
  \centering
  \includegraphics[width=\linewidth]{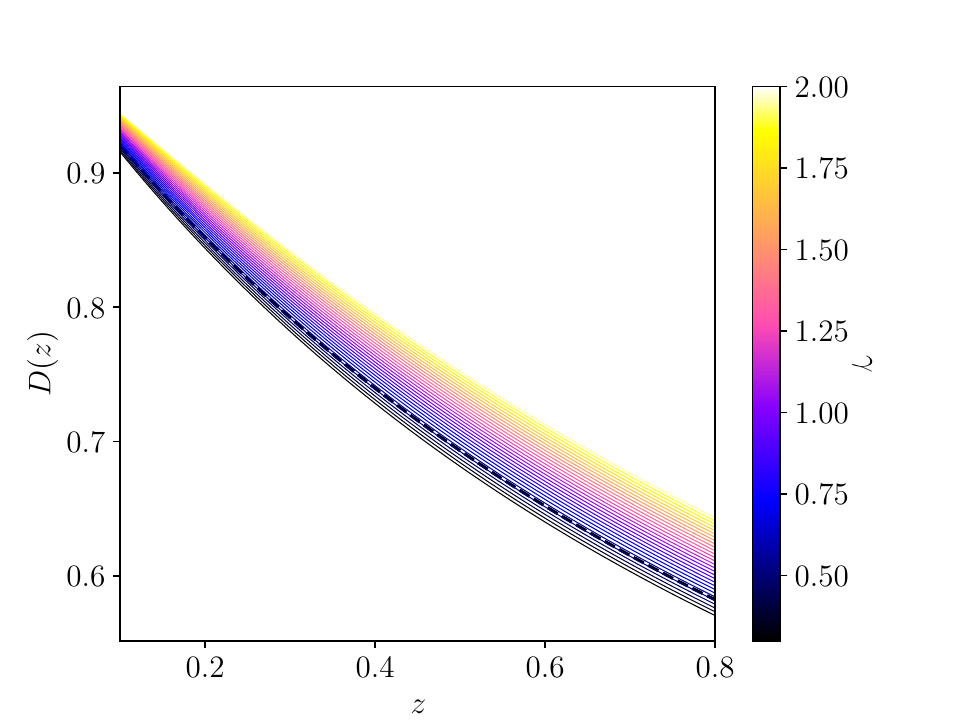}
\end{minipage}
\begin{minipage}{.35\textwidth}
  \centering
  \includegraphics[width=\linewidth]{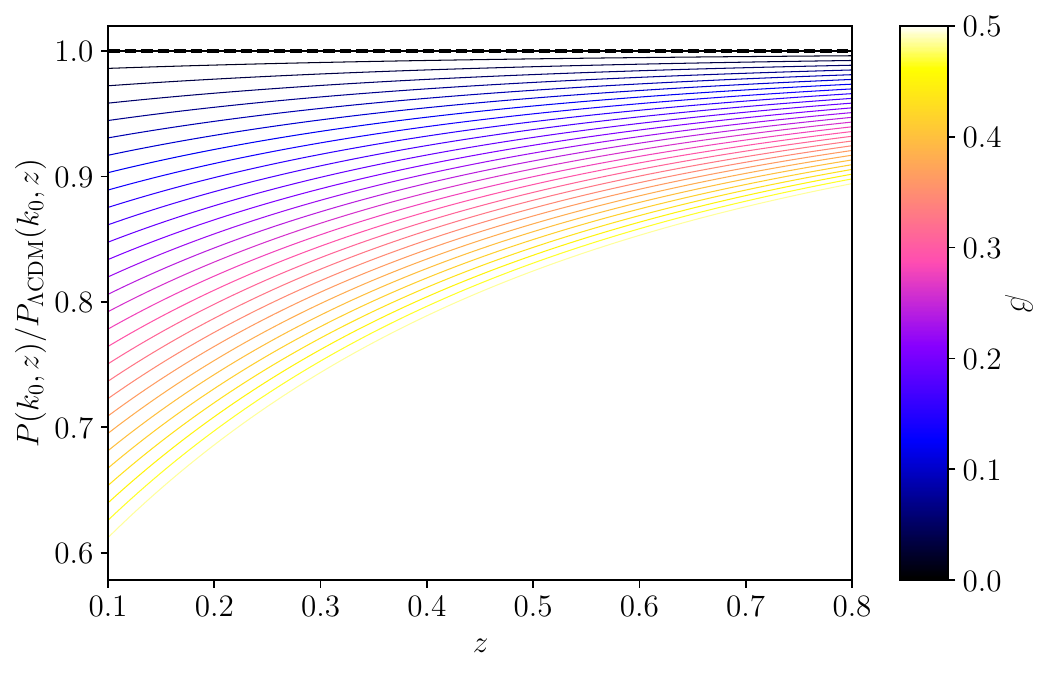}
\end{minipage}
}
      \caption{Evolution of the growth factor with the cosmic linear growth index $\gamma$ is shown in the left panel. The bold black dashed line is the growth factor in Equation~\eqref{eq:growth_factor_stand}. Increasing $\gamma$ reduces the slope of the growth factor, thus damping structure formation. The right panel shows the power spectrum reduction model described in Section~\ref{sec:power_suppression}. We show the ratio of the modified power spectrum to the regular $\mathrm{\Lambda CDM}$ one for different values of the parameter $\beta$. The index is fixed to $p=1$.  Increased values of $\beta$ reduce the structure formation at low redshift. These figures were produces using a \citetalias{PlanckCollaboration2020} cosmology (TT,TE,EE+lowE+lensing+BAO)}
    \label{fig:growth_comp}
\end{figure*}
A standard method of probing structure formation physics is to consider the cosmic linear growth index. Initially, it was introduced in the context of observational cosmology as a test for departures from general relativity at cosmological scales \cite{Nesseris2008}. 
Using classical gravity\footnote{Relativistic corrections on scales relevant to galaxy cluster formation and evolution are negligible. Thus, Newtonian gravity is an excellent approximation to GR \citep{Hwang2006}.}, the linear theory of the growth of the density field perturbations in a spatially flat universe predicts that the density contrast $\delta(\boldsymbol{x},t)$ is,

\begin{equation}
    \label{eq:density_constrast}
    \frac{\partial^2 \delta(\boldsymbol{x},t)}{\partial t^2} + 2H(t)\frac{\partial\delta(\boldsymbol{x},t)}{\partial t} = 4\pi G \rho_\mathrm{m}(t)\delta(\boldsymbol{x},t),
\end{equation}

\noindent where $H(t)$ is the Hubble parameter at time $t$, $G$ is the universal gravitational constant, and $\rho_\mathrm{m}(t)$ is the mean matter density at time $t$. We highlight that the derivatives in this equation and its coefficients depend only on the time. Thus, the solutions are the sum of two products of a spatial function and a time dependence. The growing mode of the time dependence is noted with $D_+(t)$. We follow the formalism of \citet{Rapetti2009}. The linear growth rate is defined as

\begin{equation}
    \label{eq:growth_rate_def}
    f(a) = \frac{\mathrm d \ln D}{\mathrm d \ln a},
\end{equation}

\noindent where $a\equiv  1/(1+z)$ is the time-dependent scale factor, and $D$ is the growth factor (equation~\ref{eq:norm_growth}). This function can be parameterized as a power-law of the matter density parameter $\Omega_\mathrm{m}(z)$ \citep[e.g.][]{Wang1998}. $f(z)$ is thus expressed as
\begin{equation}
    \label{eq:growth_gr}
    f(z)=\Omega_\mathrm{m}(z)^\gamma,
\end{equation}
where $\gamma$ is the cosmic linear growth index. In general, $\gamma\equiv\gamma(z)$ can be expressed as a function of redshift \citep{Linder2007, Batista2014}. In the literature, various methods are used to characterize the redshift evolution of the growth index, including linear analytical relation with redshift \citep[e.g.][]{Polarski2008}. Many theoretical frameworks also imply an evolving growth index over cosmic time, including dark energy models in GR \citep{Polarski2016}. However, most of these theories predict that the cosmic linear growth index is quasi-constant between $z=0$ and the beginning of the matter-dominated era at $z\sim 3000$. As cluster counts experiments probe redshift well after this epoch, we can safely assume a constant growth index of $\gamma(z)\equiv\gamma$.

In the absence of neutrinos (see Section \ref{sec:discussion}), which would introduce a scale dependence, we can express the growth factor $D_+(z)$ as

\begin{equation}
\label{eq:growth_factor}
 D_+(z) = \frac{\delta(z)}{\delta(z_\mathrm{init})} = \exp\left(\int_{a_\mathrm{init}}^a f(a')\;\mathrm{d}\ln a'\right)
\end{equation}

\noindent which can be expressed for a standard GR + $\Lambda\mathrm{CDM}$ cosmology as (up to a proportionality constant)

\begin{equation}
    \label{eq:growth_factor_stand}
    D_+(z) = \frac{5\Omega_\mathrm{m}}{2}\frac{H(z)}{H_\mathrm{0}}\int_z^{z_\mathrm{init}} \frac{(1+z')}{[H(z')/H_\mathrm{0}]^3} \;\mathrm{d}z'.
\end{equation}

\noindent Figure~\ref{fig:growth_comp} illustrates how this expression differs from the $\gamma$-parametrization. In this case, Equation~\eqref{eq:growth_factor}, with the linear growth rate provided by Equation~\eqref{eq:growth_gr}, should be used. We can then define the normalized growth as 

\begin{equation}
    \label{eq:norm_growth}
    D(z) = \frac{D_+(z)}{D_+(z=0)},
\end{equation}
and we thus have the usual power spectrum evolution 

\begin{equation}
    \label{eq:power_evol}
    P(k,z) = P(k,z=0)D^2(z),
\end{equation}

\noindent which in turn yields the standard root mean square (r.m.s.) density fluctuation definition:
\begin{align}
    \label{eq:sigma2_definition}
    \sigma^2(R, z) &= \frac{1}{2 \pi^2}  \int_0^\infty k^2 P(k, z) \left( \frac{3 j_1(k R)}{k R} \right)^2 \mathrm{d}k \nonumber \\
    &= \sigma^2(R, z=0) D^2(z),
\end{align}

\noindent where $j_1(x)=(\sin(x)-x\cos(x))/x^2$ is the spherical Bessel function of the first kind of order one. \citet{Wang1998} reports that for the case of a slowly varying dark energy equation of state, $\gamma \approx 6/11\sim 0.55$ for the standard linear perturbation theory approach within the context of GR, with a weak dependence on dark energy equation of state $w$,

\begin{align}
\label{eq:alpha_f}
  \gamma= & \frac{6-3(1+w)}{11-6(1+w)} + \nonumber\\
     & \frac{3}{125}\frac{(1-w)(1-3w/2)}{(1-6w/5)^3}(1-\Omega_m)+{\cal O}((1-\Omega_m)^2).
\end{align}

\noindent Different dark energy equation-of-state parameters correspond to varying growth indices. For example, $w=-1.5$ yields $\gamma \approx 0.54$ and $w=-0.5$ yields $\gamma \approx 0.57$. Consequently, departures larger than a few percent from this value of the growth index $\gamma_\mathrm{GR} = 0.55$ may be a hint for direct measurement of potential deviations from GR. 
Although Equation~\eqref{eq:alpha_f} suggests a degeneracy, we independently fit for $\gamma$ and $w$ in the rest of the paper (see Section~\ref{subsec:gamma_wcdm} for the constraints from the data).

\section{Constraints on the structure growth through the $\gamma$ parametrization}
\label{sec:res}
\begin{table*}[h]
 \caption{Definition of the parameters and priors used for the analysis for all the different parameterizations and models considered. $\mathcal{U}({\rm min}, {\rm max})$ indicates a uniform distribution between `min' and `max' and $\mathcal{N}(\mu, \sigma^2)$ indicates a normal distribution centered on $\mu$ and with standard deviation $\sigma$. We only represent the sets of parameters for which differences exist between this work and \citetalias{Ghirardini2023}. We do not represent the additional parameters related to the uncertainties on the weak lensing mass calibration, the redshift uncertainties, and the contamination modeling. See \citetalias{Ghirardini2023} for the details on these parameters.}
 \label{tab:parameters_priors}
 \centering
 \begin{tabular}{lll}
 \hline
 \hline
 \multicolumn{2}{l}{Cosmological parameters} & Priors \\
 \hline
 \hline
 $\Omega_{\mathrm{m}}$  & Mean matter density at present time & $\mathcal{U}(0.05, 0.95)$\\
 && + $\theta^*$ priors (See Appendix~\ref{app:sound_horizon}) \\
 $\log_{10} A_\mathrm{s}$  & Amplitude of the primordial power spectrum & $\mathcal{U}(-10, -8)$ \\
 $H_0$  & Hubble expansion rate ($\mathrm{km}\mathrm{s}^{-1}\mathrm{Mpc}^{-1}$) at present time & $\mathcal{N}(70, 5^2)$ \\
 $\Omega_{\mathrm{b}}$  & Mean baryon density at present time & $\mathcal{U}(0.046, 0.052)$ \\
 $n_\mathrm{s}$  & Spectra index of the primordial power spectrum & $\mathcal{U}(0.92, 1.0)$ \\
 $w$  & Dark energy equation of state. \textcolor{blue}{Fixed to -1 in $\Lambda$CDM} & $\mathcal{U}(-2.5, -0.33)$ \\
 \hline
 \hline
 \multicolumn{2}{l}{Power spectrum suppression} & Priors \\
 \hline
 \hline
 $\gamma$ & Cosmic linear growth index & $\mathcal{U}(0.05, 0.95)$ \\
 $\beta$  & Power spectrum reduction parameter & $\mathcal{U}(-5,5)$ \\
 $p$  & Power spectrum reduction index & Fixed to $1$ \\
 \hline
 \hline
 \multicolumn{2}{l}{X-ray scaling relation} & Priors \\
 \hline
 \hline
 $A_\mathrm{X}$  & Normalization of the $M - C_R$ scaling relation & $\mathcal{U}(0.01, 16)$ \\
 $B_\mathrm{X}$  & Mass slope of the $M - C_R$ scaling relation & $\mathcal{U}(0.1, 5)$ \\
 $D_\mathrm{X}$  & Luminosity distance evolution of the $M - C_R$ scaling relation & Fixed to -2 \\
 $E_\mathrm{X}$  & Scale factor evolution of the $M - C_R$ scaling relation & Fixed to 2 \\
 $F_\mathrm{X}$  & Redshift evolution of the mass slope of the $M - C_R$ scaling relation & $\mathcal{U}(-10, 10)$ \\
 $G_\mathrm{X}$  & Redshift evolution of the normalization of the $M - C_R$ scaling relation & $\mathcal{U}(-10, 10)$ \\
 $\sigma_\mathrm{X}$  & Intrinsic scatter of the $M - C_R$ scaling relation & $\mathcal{U}(0.01, 8)$ \\
 \hline
 \hline
 \multicolumn{2}{l}{Richness mass calibration} & Priors \\
 \hline
 \hline
 $ \ln A_{\lambda}$  & Normalization of the $M - \lambda$ scaling relation & $\mathcal{U}(1, 6)$ \\
 $B_{\lambda}$  & Mass slope of the $M - \lambda$ scaling relation & $\mathcal{U}(0, 2)$ \\
 $C_{\lambda}$  & Redshift evolution of the normalization of the $M - \lambda$ scaling relation & $\mathcal{U}(-10, 10)$ \\
 $D_{\lambda}$  & Redshift evolution of the mass slope of the $M - \lambda$ scaling relation & $\mathcal{U}(-10, 10)$ \\
 $\sigma_{\lambda}$  & Intrinsic scatter of the $M - \lambda$ scaling relation & $\mathcal{U}(0.01, 5)$ \\
 $\rho_{\lambda, C_R}$  & Intrinsic correlation between richness and count rate & $\mathcal{U}(-0.9, 0.9)$ \\
 \hline
 \end{tabular}
\end{table*}

In this work, we aim to constrain the growth of structure utilizing the eRASS1 cosmology sample with weak lensing mass calibration from the DES, KIDS, and HSC surveys. The results presented here are obtained by combining cluster counts with priors from the sound horizon scale at recombination ($\theta^*$) indicated by the CMB measurements presented in \citetalias{PlanckCollaboration2020}, as this represents one of the most robust cosmological measurements to date. Details on the sound horizon priors used can be found in Appendix~\ref{app:sound_horizon}.
Additionally, differently than the \citetalias{Ghirardini2023} framework, we use a conservative $H_\mathrm{0}\sim\mathcal{N}(70,5^2)$. A comprehensive description of the different priors can be found in Table~\ref{tab:parameters_priors}.

Here, we present constraints on the cosmic linear growth index for the parametrizations described in Section~\ref{sec:mg_framwork}. In particular, we focus on two different models: the canonical $\Lambda\mathrm{CDM}$ model with the cosmic linear growth index parameterized as in Equation~\eqref{eq:growth_gr} ($\gamma\Lambda\mathrm{CDM}$ model) and the $w\mathrm{CDM}$ model with the dark energy equation of state parameter, $w$, free ($\gamma w\mathrm{CDM}$ model). The standard $\gamma\Lambda\mathrm{CDM}$ is shown in Figure~\ref{fig:gamma_lcdm_all}.
Neutrinos suppress the amplitude of the power spectrum, introducing a scale dependence of the growth factor (Equation~\ref{eq:growth_factor}), which leaves imprints on the halo mass function. Although \cite{Costanzi2013} presents a method to account for the effect of neutrinos on the halo mass function by replacing the total matter power spectrum with the cold dark matter power spectrum, this method is incompatible with the growth parameterization using the cosmic linear growth index employed in this work. Therefore, we do not consider the impact of neutrinos and assume that neutrinos are massless throughout this work. At our current statistical constraining power, this modeling choice does not affect our results, as shown in section \ref{sec:discussion}. 

We also emphasize that a higher value of the cosmic linear growth index leads to a flattening of the growth factor, which in turn yields a lower measurement of $\sigma_\mathrm{8}\equiv\sigma_\mathrm{8}(z=0)$. This definition does not imply that we expect a lower $\sigma_\mathrm{8}(z)$ for all redshifts, as found with probes that include non-linear physics \citep{Abbott2022, White2022}.

\subsection{Constraints on the $\gamma\Lambda\mathrm{CDM}$ Model}
\label{subsec:gamma_lcdm}
\begin{figure}
    \centering
    \includegraphics[scale=0.6]{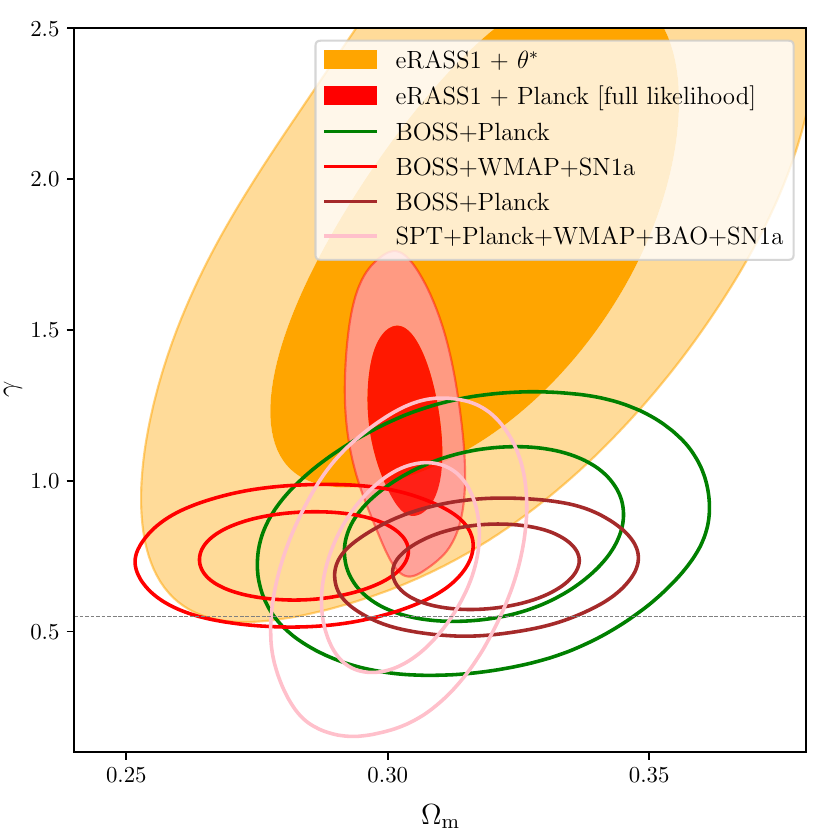}
    \caption{Constraints on the growth index $\gamma$ for the $\Lambda$CDM cosmological model. We present the confidence ellipses obtained in the $\Omega_\mathrm{m}-\gamma$ plane with eRASS1 clusters combined with sound horizon priors (in orange) and with the full likelihood from \citetalias{PlanckCollaboration2020} (in red).  
     The gray dashed line represents the prediction of  $\gamma=6/11$ from general relativity. The contours are the 68 and 95\% confidence levels. The eRASS1 contours are compared with the galaxy clustering constraints from the Baryon Oscillation Spectroscopic Survey (BOSS) combined with CMB from WMAP and SNIa data (\cite{Samushia2013} in red), or combined with \cite{Planckcollaboration2016b} \citep{Beutler2014, Gil-Marin2017}. We also show the SPT cluster counts results combined with BAO, SN, and CMB data \citep{Bocquet2015} in pink.
      }
    \label{fig:gamma_lcdm_all}
\end{figure}

\begin{figure*}
     \centering
     \includegraphics[width=\textwidth]{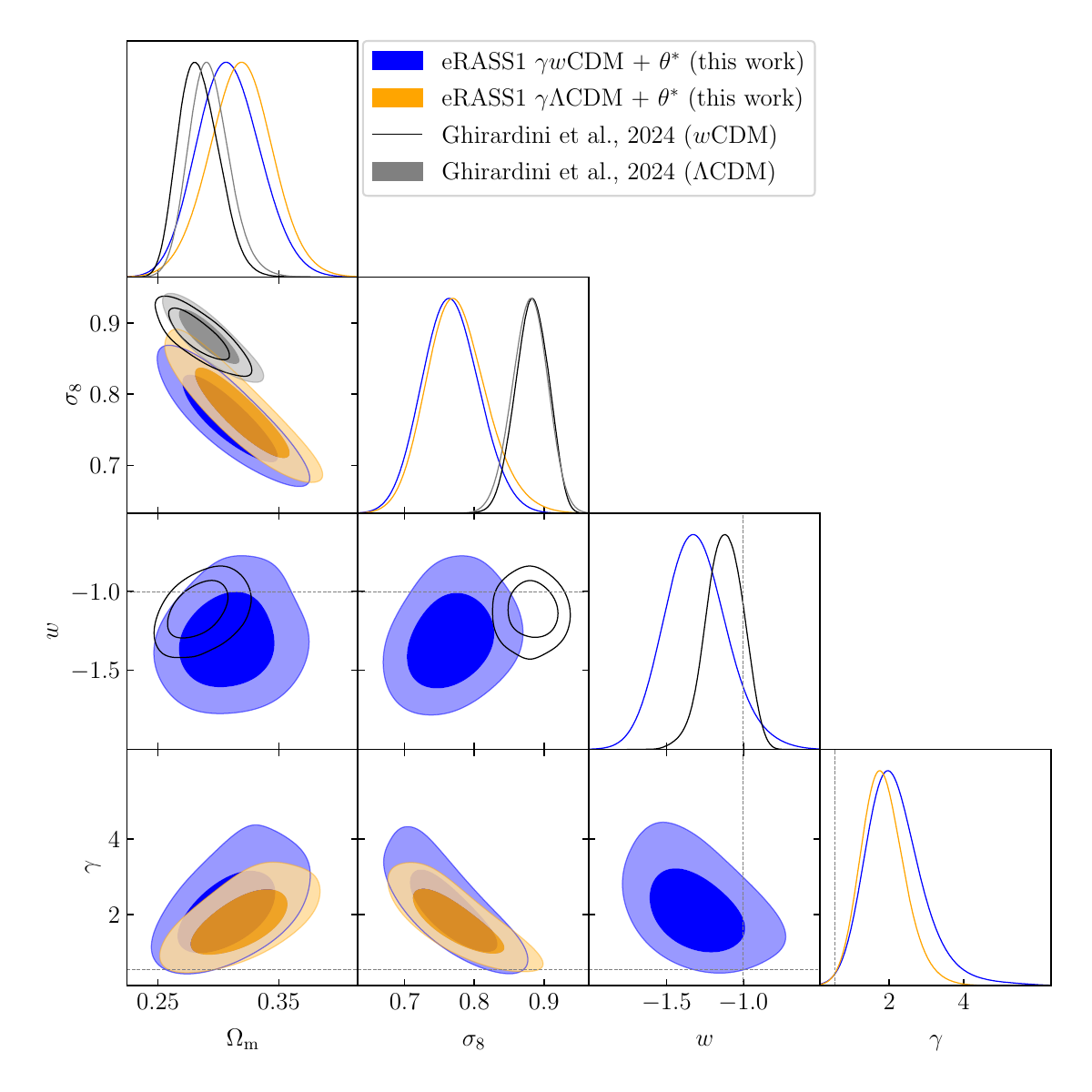}
     \caption{Comparison of the posteriors of cosmological parameters for the $\gamma\Lambda\mathrm{CDM}$ model shown in orange, and $\gamma w\mathrm{CDM}$ shown in blue. The constraint from the standard $\mathrm{\Lambda CDM}$ model presented in \citetalias{Ghirardini2023} ($\Lambda\mathrm{CDM}$ in grey, and $w\mathrm{CDM}$ in black) are plotted for comparison. The $\gamma\Lambda\mathrm{CDM}$ model prefers a lower value of $\sigma_\mathrm{8}$ and a higher value for $\Omega_\mathrm{m}$. This is explained by the anti-correlation between $\sigma_\mathrm{8}$ and $\gamma$, and the positive correlation between $\Omega_\mathrm{m}$ and $\gamma$.} 
     \label{fig:comp_main_cosmo}
 \end{figure*}

We first fit the cluster number density to the \erass\ cosmology sample with a free cosmic linear growth index $\gamma$ in the $\Lambda\mathrm{CDM}$ model. In the case of cluster counts combined with sound horizon priors, we find

\begin{equation}
    \label{eq:cosmo_glcdm_erass1}
    \begin{array}{lcl}
     \Omega_\mathrm{m} &= & 0.32 \pm 0.02  \\
     \sigma_\mathrm{8} &= & 0.77 \pm 0.04 \\
     \Sh & = & 0.79 \pm 0.02\\
     \gamma & = &1.77\pm0.55 \\
     \log A_\mathrm{s}  & = & -8.75\pm0.06 
\end{array} ,
\end{equation}

\noindent at the 68\% confidence level. We compare our results in the baseline concordance $\gamma\Lambda\mathrm{CDM}$ cosmology with the primary \erass $\Lambda\mathrm{CDM}$ cosmology analysis presented in \citetalias{Ghirardini2023}, shown in Figure~\ref{fig:comp_main_cosmo}. Unlike the case of galaxy clustering, cluster counts exhibit a clear anti-correlation between $A_\mathrm{S}$, the amplitude of the primordial power spectrum, and $\gamma$, the cosmic linear growth index. This is explained in Figure~\ref{fig:growth_comp}. As such, increasing $\gamma$ leads to a flattening of $\sigma_\mathrm{8}(z)$, thus increasing the number of high-redshift clusters for a fixed amplitude.
Freeing the linear growth index, $\gamma$, decreases the value of $\sigma_\mathrm{8}$ measured by cluster number counts, based on the following equation.

\begin{equation}
 \label{eq:sigma_8}
  \sigma_8^2 = \int_0^\infty A_\mathrm{s}k^{n_\mathrm{s}} T^2(k) \left( \frac{3 j_1(kR)}{kR} \right)^2 \mathrm{d}\,k , 
\end{equation}
where $T(k)$ is the matter transfer function. This effect is significantly reflected in a decrease in the $ A_{s}$ value measured by clusters to the abovementioned constraint. 

Overall, although eRASS1 cluster counts tend to prefer higher values of $\gamma$, which could hint at a weakening of gravity at low redshifts ($0.1<z<0.8$) considered in this work, our posteriors are in broad agreement with the value expected from GR ($\gamma = 0.55$) in statistical uncertainties and consistent with the results from galaxy clustering and cluster counts in the literature when combined with other probes (see the contours in orange in Figure~\ref{fig:gamma_lcdm_all} for comparison). The corner plot of the main cosmological parameters constrained with our cluster number counts is shown in Appendix~\ref{app:glcdm}. The primary conclusion of this analysis is that adding a degree of freedom to the growth of structure, in this case, through the cosmic linear growth index $\gamma$, decreases the value of $\sigma_{8}$, and therefore $\Sh$, constrained by cluster number counts.

In these fits, as explicitly demonstrated in the corner plots in Appendix~\ref{app:glcdm}, the cosmic linear growth index exhibits a strong degeneracy with $\sigma_\mathrm{8}$ and $\Omega_\mathrm{m}$. It requires robust external priors to be constrained. To break this degeneracy, we combine our eRASS1 measurements with the \citetalias{PlanckCollaboration2020}'s CMB measurements, specifically TTTEEE+lowl+lowE. This time, we are considering the full likelihood and not only the sound horizon measurements. In this case, we find

\begin{equation}
\label{eq:res_gammalcdm}
\gamma=1.19\pm0.2 \; \mathrm{at}\; 68\%\;\mathrm{level}\; \mathrm{confidence}.
\end{equation}

\noindent As the combination with the CMB data increases the precision, our combined results are in $3.7\sigma$ tension with the standard of $\gamma=0.55$ in the GR framework. However, it is in statistical agreement with the other probes, as shown in Figure~\ref{fig:gamma_lcdm_all}. Overall, this hints again that structure growth might undergo a late-time reduction, which only reflects on the data from the late-time probes, such as clusters of galaxies. 
This conclusion is strengthened by other probes, including other cluster count experiments, which remains consistent with $\gamma=0.55$, but skew to higher values. Indeed, \cite{Bocquet2015} used SZ-detected clusters from SPT in combination with CMB, SNIa, and BAO data to find $\gamma= 0.72\pm0.24$. In comparison, earlier constraints from \cite{Rapetti2013} report a  $\gamma$ value of $0.616 \pm 0.061$ when combining number counts from 238 clusters detected by \textit{ROSAT} with CMB data, galaxy clustering, SNIa, and BAO, as well as the cluster gas fraction. Additionally, \cite{Bocquet2015} finds $\gamma = 0.72\pm 0.24$ using a sample of 100 SPT-SZ selected clusters, where cluster abundance is combined with BAO, supernovae, and CMB data. Constraints on the cosmic linear growth index are also found to be systematically higher by galaxy clustering constraints. Combining data from the Baryon Oscillation Spectroscopic Survey (BOSS hereafter) with WMAP and SNIa data, \cite{Samushia2013} found $\gamma=0.64\pm0.05$, while combining BOSS and CMB data from \cite{Planckcollaboration2016b}, \cite{Beutler2014} found $\gamma=0.772^{+0.124}_{-0.97}$, consistently higher. Other probes also find similar results. With the combination of redshift space distortion and cosmic chronometers, \citep{Moresco2017} obtained $\gamma=0.65^{+0.05}_{-0.04}$. Using the clustering of luminous red galaxies from DES \citep{Basilakos2020} found $\gamma = 0.65\pm 0.063$. Finally, combining CMB data from
\citetalias{PlanckCollaboration2020} and weak lensing, galaxy clustering, and cosmic velocities
 \cite{Wen2023} found $\gamma=0.633^{+0.025}_{-0.024}$. These data sets favor a higher cosmic linear growth index value in all these cases. 
Our best-fit value is also higher than the one obtained in GR, which may hint that cluster number counts tend to favor suppression of structure growth at the low-$z$ regime. 
A significant finding of this section is the fact that we find a significant reduction of the $\Sh$ parameter compared to the one obtained in \citetalias{Ghirardini2023}. This result is extensively discussed in Section \ref{sec:discussion}.

\subsection{Constraints on the $\gamma w\mathrm{CDM}$ Model}
\label{subsec:gamma_wcdm}

In addition to the $\gamma$ parameter, we vary the dark energy equation of state parameter, $w$, as the most straightforward extension of the $\gamma\Lambda\mathrm{CDM}$ model. $\gamma$ and $w$ are fit as independent parameters. The constraints provided by the \erass cluster counts with the sound horizon priors produce the following best-fit results:
\begin{equation}
\label{eq:res_wgcdm}
\begin{array}{lcl}
     \Omega_\mathrm{m} &= & 0.31 \pm 0.03  \\
     \sigma_\mathrm{8} &= & 0.76 \pm 0.04 \\
     \Sh &= & 0.77 \pm 0.02 \\
     w &=& -1.32\pm 0.2\\
     \gamma &=&2.01\pm0.75 \;
\end{array} .
\end{equation}
The results of this fit are shown in Figure~\ref{fig:comp_main_cosmo}, where the uncertainties are plotted as contours at the 68\% and 95\% levels. The dark energy equation of state posterior distribution is consistent with the $\Lambda\mathrm{CDM}$ value of $w = -1$ within the statistical uncertainties. Our value also agrees with the primary eROSITA cosmology analysis presented in the \citetalias{Ghirardini2023}, which reports a best-fit value of $w = -1.12\pm0.12$ in their $w\Lambda\rm CDM$ model. 
Our results favor a higher value of $\gamma=2.01\pm0.75$, consistent with $\gamma\Lambda\rm CDM$ model, which once again hints at a modification of the growth of structures and potential damping at the redshift range considered in this work (although we remain compatible with the standard $\gamma=0.55$ at the $2\sigma$ level due to the large error bars). Appendix~\ref{app:wcdm} presents the corner plot of the main cosmological parameters constrained in this fit. We note that the dark energy equation of state parameter, $w$, is positively correlated with the normalization of the power spectrum $\log A_\mathrm{S}$. This potentially increases the uncertainties on the magnitude of the power spectrum.

\begin{figure}
    \centering
    \includegraphics[scale=0.6]{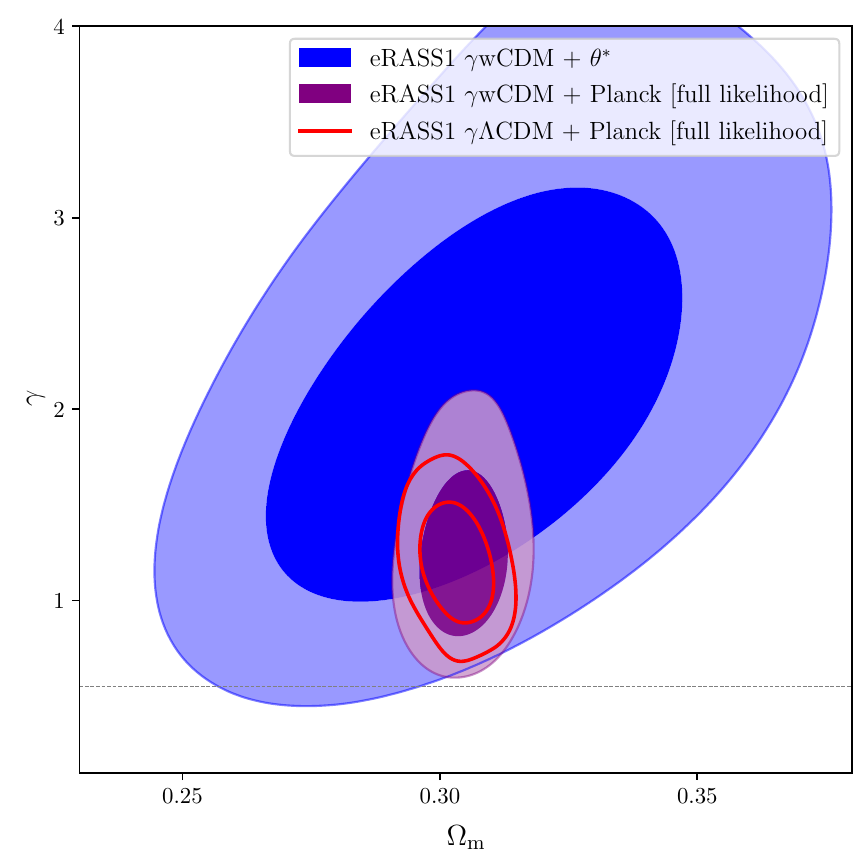}
    \caption{Constraints on the growth index $\gamma$ for a $w$CDM cosmology. Confidence ellipses are obtained in the $\Omega_\mathrm{m}-\gamma$ plane. The gray dashed line represents the GR prediction $\gamma=6/11$. The contours are the 68 and 95\% confidence levels.}
    \label{fig:gamma_wcdm_comp}
\end{figure}

We proceed by combining our cluster number density constraints with the full likelihood from the \citetalias{PlanckCollaboration2020} CMB (TT-TEEE+lowl+lowE) chains. The results are shown in Figure~\ref{fig:gamma_wcdm_comp}. We find 
\begin{equation}
\label{eq:res_gammawcdm}
\gamma=1.22\pm0.29 \;.
\end{equation}
Including the \citetalias{PlanckCollaboration2020} CMB results in our chains lowers the best-fit value of the $\gamma$ parameter similar to the $\gamma\Lambda\rm CDM$ case presented in the previous section. The result is in $3.6\sigma$ tension with the canonical value 0.55 predicted by the GR model. However, in this case, similarly to the results presented in Section~\ref{subsec:gamma_lcdm}, finding a higher value of the cosmic linear growth index is consistent with earlier works presented in the literature. When combining \textit{ROSAT} cluster counts with CMB data, galaxy clustering, SNIa, and BAO, as well as the cluster gas fraction, \cite{Rapetti2013} found $\gamma = 0.604 \pm 0.078$, favoring higher values, although the measurement remains in perfect agreement with the GR value. Additionally, \cite{Bocquet2015} found, when freeing the dark energy equation of state $\Sh = 0.73\pm0.28$.
Similarly to Section~\ref{subsec:gamma_lcdm}, the parameter $\Sh$ is reduced to a value similar to the one expected from cosmic shear surveys. The new values that appear to reconcile the linear regime measurements and the ones inferred from cosmic shear surveys are discussed in Section~\ref{sec:discussion}.

%
%\subsection{Power spectrum suppression}
\section{Constraints on the structure growth through the power spectrum suppression model}
\label{sec:power_suppression}
%

%
%\subsection{Late-time modification of structure growth}
%\label{subsec:late_time_mod}
%

    Previous sections present hints for departures from the standard structure formation scenario and potential damping in the formation of halos indicated by the eRASS1 cluster counts in the redshift range $0.1<z<0.8$. It is thus interesting to explore if these findings are consistent with other models describing a reduction of the power spectrum at low redshift. Following the results, we implement the late-time power spectrum reduction model proposed by \cite{Lin2024} as an additional effective parameterization of the observed deviations from the standard structure formation scenario. We stress that this model is a new parameterization of the modification of the linear growth factor, and we do not use the framework described in Section~\ref{sec:mg_framwork}. Here, we use the standard growth factor introduced in equation~\ref{eq:growth_factor_stand}, and thus the cosmic linear growth index $\gamma$ is irrelevant in this section. However, both models describe a reduction of structure formation as described in Section~\ref{sec:discussion}, and are thus complementary. The modified matter power spectrum is defined as 

\begin{equation}
    \label{eq:ps_reduction}
    P(k,z) = \alpha(z)P(k,z)_\mathrm{\Lambda CDM}
\end{equation}

\noindent where $P(k,z)_\mathrm{\Lambda CDM}$ is the standard linear matter power spectrum expected in $\mathrm{\Lambda CDM}$, and $\alpha(z)$ is a correction factor following

\begin{equation}
    \label{eq:ps_correct_fact}
    \alpha(z) = 1 - \beta\left(\frac{\Ode(z)}{\Ode}\right)^p ,
\end{equation}

\noindent where $\Ode(z)$ is the energy density of dark energy at a given redshift, and $\Ode$ is the same quantity as $z=0$. $\beta$ and $p$ are two constants quantifying the redshift dependence of the power modification. For $\beta>0$, a suppression of the matter power spectrum occurs at later times if $p>0$. In this work, we fix the index to $p=1$, as the most direct power spectrum reduction. 
 The $\gamma$ parameterization is linked to the parameters of the late-time suppression of the power spectrum through:
\begin{equation}
    \label{eq:power_gamma}
    \frac{\mathrm{d}\ln \alpha (a)}{\mathrm{d}\ln a} + \frac{\mathrm{d}\ln D_\mathrm{+}(a)}{\mathrm{d}\ln a} = \Om(a)^\gamma -1 , 
\end{equation}
where $\alpha$ is presented in Equation~\ref{eq:ps_reduction} and $a=1/(1+z)$ is the scale factor. The two approaches are thus related, and we expect any potential deviation from the standard scenario appearing in one to be reflected in the other.
Thus, we expect $\beta>0$ to confirm the results obtained in Section \ref{sec:res}.  
Additionally, as reported in Section~\ref{sec:intro}, there is good agreement between the $\Sh$ measurements from the primary anisotropies of the CMB ($z\sim 1100$) and the CMB lensing, which probes a redshift range of $0.5<z<5$. Consequently, a reduction of the power spectrum should be investigated in the low redshift Universe. Quantitatively, the best-fit energy density of dark energy reported in the last CMB anisotropy results of \citetalias{PlanckCollaboration2020} is $\Omega_{\Lambda,\mathrm{Planck}} =0.6834 \pm 0.0084$. This value is significantly reduced when the redshift increases.  Consequently, $\alpha(z)$ significantly differs from 1 only at low redshift, indicating that the power spectrum is left unchanged at early times. 

Fitting the eRASS1 cluster number counts with the sound horizon prior to this model, we find 
 \begin{figure}
     \centering
     \includegraphics[scale=0.55]{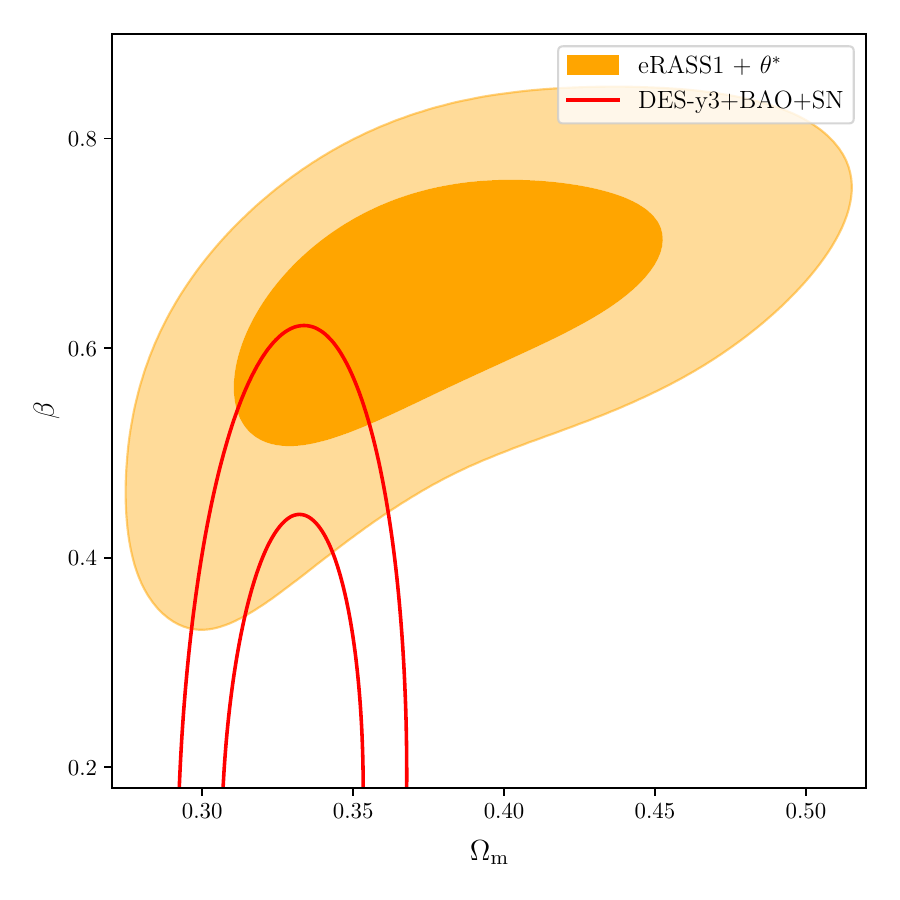}
     \caption{Constraints obtained on the power spectrum reduction model from eRASS1 cluster counts combined with the sound horizon priors from the CMB \citetalias{PlanckCollaboration2020}. These are compared with the prediction from \cite{Lin2024} when the authors used DES-Y3 data to constrain the model. The two surveys are in agreement, although eRASS1 data points to a reduction of structure formation at low redshift.} 
     \label{fig:betamodel}
 \end{figure}
\begin{equation}
\label{eq:res_betagcdm}
\begin{array}{lcl}
     \Omega_\mathrm{m} &= & 0.37 \pm 0.05  \\
     \sigma_\mathrm{8} &= & 0.66 \pm 0.08 \\
     \beta &=& 0.62\pm 0.14\\
     \Sh&=& 0.73\pm0.04 \;
\end{array} ,
\end{equation}
at 68\% confidence level, and where we fixed $p=1$. We recall that the priors on all the parameters can be found in Table~\ref{tab:parameters_priors}, and especially that the priors on $\beta$ are $\mathcal{U}(-5,5)$. Since we are measuring a non-zero $\beta$ parameter consistently with the findings of \cite{Lin2024} and Section \ref{sec:res}, these results suggest that a significant reduction of structure formation happens at low redshift. Although significant, this finding must be interpreted in cluster count cosmology, where the density of dark matter halos hosting clusters is computed through the halo mass function. The parameters of the HMF are assumed to be universal in all the models considered (see Section~\ref{subsec:cluster_counts}). The impact of the change that we apply to the power spectrum on these parameters should be further investigated in the future. Additionally, another main result is that the best-fit value of the $\Sh$ parameter decreases when we add the new degree of freedom to the power spectrum, consistent with the results presented in Section~\ref{sec:res}. The $\Sh$ measurements become consistent with the cosmic shear inferred values. For example, \citep{Li2023} reports a value of $\Sh = 0.776 ^{+0.029}_{-0.027}$ obtained from the two-point correlation functions from the HSC data. 

Galaxy clusters probe the linear regime of structure formation and involve only the linear matter power spectrum. Remarkably, adding a degree of freedom to the matter power spectrum suggests a reduction of structure formation and decreases the $\Sh$ value. This finding is discussed in Section~\ref{sec:discussion}.
Overall, the parameter $\beta$ quantifies how the power spectrum could be reduced due to unknown physical processes or systematic effects at low redshift. Our results suggest that such reduction exists, although its nature remains to be identified.

\section{Direct measurements of the structure growth through redshift evolution}
\label{sec:direct_strucutre_growth}

The frameworks and results presented in Section~\ref{sec:mg_framwork}, Section~\ref{sec:res}, and Section~\ref{sec:power_suppression} assume a functional form of the growth rate presented in Equation~\eqref{eq:growth_gr}, or the power spectrum in Equation~\eqref{eq:ps_correct_fact}. To independently probe the structure growth free of any functional parametrization and test the potential deviations from the standard structure formation scenario, we use a different method to measure the structure evolution using eRASS1 cluster counts. 

In Section~\ref{sec:res}, we assume a value of the growth index $\gamma$, we then compute the associated growth factor $D_\gamma$ with equations \ref{eq:growth_factor} and \ref{eq:norm_growth} and obtain the evolution of the r.m.s density fluctuations through $\sigma_\mathrm{8}(z)=\sigma_\mathrm{8}D_\gamma(z)$. However, the cosmic linear growth index lacks an immediate physical interpretation. Indeed, even though a redshift evolution of the cosmic linear growth index can be expected in some theoretical models \citep[][]{Linder2007, Wen2023}, the cosmic linear growth index remains an effective parameterization. The same issue exists with the power spectrum reduction model we used in Section~\ref{sec:power_suppression}, also a phenomenological parameterization. 
To avoid these issues, another method commonly used in the literature for probing the growth of structure is to divide the cluster sample into multiple bins and to compute the cosmological parameters directly in those subsamples employing the standard $\Lambda$CDM parametrization in each bin \citep[e.g.,][]{Bocquet2024}.

The fluctuations of the density field probed by clusters number count over comoving spheres with a radius of $8~h^{-1}\mathrm{Mpc}$ ($\sigma_8(z)$) in several redshift bins traces the evolution of structure growth with redshift. This is possible since cluster abundance in each redshift bin is indeed a direct probe of the halo mass function, defined in Equation~\ref{eq:halo_mass_function}. Although we can express $\sigma(R, z)$ using its value at redshift zero multiplied by the normalized growth rate, the halo mass function depends only on the value of $\sigma$ at the redshift of interest, and therefore $\sigma(R, z)$ is probed quasi-directly. In practice, the difference in growth rate modeling between the $\gamma\Lambda\mathrm{CDM}$ model and the canonical $\Lambda\mathrm{CDM}$ model with the standard GR will act just as a normalization difference in $\sigma(R, z)$, given that the growth rate is a function of redshift. This method works independently of modeling assumptions as long as the growth rate is only a function of redshift, $D = D(z)$. 
 Although this method is assumed to be model-independent compared to the $\gamma\Lambda\mathrm{CDM}$ model, it still relies on the assumptions of linear perturbation theory with a conventional growth within the corresponding redshift bins. For this reason, we consider both approaches (the $\gamma\Lambda\mathrm{CDM}$ model and the binning method in the redshift) jointly and test them using the \erass observations. In the case of damped structure growth at lower redshifts, we expect that the normalization of the power spectrum is affected, creating discrepancies between our measurement and values inferred from the CMB measurements.

\begin{figure*}
   \resizebox{\hsize}{!}
   {\begin{minipage}{.35\textwidth}
  \centering
  \includegraphics[width=\linewidth]{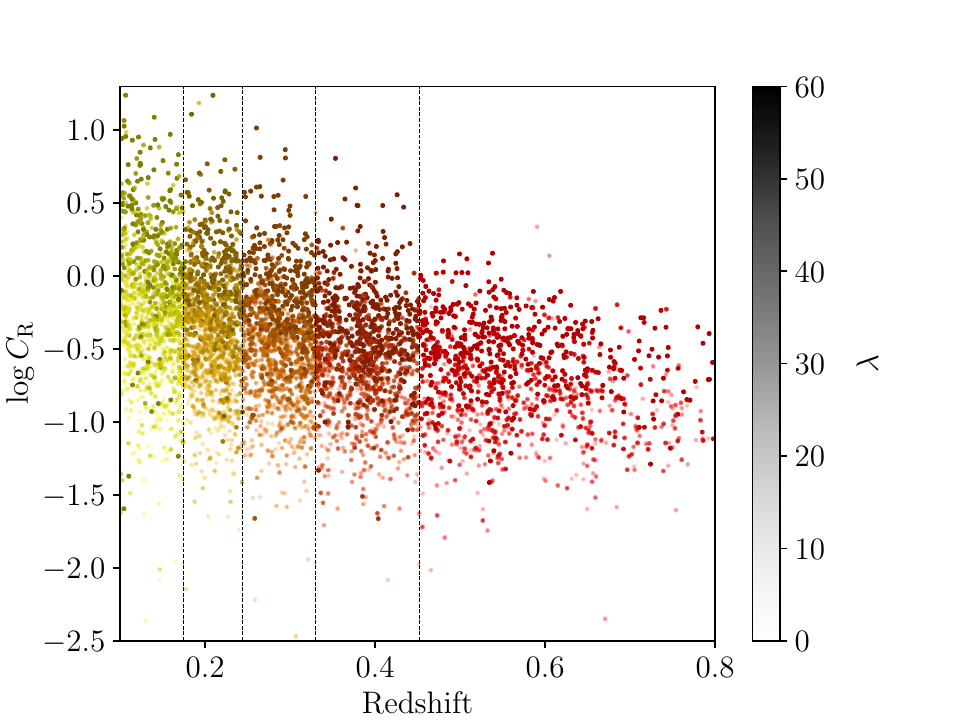}
\end{minipage}
\begin{minipage}{.35\textwidth}
  \centering
  \includegraphics[width=\linewidth]{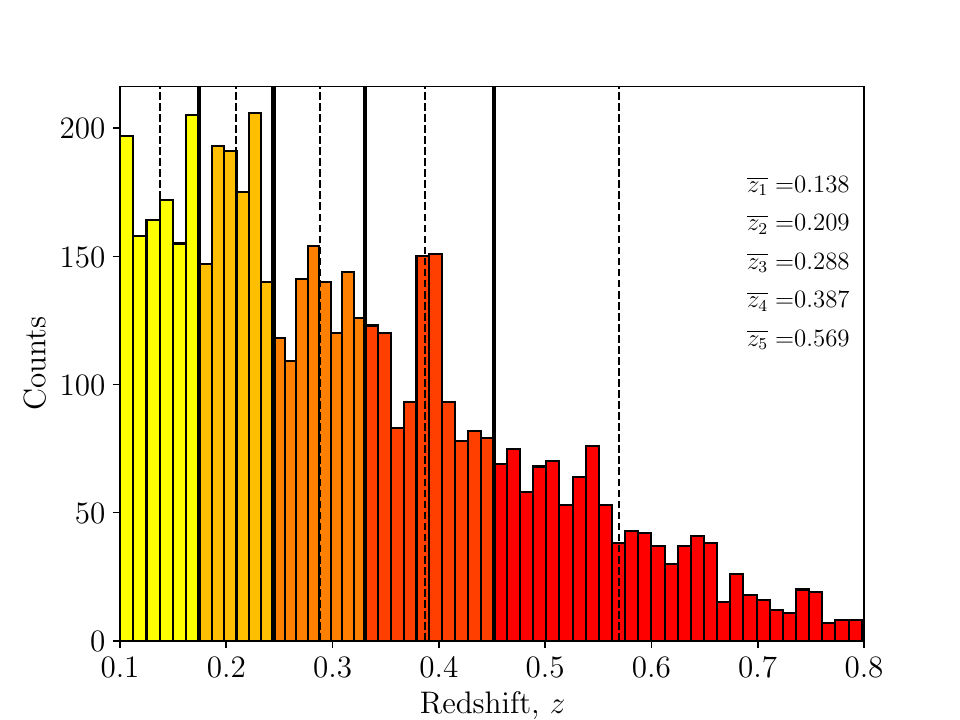}
\end{minipage}
}
      \caption{The X-ray count rate distribution as a function of redshift is shown for the redshift bins used to assess the cosmological growth of structure evolution using the same $\Lambda$CDM parametrization in each bin on the left panel. The different colors represent the separation. The intensity of the colors represents the richness. 
     The observable distribution in the five subsamples differs as expected; in particular, count rates are higher at low redshift for fixed richness. On the right is the redshift distribution of the clusters in each bin. The mean redshift of each sub-sample is represented with an overline. Although the volume probed increases with the redshift, the number of objects decreases due to the \erass selection function.}
    \label{fig:sample_bin}
\end{figure*}%

Our approach to measuring the redshift evolution in structure growth is based on binning the cluster number counts in redshift to measure the evolution of structure growth. The eRASS1 cosmology sample, comprising 5,259 ICM-selected and optically confirmed clusters of galaxies, is the largest ICM-selected sample to date. The cosmological parameter inference reaches a statistical precision similar to CMB experiments (see \citetalias{Ghirardini2023}).  We determine the number of clusters in each bin such that the matter density parameter $\Omega_\mathrm{m}$ is measured with at least $\sim$15\% precision and has a similar number of clusters. With this requirement, the sample provides sufficient statistical power to generate five independent\footnote{Although the bins are never totally uncorrelated, they are large enough so that we can neglect the covariance.} redshift bins (see Table \ref{tab:results_in_bins}).  The first four bins have 1,052 clusters each with this binning scheme, leaving 1,051 clusters for the last highest redshift bin ($0.452<z<0.8$). The corresponding redshift ranges are  $z = 0.1-0.175, 0.175-0.244, 0.244-0.330, 0.330-0.452$, and $0.452- 0.8$. The X-ray count-rate distribution as a function of redshift and richness of the subsamples are displayed in Figure~\ref{fig:sample_bin}. 

\begin{figure}
    \centering
    \includegraphics[scale=0.9]{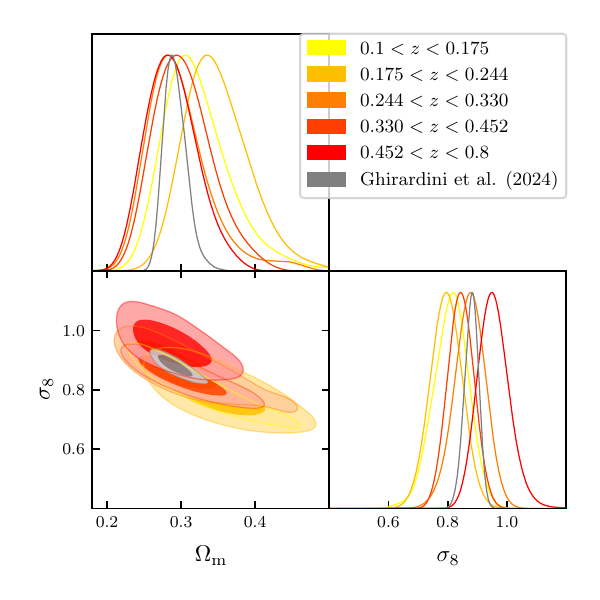}
    \caption{Cosmological posteriors obtained on the main parameters. The color scheme follows the one in Figure~\ref{fig:sample_bin}: blue ($z = 0.1-0.175$), red ($0.175-0.244$), black ($z=0.244-0.330$), green ($z=0.330-0.452$) and purple ($z=0.452- 0.8$). All the posteriors are fully compatible at the $1\sigma$ level, showing the robustness of the eRASS1 analysis. }
    \label{fig:bin_om_s8}
\end{figure}

Our sample has sufficient statistical constraining power to measure the scaling relation parameters independently in each redshift bin jointly with the cosmological parameters shown in Table~\ref{tab:parameters_priors}, unlike the previously published ICM-selected samples \citep[e.g.][]{Bocquet2024}. Consequently, we fix the pivot redshift $z_\mathrm{p}$ of the count-rate to mass ($C_\mathrm{R}-M$) and richness to mass ($\lambda-M$) scaling relations as described in \citetalias{Ghirardini2023} to the mean redshift in each bin $z_{\mathrm{p},i} = 0.138, 0.209, 0.288, 0.387, 0.569$. 
Particularly relevant measurements are the normalization of the power spectrum $A_{\mathrm{s},i}$, and the matter density $\Omega_{\mathrm{m},i}$. These parameters are combined to recover the $\sigma_{8,i}$ parameter in the redshift bin space specified above. We utilize the Boltzmann solver \texttt{CAMB} \citep{Lewis2011CAMB} to compute the power spectrum for each redshift bin and infer the cosmological parameters. This process additionally provides a consistency check for our standard cosmology analysis presented in \citetalias{Ghirardini2023}, also shown in Figures~\ref{fig:bin_om_s8} and ~\ref{fig:s8_z}. For instance, our posteriors on the cosmological parameters, $\Omega_M$, $\sigma_8$, and $S_\mathrm{8}$, are consistent with each other within the five redshift bins, and the results in \citetalias{Ghirardini2023} at the $2\sigma$ confidence level.  Our results are summarized in Table~\ref{tab:results_in_bins}. 

\begin{table*}
\caption{Constraints on the cosmological parameters obtained in the different bins. The confidence intervals presented are the $1\sigma$ errors. The differences in confidence intervals are due to the redshift and mass distributions of clusters in each bin. Since we used different pivot redshifts for each bin, the normalization of the scaling relation is not consistently measured. We thus introduce the rescaled amplitude parameter  $\ln A = \ln A_\mathrm{X} + F_\mathrm{X}\ln(1+z_{\mathrm{p},i})\ln(2) +2\ln(D_\mathrm{L}(z_{\mathrm{p},i}))-2\ln(E_\mathrm{X}(z_{\mathrm{p},i}))-G_\mathrm{X}\ln(1+z_{\mathrm{p},i})$, where $z_{\mathrm{p},i}$ represent the pivot redshift in each redshift bin.}
\label{tab:results_in_bins}
\begin{center}

\begin{tabular}[width=0.5\textwidth]{c| c c c c c c}
\hline\hline 
            & \cite{Ghirardini2023} & $z = 0.1-0.175$ & $0.175-0.244$ & $z=0.244-0.330$ & $z=0.330-0.452$ & $z=0.452- 0.8$ \\

\hline 
    $\Omega_\mathrm{m}$    & $0.29\pm0.02$ & $0.31\pm 0.05$ & $0.34\pm 0.05$ & $0.28\pm 0.04$ & $0.30 \pm 0.04$ & $0.29\pm 0.04$ \\
    $\sigma_\mathrm{8}$ & $0.88\pm 0.02$ & $0.82\pm0.06$ & $0.80\pm 0.08$ & $0.89\pm 0.07$ & $0.84\pm 0.05$ & $0.95\pm 0.05$\\
    $S_\mathrm{8}$ & $0.86\pm 0.01$ & $0.83\pm 0.02$ & $0.85\pm 0.07$ & $0.86\pm 0.06$ & $0.84\pm 0.03$ & $0.94\pm 0.04$\\
    $A_\mathrm{X}$ & $0.64\pm 0.05 $ & $5.33\pm 0.99 $ & $2.27\pm0.46 $ & $1.52\pm0.29 $ & $0.93\pm0.18 $ & $0.18\pm 0.32 $\\
    $B_\mathrm{X}$ & $1.38\pm 0.03 $ & $1.51\pm 0.07 $ & $1.51\pm0.08 $ & $1.45\pm0.07 $ & $1.52\pm0.08 $ & $1.35\pm0.08 $\\
    $F_\mathrm{X} $ & $-0.3\pm 0.1$ & $-1.7\pm1.9 $ & $-0.8\pm 2.3$ & $-1.7\pm1.5 $ & $-0.7\pm1.4 $ & $0.14\pm 0.6 $\\
    $G_\mathrm{X} $ & $0.29\pm 0.12 $ & $1.3\pm 1.70 $ & $2.7\pm1.60 $ & $2.5\pm 1.00$ & $1.0\pm0.91 $ & $-0.1\pm0.40 $\\
    $\ln A $ & $14.1\pm 0.08$ & $14.2\pm 0.23$ & $13.9\pm 0.37$ & $13.8\pm0.35 $ & $14.3\pm0.51 $ & $14.0\pm0.32 $\\
    $\sigma_\mathrm{X}$ & $  0.98 \pm 0.04 $ & $ 0.72\pm0.09 $ & $0.74\pm 0.12$ & $0.67\pm 0.13 $ & $0.60\pm0.11 $ & $0.97\pm0.06 $\\
\hline\hline
\end{tabular}

\end{center}
\end{table*}
\begin{figure*}
   \includegraphics[width=\linewidth]{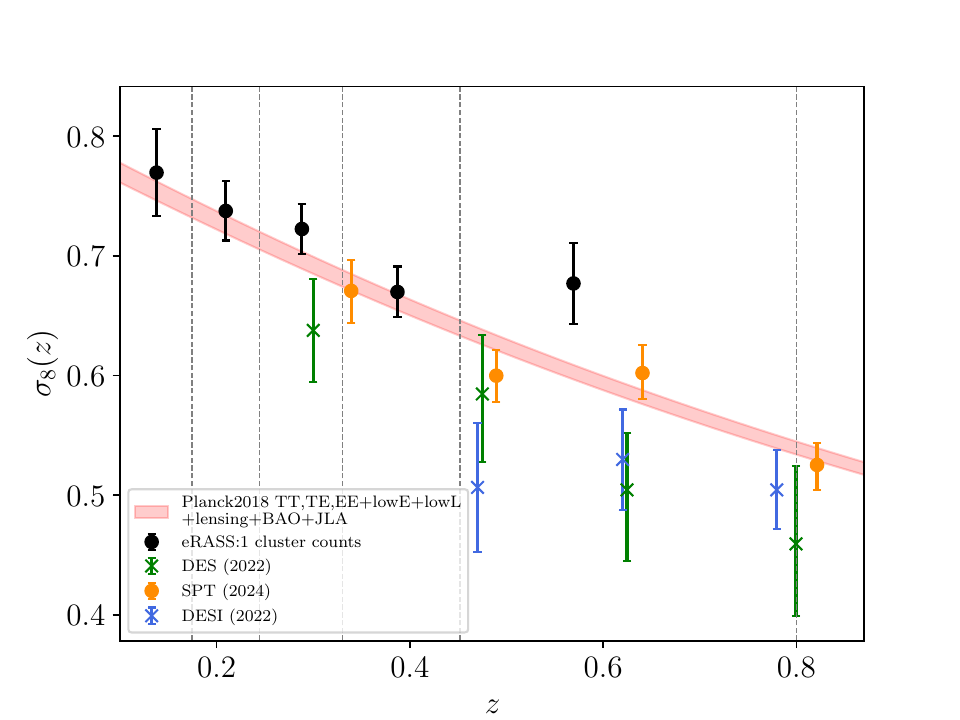}
      \caption{
      Redshift evolution of the r.m.s. density fluctuation in 8$\; h^{-1}\mathrm{Mpc}$, $\sigma_8(z)$. The red curve is the 1$\sigma$ model obtained from \textit{Planck}, \textit{galaxy clustering}, and supernovae data, assuming linear perturbation theory.
      The black dots are obtained with CMB sound horizon priors from the \textit{Planck} satellite. The orange points are the results of SPT cluster abundance from \cite{Bocquet2024} when the cluster data are combined with the horizon scale of the cosmic microwave background. We also provide the latest Dark Energy Survey cosmic shear results presented in \cite{Abbott2023} in green. Finally, the results of the DESI legacy imaging survey combined with CMB lensing are shown in blue \citep{White2022}. Overall, probes that include strong non-linear effects find a lower normalization of the growth of structure. We note that the highest redshift bin probed by \erass cluster count is significantly higher than the prediction of the CMB. We investigate this result by removing the highest bin and reprocessing all the models presented in this work, as well as the $\Lambda\mathrm{CDM}$ standard model. We find that our conclusions are robust against removing this bin (see Figure~\ref{fig:comp_lowz_highz} and Appendix~\ref{app:red_dep}).}
    \label{fig:sigma8z}
\end{figure*}

Finally, we can use these posteriors to recover the redshift evolution of the r.m.s of the density fluctuations, i.e., $\sigma_{8}$ parameter. Similarly, we present our results that combine the sound horizon priors with eRASS1 cluster number counts in data points in black in Figure~\ref{fig:sigma8z}. The eRASS1 cosmology sample, when divided into five redshift intervals, shows hints of a reduction of structure formation at low redshifts. Although all the data points are compatible with the combined analysis of \textit{Planck}, BAO, and JLA results (see \citetalias{PlanckCollaboration2020}) in Figure~\ref{fig:sigma8z}, they all remain above the CMB expectations. Although not significantly significant, this trend is visible in Figure~\ref{fig:sigma8z}. Consequently, we are compatible with the results provided in Section~\ref{sec:res} and Section~\ref{sec:power_suppression}.

Our measurements broadly agree with the cluster abundance analysis from the SPT SZ experiment combined with the same sound horizon priors used in this work \citep{Bocquet2024}.
However, it is worth noting that, unlike our approach, the analysis presented in \citet{Bocquet2024} uses the same scaling relation parameters for all redshift bins, as well as a unique $\Omega_\mathrm{m}$ at all redshift. Although their sample size is significantly smaller (1005 clusters of galaxies as opposed to 5259 clusters used in this work), this assumption is reflected in the uncertainties in the quoted results. This might explain why they do not recover the power suppression trend found in this work, as their sample covers a larger and broader redshift range. 

We also compare our result with the DES Y3 combined cosmic shear and galaxy clustering results \citep{Abbott2023}, 
who provided results with the same approach using cosmic shear and galaxy clustering. They divide their sample in redshift bins in a range of $0<z<1.5$, similar to the redshift coverage of \erass clusters. 
The observed trend of lower structure formation in DES-Y3 data reported in the low redshift regime is in tension with our and CMB measurements.
The same trend is found in the DESI Legacy Imaging Survey Luminous Red Galaxies (LRG) combined with CMB lensing \citep{White2022}.

In this work, we first defined a reduction of structure formation as a higher value of the cosmic linear growth index $\gamma$ (see Section \ref{sec:mg_framwork}). Indeed, equation~\ref{eq:growth_rate_def} and equation~\ref{eq:growth_gr} show that a higher $\gamma$ leads to a lower $f(z)$, and thus, the clustering of matter with time occurs at a slower rate. Consequently, we show that the growth factor is flattened, as shown in Figure~\ref{fig:growth_comp}. All the models we consider show the same trend, including the last measurements through the sample binning method.
As a consequence, the flattening of the growth factor leads to a lower measurement of $\sigma_\mathrm{8}\equiv\sigma_\mathrm{8}(z=0)$. We emphasize that this definition does not imply that we expect a lower $\sigma_\mathrm{8}(z)$ for all redshifts, as found with probes that include non-linear physics \citep{Abbott2022, White2022}. 

Overall, our model agnostic method agrees well with the CMB extrapolated values while hinting at a reduction of structure growth at low-$z$. We note that the fifth bin ($0.452<z<0.8$) has a higher inferred $\sigma_\mathrm{8}(z)$ compared to the trend observed in the previous ones. We investigate the impact of this specific bin in our analysis and the origin of this systematic effect in Section~\ref{sec:robust}. 

\section{Systematic effects and robustness tests of the cosmology analysis}
\label{sec:robust}

In this section, we evaluate the effect of the possible systematic effects on the results presented in this work, scaling relations, and cosmological constraints in general, shown \citetalias{Ghirardini2023} and \citet{Artis2024}. Among those is the impact of neutrinos (see Section \ref{sec:discussion}) and the large scatter observed in the scaling relations between the X-ray observable, i.e., count-rate, in the high redshift bin used to infer cluster mass. We examine the redshift-dependent scatter (the count rates follow a log-normal probability distribution function with scatter $\sigma_\mathrm{X}$) in the count-rate mass scaling relation ($\sigma_{X}=0.98_{-0.04}^{+0.05}$)  to understand the effect of it in the subsequent cosmology analyses using the binned analysis presented in Section~\ref{sec:direct_strucutre_growth}.  

\subsection{Impact of the high redshift clusters on cosmology and observed scatter}
\label{sec:comparison}

Motivated by the results obtained in Section~\ref{sec:direct_strucutre_growth}, we examine the cosmology results obtained from the standard analysis presented in \citetalias{Ghirardini2023} and the results of the binned analysis presented in this work further (see the comparison shown in Figure~\ref{fig:bin_om_s8}). As noted earlier, the cosmological parameters primarily constrained by cluster counts, ${\sigma_8}$, and $\Omega_{M}$, are consistent at the $2\sigma$ level in each redshift bin with the full sample analysis. In this subsection, we further investigate the robustness of our results when the analysis is limited to only lower redshift clusters at $0.1<z<0.45$ and explore the dependence on the assumptions on the mass scaling relations.

As a first test, we repeat the standard cosmology analysis performed in \citetalias{Ghirardini2023} while considering the sample with $0.1<z<0.45$. The catalog in this redshift range comprises a sample of 4196 clusters, which represent 79.8\% of the cosmology sample provided in \citet{Bulbul2023}. The best-fit parameters of the concordance $\Lambda\mathrm{CDM}$ model in the redshift interval of $0.1<z<0.45$ are;

\begin{equation}
    \label{eq:lowz_ghirardini}
    \begin{array}{lcll}
     \Omega_\mathrm{m} &= & 0.30^{+0.02}_{-0.03} & (\mathrm{Full ~sample} : 0.29^{+0.01}_{-0.02})\\
     \sigma_\mathrm{8} &= & 0.87 \pm 0.03 & (\mathrm{Full ~sample} : 0.88\pm 0.02)\\
     \Sh & = & 0.86 \pm 0.01 & (\mathrm{Full ~sample} : 0.86\pm 0.01)
\end{array}.
\end{equation}

\begin{figure}
    \centering
    \includegraphics[scale=0.9]{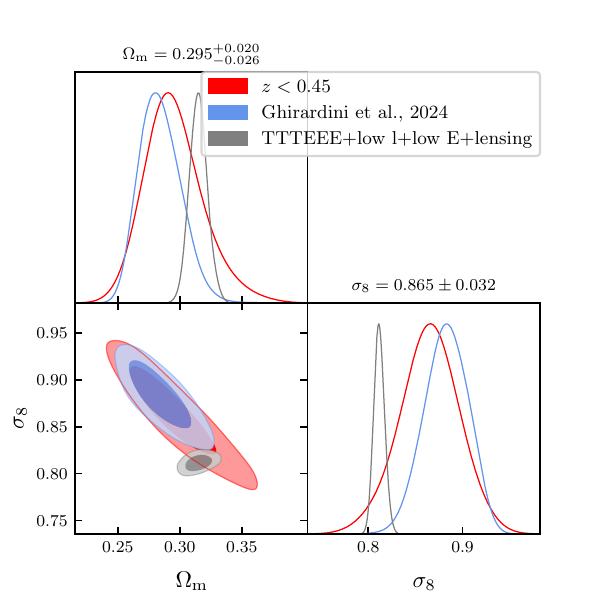}
    \caption{In $\Lambda\mathrm{CDM}$, comparison of the low-redshift sample ($z<0.45$) in red, with the full cosmology sample results in blue. The results from the CMB (\citetalias{PlanckCollaboration2020}) are in grey. The agreement between the standard analysis and the low-redshift analysis shows the robustness of the eRASS1 cosmology analysis. }
    \label{fig:comp_lowz_highz}
\end{figure}

\noindent In parentheses, the best-fit parameters of the $\Lambda\mathrm{CDM}$ model from the full sample analysis reported in \citetalias{Ghirardini2023} are provided. 
Although the statistical precision is diminished due to the lower number of clusters in the analysis, we do not observe any departures from the full sample analysis of \citetalias{Ghirardini2023} within the $1\sigma$ confidence level. The results are displayed in Figure~\ref{fig:comp_lowz_highz}. Consequently, despite the fifth bin ($0.452<z<0.8$) having a best-fit value higher than those of the low-redshift regime, we conclude that the cosmological parameter inference is, for the most part, insensitive to the subdivision of the full sample to high and low-redshift clusters. A similar analysis has been performed for a full sample of clusters in a redshift range of $0.1<z<0.6$ in \citetalias{Ghirardini2023}; the authors reached the same conclusion that the exclusion of the highest redshift clusters in the sample has minimal effect on the best-fit parameters $\sigma_8$ and $\Omega_M$ of the concordance $\Lambda\rm CDM$ model. 

We repeat the same exercise for the $\gamma\Lambda \rm CDM$ model to test if an impact could be detected when we free the structure growth parameters. We reach a similar conclusion that excluding the highest redshift clusters above a redshift of $z>0.45$ does not produce significant tension on the constraints on the $\gamma$, $\sigma_8$, and  $\Omega_M$ parameters in the $\gamma\Lambda\rm CDM$ case, they remain consistent at $1\sigma$ level. The comparisons of the results based on both samples are shown in Figure~\ref{fig:low_z_impact}. However, the tension between the $\gamma$ measurements of eROSITA combined with the CMB and the standard GR inferred value $\gamma=0.55$ is reduced as we remove the high redshift clusters from our sample; the  $3.7\sigma$ tension obtained with the full sample analysis is reduced to $2.4\sigma$ confidence level due to the increase in the uncertainties. The results of this analysis are presented in Appendix~\ref{app:red_dep}. We conclude that while high redshift clusters might have a small impact on the analysis, our main conclusions are preserved.

\begin{figure}
    \centering
    \includegraphics[scale=0.9]{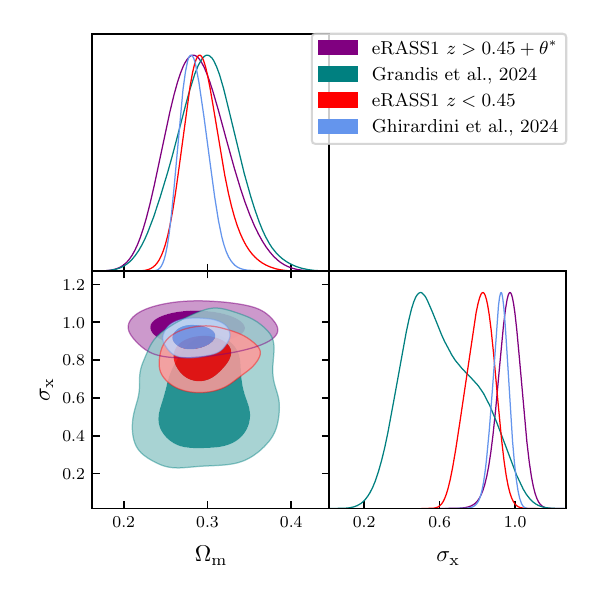}
    \caption{Comparison of the posteriors of the intrinsic scatter of the $C_\mathrm{R}-M$ scaling relation obtained for the $\Lambda$CDM analysis of \citetalias{Ghirardini2023} in blue, the low-redshift sample in red, and the weak-lensing mass calibration information only from DES \citep{Grandis2023} in teal, and the high redshift sample $z>0.45$ combined with the sound horizon priors. All three results are in good statistical agreement, although the higher value in the high-redshift sample hints at a potential selection effect.}
    \label{fig:comp_scatter}
\end{figure}

One interesting outcome of this test is the potential systematics related to the high redshift cluster sample at $z>0.45$. 
The posterior value of $\sigma_\mathrm{8}$ is slightly higher in the highest redshift bin ($0.45<z<0.8$), consequently increasing the $\Sh$ value compared to the lower redshift bins as clearly seen in Figure~\ref{fig:sigma8z}.
Although excluding the highest redshift range produces consistent results with the full sample analysis as presented in the previous section and proves that our standard cosmology analysis is robust against sample selection in a wide redshift interval utilized in the consequent cosmology analyses, we further investigate the effect of the high redshift clusters on the scatter of the count-rate and mass scaling relation. The observed large scatter on the X-ray count-rate scaling relations of the full sample decreases from $\sigma_{X}=0.98_{-0.04}^{+0.05}$ to  $\sigma_{X} = 0.81^{+0.09}_{-0.06}$ when the highest redshift bin excluded in the analysis. As seen in Figure~\ref{fig:comp_scatter}, this result indicates that the high redshift clusters are mostly responsible for the high observed scatter in the scaling relations utilizing the full sample in the redshift range of $0.1<z<0.8$, presented in. This result can also be noticed in Table~\ref{tab:results_in_bins}. This high scatter is likely due to a higher contamination fraction of the sample at high redshifts. Our implementation of the mixing model partially mitigates the effect of contamination in the results presented on eRASS1 data in \citetalias{Ghirardini2023}. However, the increased number of AGN in clusters of galaxies may lead to skewness and high scatter in the count rate mass scaling relation \citep{Biffi2018}. Future work on the deeper eROSITA and DES data will carefully investigate and model the high redshift AGN contamination in clusters with a reduced scatter in scaling relations. It is important to note that the tests performed in this work prove that our modeling through scaling relations accurately represents the underlying cluster population regardless of the observed large scatter due to potential observational effects or cluster physics. Therefore, we conclude that the observed large scatter does not significantly affect our results presented in \citetalias{Ghirardini2023}, \cite{Artis2024}.

\begin{figure}
    \centering
    \includegraphics[scale=0.7]{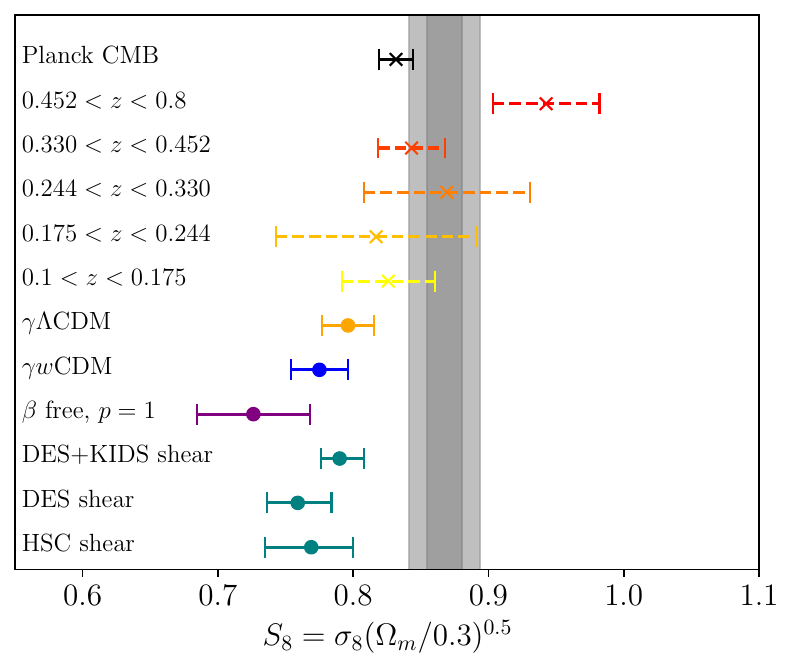}
    \caption{Comparison of the $S_\mathrm{8}$ parameter in the different bins, and different models. The color scheme follows the one introduced in Figure~\ref{fig:sample_bin} for the cosmological parameter inference. Different cosmic shear results are shown in teal: DES + KIDS \citep{Abbott2023b}, DES \citep{Amon2022}, and HSC \citep{Dalal2023}.  Results from the cosmic microwave background (\citetalias{PlanckCollaboration2020}) are shown in black. The grey area represents the 68\% and 95\% confidence levels from \citetalias{Ghirardini2023}. Our main result is that no matter what the considered model is, adding a degree of freedom in the power spectrum decreases the value of $\Sh$ to make it compatible with the cosmic shear value.}
    \label{fig:s8_z}
\end{figure}
\subsection{Consistency of the mass scale}
\label{subsec:mass_scale}

Another key ingredient of cluster abundance cosmology is the scaling relation between the observable used in the detection chain and the underlying cluster/dark matter halo mass. Our cosmological framework uses the X-ray count rate measured within R$_{500}$ as the primary observable. The parameters of this count rate to mass ($C_\mathrm{R}-M$) relation are fitted jointly with the cosmological parameters.
The X-ray scaling relation is described as follows:
\begin{equation}
\left\langle \ln \frac{C_\mathrm{R}}{C_\mathrm{R,p}} \bigg| M, z \right\rangle = 
\ln A_\mathrm{X} + 
b_\mathrm{X}(z) \, \ln \frac{M}{M_\mathrm{p}} + e_\mathrm{x}(z) .
\label{eq:cr_m}
\end{equation}
As in \citetalias{Ghirardini2023}, $C_\mathrm{R,p} = 0.1$ cts~/~s, $M_\mathrm{p} = 2 \times 10 ^{14} M_\odot$, and $z_\mathrm{p} = 0.35$. Note that the pivot value is changed accordingly in each redshift bin in the binned analysis.  The other terms follow
\begin{equation}
b_\mathrm{X}(z) = B_\mathrm{X} + F_\mathrm{X} \, \ln \frac{1+z}{1+z_\mathrm{p}}.,
\end{equation}
and
\begin{equation}
e_\mathrm{x}(z) = D_\mathrm{X} \, \ln \frac{d_\mathrm{L}(z)}{d_\mathrm{L}(z_\mathrm{p})} + E_\mathrm{X} \, \ln \frac{E(z)}{E(z_\mathrm{p})} + G_\mathrm{X} \, \ln \frac{1+z}{1+z_\mathrm{p}}.
\label{eq:red_evol}
\end{equation}

Since the modifications of the power spectrum applied here do not concern mass estimation, we should find consistent scaling relations in the different models. 
In all this work, we fix the luminosity distance dependence $D_\mathrm{X}$ to $-2$ and normalized Hubble parameter dependence $E_\mathrm{X}$ to 2, which are the values expected in the self-similar model \citep{Kaiser1986}. In Appendix~\ref{app:scaling_relations}, we show the corner plot obtained for the scaling relation parameters for the two models. We present the $\gamma\Lambda\mathrm{CDM}$, $\gamma w\mathrm{CDM}$, and the $\Lambda\mathrm{CDM}$ models as reference cases.
Overall, there is good agreement between our different models whenever we consider modifications of the power spectrum. 
Thus, the shape of the scaling relation remains consistent in different cosmological models.
This means that the mass calibration is robust and that changes in the cosmological model primarily affect the underlying distribution of clusters represented by the halo mass function.
One exception to this general agreement is the parameters $G_\mathrm{X}$, which encapsulates the redshift dependence of the scaling relation. It is significantly lower at the 2$\sigma$ level in $\gamma\Lambda\mathrm{CDM}$ than for our standard $\Lambda$CDM case.  Indeed, in $\Lambda\mathrm{CDM}$, the best-fit value of this parameter is consistent with $0$. Since we modify the growth rate of structures in the different considered models, we expect a propagation on this parameter without it being related to a specific systematic effect.
However, it seems that for the $\gamma\Lambda\mathrm{CDM}$ case, the preferred value tends to be lower.

We also note that when freeing the dark energy equation of state, we recover a good agreement between the different posteriors of the scaling relations.
Overall, the scaling relations parameter estimates are robust against our modeling at the $2\sigma$ level. 

We also directly estimate a possible redshift evolution of the scaling relation, thanks to the framework described in Section~\ref{sec:direct_strucutre_growth}. We provide a corner plot of all the X-ray scaling relation parameters compared in Appendix~\ref{app:scaling_relations}. 
We do not observe a statistically significant difference between the scaling relation parameters in our redshift bins.
We defined the redshift bins to possess the same number of observed clusters. As fewer clusters are observed at high redshift (due to the halo mass function and the \erass selection function), high redshift bins cover a larger redshift space. As a consequence,  the parameters describing the redshift evolution of the X-ray scaling relation $F_\mathrm{X}$ and $G_\mathrm{X}$ (see \citetalias{Ghirardini2023}) are better constrained using the last and largest redshift bins. 
 This also confirms that our weak-lensing mass calibration performs well at all redshifts.

\subsection{Goodness of fit}
\label{subsec:goodness_of_fit}

\begin{figure}
 \centering
 \includegraphics[width=0.5\textwidth]{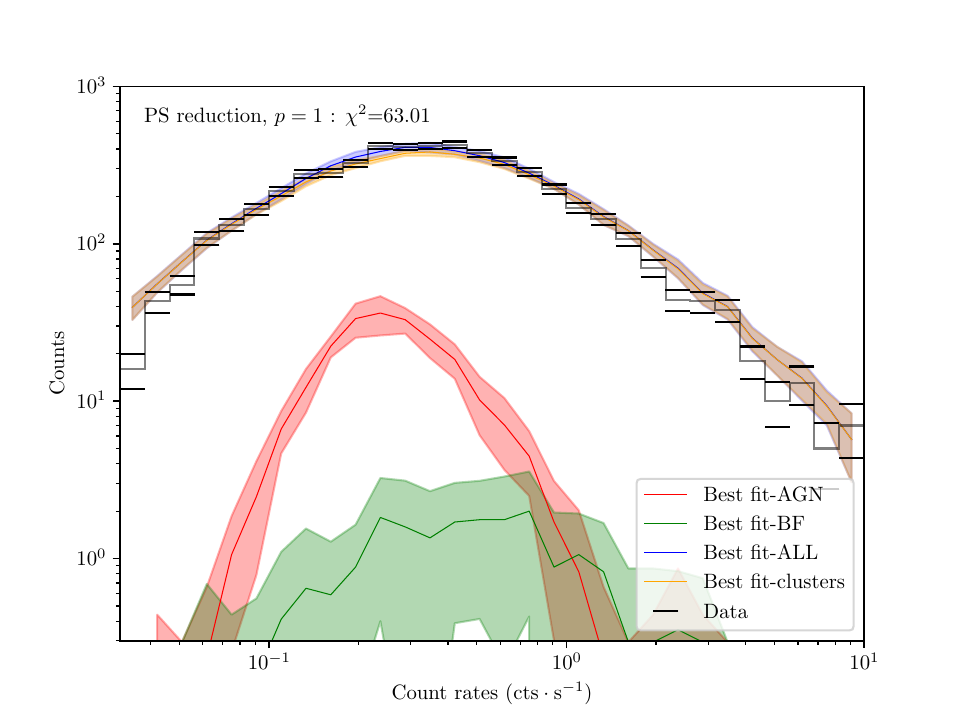}

 \includegraphics[width=0.5\textwidth]{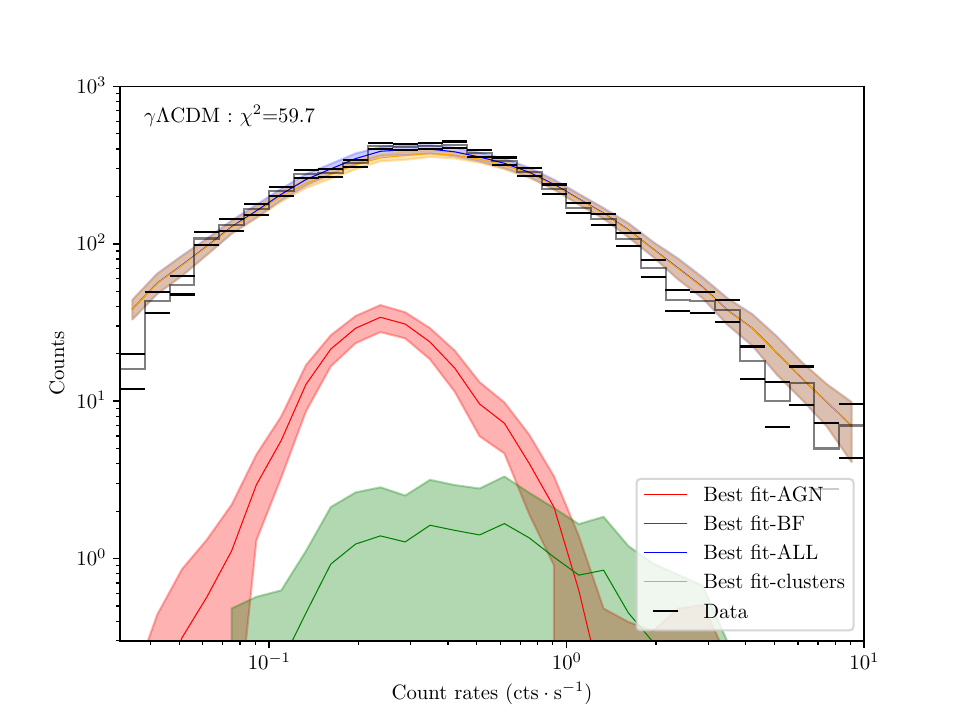}

 \includegraphics[width=0.5\textwidth]{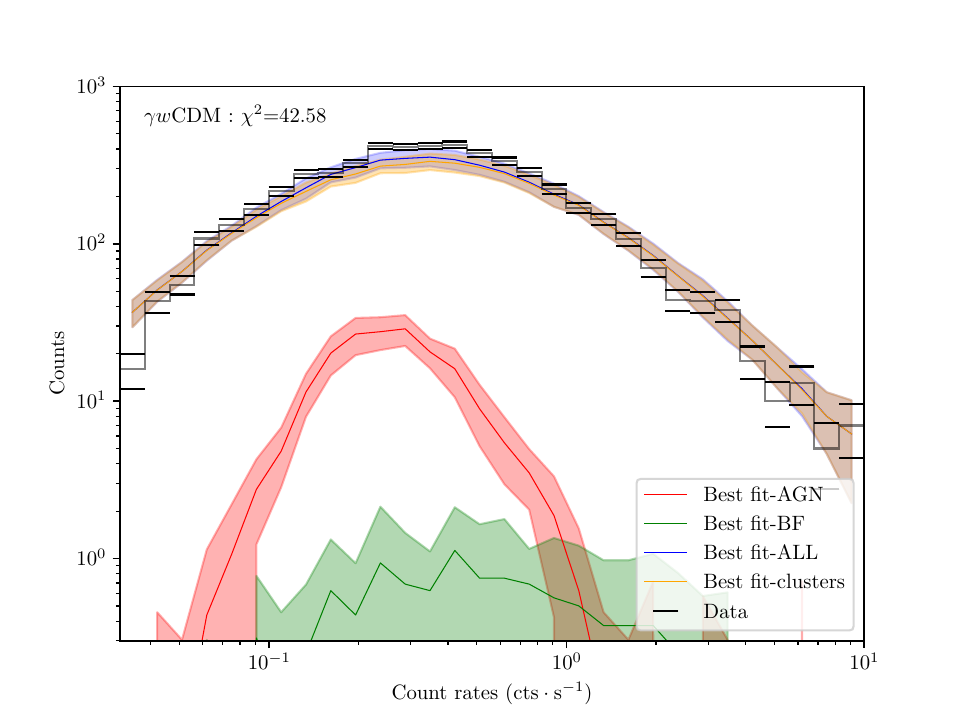}
 
 \caption{Goodness of fit for the three models fitted in this analysis: the power spectrum reduction model \cite{Lin2024} with $p =1$ fixed, $\gamma\Lambda$CDM and $\gamma w$CDM. The combination of all the model components is shown in blue, while the cluster is in orange. We also show the contaminants with the active galactic nuclei (AGN, in red) and false detections due to background fluctuations (BF, in green). The filled areas represent the 68\% errors on each component. The grey solid line represents the distribution of the data, and the black line is the Poisson uncertainties in each bin.}
 \label{fig:fit_goodness}
\end{figure}

This work uses the formalism described in \citetalias{Ghirardini2023} to fit the cosmological parameters. In particular, we are using a Poison likelihood with a complex underlying model. Consequently, it is not straightforward to produce a goodness-of-fit metric for our measurements. Nevertheless, since we fit multiple models, we estimate how successfully our model reproduces the data distribution. We proceed as follows: for each of the three models, we produce 500 mock observations from the posterior distribution of the model parameters obtained in Section~\ref{sec:res} and Section~\ref{sec:power_suppression}. We produce the mock observation for the galaxy clusters and the contaminant part, composed of active galactic nuclei (AGN) and false detection due to background fluctuations  (BF). From these mocks, we produce the expected distribution of objects in the catalog, i.e., the number density of objects per unit of  X-ray count rates. We then compare this distribution with the observed distribution of clusters. The results are shown in Figure~\ref{fig:fit_goodness}. In all cases, the fit is visibly a good data reproduction. 
For the case of the power spectrum reduction model, we find $\chi^2 = 63.01$, while for $\gamma\Lambda\mathrm{CDM}$ $\chi^2 = 59.7$ and for $\gamma w\mathrm{CDM}$,\, $\chi^2 = 42.58$, all of these obtained with 30 d.o.f.

We conclude that our model reliably reproduces the data distribution, with a preference for the $\gamma w\mathrm{CDM}$. We emphasize that this quantitative estimation primarily verifies that our model performs well. The next data releases will provide a full goodness-of-fit analysis for our cluster abundance Poisson statistics.

\section{Discussion and conclusions}
\label{sec:discussion}

In this work, we present results on the measurement of the growth of structure using X-ray detected \erass clusters, in combination with optical surveys, such as LS DR10-South for optical confirmation and redshift measurements, the DES, KiDS, and HSC for weak lensing mass calibration. Specifically, the $\gamma$ parameterization of the growth factor presented in equation~\ref{eq:growth_gr} modifies the structure growth rate (Section \ref{sec:res}) in connection with the power spectrum reduction model (Section \ref{sec:power_suppression}). Secondly, a model agnostic strategy aiming at measuring the cosmological parameters at different redshifts to reconstruct the history of the evolution in density fluctuations (Section \ref{sec:direct_strucutre_growth}).

We also probe the evolution of structure formation with the late-time power suppression presented in Section~\ref{sec:power_suppression}. The two approaches are thus related, and we expect any potential deviation from the standard scenario appearing in one to be reflected in the other. 
In all three cases, we combine our structure growth constraints with the sound horizon priors from the cosmic microwave background $\theta^*$ as it is one of the most robust cosmological measurements available. In all three cases, we find observations that can be interpreted as a low redshift reduction of structure formation. First, the $\gamma$ parameterization of the growth factor, while still compatible with the standard $\gamma = 0.55$ value at $2\sigma$ both in $\gamma\Lambda\mathrm{CDM}$ and $\gamma w\mathrm{CDM}$, has a best-fit value that favors higher values of the cosmic linear growth index. Additionally, the model agnostic redshift splitting of the eRASS1 cosmological catalog shows a similar pattern. Indeed, Figure~\ref{fig:growth_comp} shows how the cosmic linear growth index increases the growth factor at low redshift. Our binned cosmological results indicate a preference for the same trend while still being statistically consistent with CMB measurements. Finally, the power spectrum reduction model shows a best-fit value that significantly excludes an absence of power reduction.

These three results indicate a deviation from the widely accepted paradigm of structure formation. Interestingly, in the case of the cosmic linear growth index and power spectrum reduction models, we find that modifying the power spectrum leads to a decrease in the $\sigma_\mathrm{8}$ value, and thus our $\Sh$ inferred value. Consequently, these observations may offer evidence that the $\Sh$ tension could be accounted for by a modification of the matter power spectrum at the scales investigated by cosmic shear and other lensing studies \citep{Nguyen2023}, provided that the explanation for the tension as a result of systematic uncertainties related to baryonic effects is ruled out. Indeed, Figure~\ref{fig:s8_z} shows that the in all of the three models used in this work ($\gamma\Lambda\mathrm{CDM}$, $\gamma w\mathrm{CDM}$, and the power spectrum reduction model) the value of the parameter $S_\mathrm{8}$ is reduced from the value found in \citetalias{Ghirardini2023} in the $\Lambda\mathrm{CDM}$ case, $\Sh=0.86\pm 0.01$ to

\begin{equation}
    \label{eq:all_s8}
    \begin{array}{lcll}     
     \Sh & = & 0.79 \pm 0.02 & (\gamma\Lambda\mathrm{CDM})\\
     \Sh & = & 0.77 \pm 0.02 & (\gamma w\mathrm{CDM})\\
     \Sh & = & 0.73 \pm 0.04 & (\mathrm{Power\, spectrum :\;} \beta\, \mathrm{free}, p=1)
\end{array} .
\end{equation}

These values are compatible with the ones routinely inferred by cosmic shear studies but in $\mathrm{\Lambda CDM}$ and at small scales. Using KIDS data, \cite{Li2023} found $\Sh = 0.776 ^{+0.029}_{-0.027}$, close to the HSC results presented in \cite{Dalal2023} of $\Sh = 0.776^{+0.032}_{-0.033}$. Using DES data and the same cosmological probe, \cite{Amon2022} found $\Sh = 0.772^{+0.018}_{-0.017}$. Finally, combining DES and KIDS data, \cite{Abbott2023b} found $\Sh = 0.790^{+0.018}_{-0.014}$. We also note that other probes affected by non-linear physical effects also find lower $\Sh$ values. For example \cite{PalanqueDelabrouille2020} used the Ly$\alpha$ forest to infer $\Sh=0.77\pm 0.04$. Given that cosmological probes sensitive to the non-linear regime yield a lower $\Sh$, our findings may imply that this discrepancy may be simply due to non-linear effects influencing those probes. Although galaxy clusters are highly non-linear structures, their number density can be described using the linear matter power spectrum. Effects related to non-linear physics are restricted to the HMF parameterization, which is assumed to be universal. Thus, when inferring cosmology using galaxy cluster number counts, we obtain values closer to the one inferred by the CMB. Introducing an additional degree of freedom in the power spectrum analysis could suggest that the reduction in structure formation is scale-dependent, potentially resolving the $\Sh$ tension.

Neutrinos smooth structure formation at small scales, thus reducing the power spectrum at high $k$. Therefore, their effect on the cosmological parameter estimation impacts the power spectrum reduction we are probing here. However, in this work, we have chosen to consider massless neutrinos. This modeling choice thus requires a very careful justification. We show that the impact of neutrinos is negligible and intrinsically different compared to the power spectrum reduction reported here. We leave the combined measurement of the neutrino masses and the power spectrum reduction parameterization to future works. 

Three main reasons explain our consideration of massless neutrinos: the lack of theoretical background for considering neutrinos together with a power spectrum reduction at low redshift, the relative size of the shift in the cosmological parameter space, and the intrinsic nature of the power impression we are investigating. 

First, since \cite{Costanzi2013}, we know that the effect of neutrinos on the halo mass function is well captured when the cold dark matter only power spectrum $P_\mathrm{cdm}(k,z)$ is considered instead of the total matter power spectrum $P_\mathrm{m}(k,z)$ in the fitting function parameterized by \cite{Tinker2008}. However, in this formalism, neutrinos also induce a scale dependence in the growth factor presented in equation~\ref{eq:power_evol}. In other words, we have 
\begin{equation}
    \label{eq:massive_pk_evol}
    P_\mathrm{cdm}(k,z) \neq  P_\mathrm{cdm}(k,z=0) \times D^2(z),
\end{equation}
as $D(z)\leftarrow D(z,k)$. Consequently, when using the cold dark matter spectrum, we need to compute the spectrum directly at the considered redshift. Analytical models exist for the linear growth factor in the presence of neutrinos \citep{Kiakotou2008}, but they are not compatible with the cosmic linear growth index. In general, we do not currently have a satisfying way of parameterizing the linear growth factor in the presence of neutrinos together with the cosmic linear growth index in the context of galaxy cluster number counts. This consideration highlights the difficulty of finding a model that fits both. The following two paragraphs, however, show that the effect of neutrinos can be neglected in this work. 

\begin{figure}
    \centering
    \includegraphics[scale=0.9]{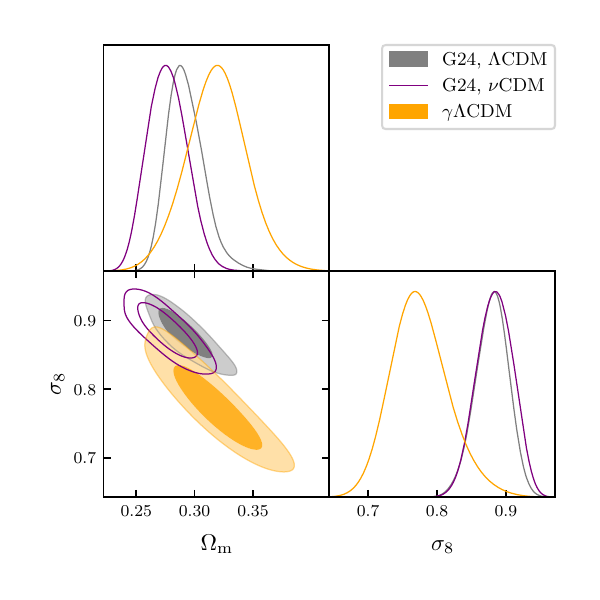}
    \caption{Comparaison of the shift in the $\Omega_\mathrm{m}-\sigma_\mathrm{8}$ plane. The results obtained while assuming massless neutrinos agree with the one probing their mass. Additionally, adding neutrinos does not impact the measurement of $\sigma_\mathrm{8}$.
   }
    \label{fig:comp_model_massive}
\end{figure}
Figure~\ref{fig:comp_model_massive} shows how neutrinos impact cosmological parameter estimation. Indeed, the shift in the cosmological parameter space produced by neutrinos compared to a model when they are massless is of $0.5\sigma$ in the $\Omega_\mathrm{m}/\sigma_\mathrm{8}$ plane. This number must be compared with the $2.4\sigma$ shift produced by the power spectrum reduction models considered in this work. In other words, neutrinos alone cannot explain the reduction observed and presented here. Although their effect is indeed degenerate with the power spectrum reduction, it is negligible compared to our results for a catalog of clusters of the size of \erass.

\begin{figure}
    \centering
    \includegraphics[scale=0.55]{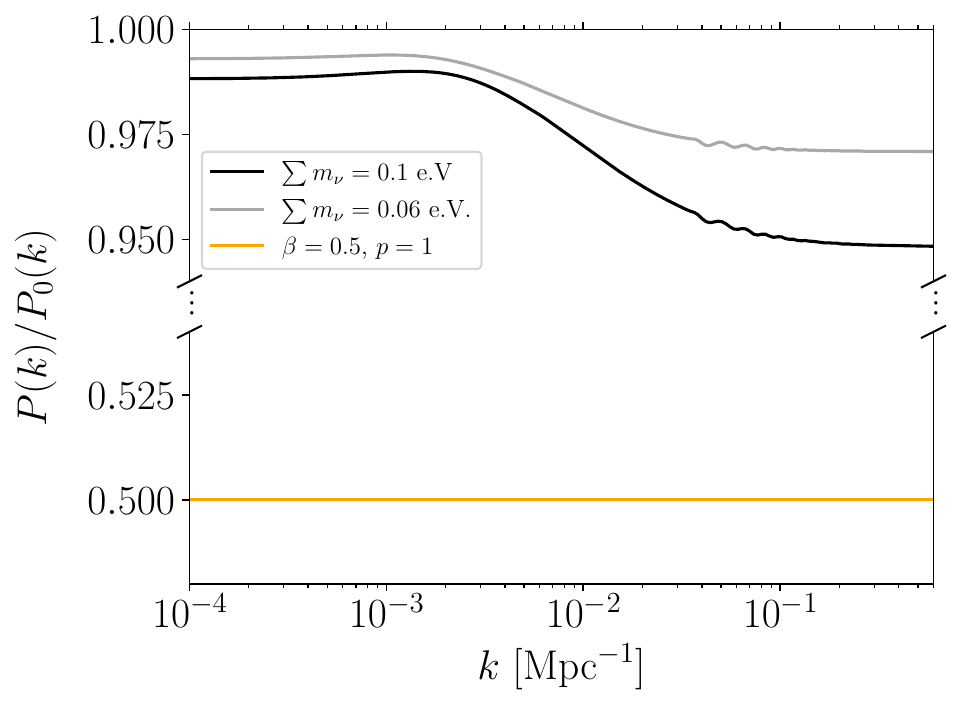}
    \caption{ Impact of neutrinos on the matter power spectrum compared to one of the redshift reduction models we probe in this work. We represent the ratio of the power spectra when compared to the $\Lambda$CDM case with massless neutrinos noted as $P_\mathrm{0}$. For readability, we have removed a large part of the y-axis. Neutrinos induce a scale reduction of the matter power spectrum, while the power spectrum reduction model is scale-independent. Additionally, the orders of magnitude are significantly different: for neutrinos with a mass of $0.1\mathrm{eV}$, the damping amplitude is about 5\% at maximum for the smallest scales. This is one order of magnitude below the effect of the power spectrum reduction.
    }
    \label{fig:neutrino_reduction}
\end{figure}

Finally, this work investigates a potential redshift reduction of the power spectrum. Neutrinos induce a scale-dependent damping of the power spectrum in the redshift range probed by \erass galaxy cluster count. Indeed, at a fixed scale $k$, the reduction induced by neutrinos on the matter power spectrum is constant in the redshift range $0.1<z<0.8$.
Figure~\ref{fig:neutrino_reduction}. On the other hand, the models we are probing only modify structure formation at low redshift but at all scales. Consequently, both effects are different and do not affect the result in a way that can be compared. 

For the aforementioned reasons, our results are primarily not affected by the absence of massive neutrinos. However, more theoretical efforts will be necessary to probe the structure reduction we detect when larger and deeper eRASS catalogs are available. At this time, it will not be possible to neglect the effect of neutrinos anymore.

Recent studies show the current implementation of baryonic feedback in numerical simulations isn't enough to account for the reduction in $\Sh$ observed in surveys sensitive to small scales, as demonstrated in \citep{Grandis2024}. However, if baryonic feedback is poorly understood and its effects are greater than anticipated, it could potentially reconcile the linear and non-linear regimes and explain the observed$\Sh$ tension \citep{Preston2023}.

Our results agree with other studies that infer a reduction of the structure formation at low redshifts \citep{Samushia2013, Beutler2014, Kazantzidis2021, Lin2024}. However, a robust conclusion can only be provided if every potential systematic effect is extensively studied through numerical simulations. We also investigate redshift-dependent potential systematic effects in this work and our $\Lambda$CDM analysis. Although statistically consistent, the fifth and highest redshift bin ($0.452<z<0.8$) has a higher $\Sh$ value than the other four at lower redshifts. Although the scatter in this bin is significantly higher, it does not affect the results of this work. However, the reason for this scatter in the high redshift regime, e.g., increased AGN contamination or photometric redshift uncertainties, will be investigated in upcoming cosmology analyses.

\begin{acknowledgement}

This work is based on data from eROSITA, the soft X-ray instrument aboard SRG, a joint Russian-German science mission supported by the Russian Space Agency (Roskosmos), in the interests of the Russian Academy of Sciences represented by its Space Research Institute (IKI), and the Deutsches Zentrum f{\"{u}}r Luft und Raumfahrt (DLR). The SRG spacecraft was built by the Lavochkin Association (NPOL) and its subcontractors and is operated by NPOL with support from the Max Planck Institute for Extraterrestrial Physics (MPE).

The development and construction of the eROSITA X-ray instrument was led by MPE, with contributions from the Dr. Karl Remeis Observatory Bamberg \& ECAP (FAU Erlangen-Nuernberg), the University of Hamburg Observatory, the Leibniz Institute for Astrophysics Potsdam (AIP), and the Institute for Astronomy and Astrophysics of the University of T{\"{u}}bingen, with the support of DLR and the Max Planck Society. The Argelander Institute for Astronomy of the University of Bonn and the Ludwig Maximilians Universit{\"{a}}t Munich also participated in the science preparation for eROSITA.

The eROSITA data shown here were processed using the \esass software system developed by the German eROSITA consortium.
\\

E. Bulbul, V. Ghirardini, A. Liu, S. Zelmer, and X. Zhang acknowledge financial support from the European Research Council (ERC) Consolidator Grant under the European Union’s Horizon 2020 research and innovation program (grant agreement CoG DarkQuest No 101002585). N. Clerc was financially supported by CNES. T. Schrabback and F. Kleinebreil acknowledge support from the German Federal
Ministry for Economic Affairs and Energy (BMWi) provided
through DLR under projects 50OR2002, 50OR2106, and 50OR2302, as well as the support provided by the Deutsche Forschungsgemeinschaft (DFG, German Research Foundation) under grant 415537506.

\\

The Innsbruck authors acknowledge the support provided by
the Austrian Research Promotion Agency (FFG) and the Federal Ministry of
the Republic of Austria for Climate Action, Environment, Energy, Mobility,
Innovation and Technology (BMK) via the Austrian Space Applications Programme with grant numbers 899537, 900565, and 911971.

\\

The Legacy Surveys consist of three individual and complementary projects: the Dark Energy Camera Legacy Survey (DECaLS; Proposal ID \#2014B-0404; PIs: David Schlegel and Arjun Dey), the Beijing-Arizona Sky Survey (BASS; NOAO Prop. ID \#2015A-0801; PIs: Zhou Xu and Xiaohui Fan), and the Mayall z-band Legacy Survey (MzLS; Prop. ID \#2016A-0453; PI: Arjun Dey). DECaLS, BASS, and MzLS together include data obtained, respectively, at the Blanco telescope, Cerro Tololo Inter-American Observatory, NSF’s NOIRLab; the Bok telescope, Steward Observatory, University of Arizona; and the Mayall telescope, Kitt Peak National Observatory, NOIRLab. Pipeline processing and analyses of the data were supported by NOIRLab and the Lawrence Berkeley National Laboratory (LBNL). The Legacy Surveys project is honored to be permitted to conduct astronomical research on Iolkam Du’ag (Kitt Peak), a mountain with particular significance to the Tohono O’odham Nation.

\\

Funding for the DES Projects has been provided by the U.S. Department of Energy, the U.S. National Science Foundation, the Ministry of Science and Education of Spain, the Science and Technology FacilitiesCouncil of the United Kingdom, the Higher Education Funding Council for England, the National Center for Supercomputing Applications at the University of Illinois at Urbana-Champaign, the Kavli Institute of Cosmological Physics at the University of Chicago, the Center for Cosmology and Astro-Particle Physics at the Ohio State University, the Mitchell Institute for Fundamental Physics and Astronomy at Texas A\&M University, Financiadora de Estudos e Projetos, Funda{\c c}{\~a}o Carlos Chagas Filho de Amparo {\`a} Pesquisa do Estado do Rio de Janeiro, Conselho Nacional de Desenvolvimento Cient{\'i}fico e Tecnol{\'o}gico and the Minist{\'e}rio da Ci{\^e}ncia, Tecnologia e Inova{\c c}{\~a}o, the Deutsche Forschungsgemeinschaft, and the Collaborating Institutions in the Dark Energy Survey.

The Collaborating Institutions are Argonne National Laboratory, the University of California at Santa Cruz, the University of Cambridge, Centro de Investigaciones Energ{\'e}ticas, Medioambientales y Tecnol{\'o}gicas-Madrid, the University of Chicago, University College London, the DES-Brazil Consortium, the University of Edinburgh, the Eidgen{\"o}ssische Technische Hochschule (ETH) Z{\"u}rich,  Fermi National Accelerator Laboratory, the University of Illinois at Urbana-Champaign, the Institut de Ci{\`e}ncies de l'Espai (IEEC/CSIC), the Institut de F{\'i}sica d'Altes Energies, Lawrence Berkeley National Laboratory, the Ludwig-Maximilians Universit{\"a}t M{\"u}nchen and the associated Excellence Cluster Universe, the University of Michigan, the National Optical Astronomy Observatory, the University of Nottingham, The Ohio State University, the OzDES Membership Consortium, the University of Pennsylvania, the University of Portsmouth, SLAC National Accelerator Laboratory, Stanford University, the University of Sussex, and Texas A\&M University.

Based on observations made with ESO Telescopes at the La Silla Paranal Observatory under program IDs 177.A-3016, 177.A-3017, 177.A-3018, and 179.A-2004, and on data products produced by the KiDS consortium. The KiDS production team acknowledges support from: Deutsche Forschungsgemeinschaft, ERC, NOVA and NWO-M grants; Target; the University of Padova, and the University Federico II (Naples).

This paper makes use of software developed for the Large Synoptic Survey Telescope. We thank the LSST Project for making their code available as free software at  http://dm.lsst.org

\\

This work made use of the following Python software packages: 
SciPy\footnote{https://scipy.org/} \citep{Virtanen2020SciPy}, 
Matplotlib\footnote{https://matplotlib.org/} \citep{Hunter2007matplotlib}, 
Astropy\footnote{https://www.astropy.org/} \citep{Astropy2022}, 
NumPy\footnote{https://numpy.org/} \citep{Harris2020},
CAMB \citep{Lewis2011CAMB},
pyCCL\footnote{https://github.com/LSSTDESC/CCL} \citep{Chisari2019},
GPy\footnote{https://github.com/SheffieldML/GPy} \citep{gpy2014},
climin\footnote{https://github.com/BRML/climin} \citep{Bayer2015},
ultranest\footnote{https://github.com/JohannesBuchner/UltraNest/} \citep{Buchner2021}

\end{acknowledgement}

\bibliography{references.bib}

\begin{thebibliography}{108}
\expandafter\ifx\csname natexlab\endcsname\relax\def\natexlab#1{#1}\fi

\bibitem[{{Abbott} {et~al.}(2020){Abbott}, {Aguena}, {Alarcon}, {Allam},
  {Allen}, {Annis}, {Avila}, {Bacon}, {Bechtol}, {Bermeo}, {Bernstein},
  {Bertin}, {Bhargava}, {Bocquet}, {Brooks}, {Brout}, {Buckley-Geer}, {Burke},
  {Carnero Rosell}, {Carrasco Kind}, {Carretero}, {Castander}, {Cawthon},
  {Chang}, {Chen}, {Choi}, {Costanzi}, {Crocce}, {da Costa}, {Davis}, {De
  Vicente}, {DeRose}, {Desai}, {Diehl}, {Dietrich}, {Dodelson}, {Doel},
  {Drlica-Wagner}, {Eckert}, {Eifler}, {Elvin-Poole}, {Estrada}, {Everett},
  {Evrard}, {Farahi}, {Ferrero}, {Flaugher}, {Fosalba}, {Frieman},
  {Garc{\'\i}a-Bellido}, {Gatti}, {Gaztanaga}, {Gerdes}, {Giannantonio},
  {Giles}, {Grandis}, {Gruen}, {Gruendl}, {Gschwend}, {Gutierrez}, {Hartley},
  {Hinton}, {Hollowood}, {Honscheid}, {Hoyle}, {Huterer}, {James}, {Jarvis},
  {Jeltema}, {Johnson}, {Johnson}, {Kent}, {Krause}, {Kron}, {Kuehn},
  {Kuropatkin}, {Lahav}, {Li}, {Lidman}, {Lima}, {Lin}, {MacCrann}, {Maia},
  {Mantz}, {Marshall}, {Martini}, {Mayers}, {Melchior}, {Mena-Fern{\'a}ndez},
  {Menanteau}, {Miquel}, {Mohr}, {Nichol}, {Nord}, {Ogando}, {Palmese},
  {Paz-Chinch{\'o}n}, {Plazas}, {Prat}, {Rau}, {Romer}, {Roodman}, {Rooney},
  {Rozo}, {Rykoff}, {Sako}, {Samuroff}, {S{\'a}nchez}, {Sanchez}, {Saro},
  {Scarpine}, {Schubnell}, {Scolnic}, {Serrano}, {Sevilla-Noarbe}, {Sheldon},
  {Smith}, {Smith}, {Suchyta}, {Swanson}, {Tarle}, {Thomas}, {To}, {Troxel},
  {Tucker}, {Varga}, {von der Linden}, {Walker}, {Wechsler}, {Weller},
  {Wilkinson}, {Wu}, {Yanny}, {Zhang}, {Zhang}, {Zuntz}, \& {DES
  Collaboration}}]{DesCollaboration2020}
{Abbott}, T.~M.~C., {Aguena}, M., {Alarcon}, A., {et~al.} 2020, \prd, 102,
  023509

\bibitem[{{Abbott} {et~al.}(2022){Abbott}, {Aguena}, {Alarcon}, {Allam},
  {Alves}, {Amon}, {Andrade-Oliveira}, {Annis}, {Avila}, {Bacon}, {Baxter},
  {Bechtol}, {Becker}, {Bernstein}, {Bhargava}, {Birrer}, {Blazek},
  {Brandao-Souza}, {Bridle}, {Brooks}, {Buckley-Geer}, {Burke}, {Camacho},
  {Campos}, {Carnero Rosell}, {Carrasco Kind}, {Carretero}, {Castander},
  {Cawthon}, {Chang}, {Chen}, {Chen}, {Choi}, {Conselice}, {Cordero},
  {Costanzi}, {Crocce}, {da Costa}, {da Silva Pereira}, {Davis}, {Davis}, {De
  Vicente}, {DeRose}, {Desai}, {Di Valentino}, {Diehl}, {Dietrich}, {Dodelson},
  {Doel}, {Doux}, {Drlica-Wagner}, {Eckert}, {Eifler}, {Elsner}, {Elvin-Poole},
  {Everett}, {Evrard}, {Fang}, {Farahi}, {Fernandez}, {Ferrero}, {Fert{\'e}},
  {Fosalba}, {Friedrich}, {Frieman}, {Garc{\'\i}a-Bellido}, {Gatti},
  {Gaztanaga}, {Gerdes}, {Giannantonio}, {Giannini}, {Gruen}, {Gruendl},
  {Gschwend}, {Gutierrez}, {Harrison}, {Hartley}, {Herner}, {Hinton},
  {Hollowood}, {Honscheid}, {Hoyle}, {Huff}, {Huterer}, {Jain}, {James},
  {Jarvis}, {Jeffrey}, {Jeltema}, {Kovacs}, {Krause}, {Kron}, {Kuehn},
  {Kuropatkin}, {Lahav}, {Leget}, {Lemos}, {Liddle}, {Lidman}, {Lima}, {Lin},
  {MacCrann}, {Maia}, {Marshall}, {Martini}, {McCullough}, {Melchior},
  {Mena-Fern{\'a}ndez}, {Menanteau}, {Miquel}, {Mohr}, {Morgan}, {Muir},
  {Myles}, {Nadathur}, {Navarro-Alsina}, {Nichol}, {Ogando}, {Omori},
  {Palmese}, {Pandey}, {Park}, {Paz-Chinch{\'o}n}, {Petravick}, {Pieres},
  {Plazas Malag{\'o}n}, {Porredon}, {Prat}, {Raveri}, {Rodriguez-Monroy},
  {Rollins}, {Romer}, {Roodman}, {Rosenfeld}, {Ross}, {Rykoff}, {Samuroff},
  {S{\'a}nchez}, {Sanchez}, {Sanchez}, {Sanchez Cid}, {Scarpine}, {Schubnell},
  {Scolnic}, {Secco}, {Serrano}, {Sevilla-Noarbe}, {Sheldon}, {Shin}, {Smith},
  {Soares-Santos}, {Suchyta}, {Swanson}, {Tabbutt}, {Tarle}, {Thomas}, {To},
  {Troja}, {Troxel}, {Tucker}, {Tutusaus}, {Varga}, {Walker}, {Weaverdyck},
  {Wechsler}, {Weller}, {Yanny}, {Yin}, {Zhang}, {Zuntz}, \& {DES
  Collaboration}}]{Abbott2022}
{Abbott}, T.~M.~C., {Aguena}, M., {Alarcon}, A., {et~al.} 2022, \prd, 105,
  023520

\bibitem[{{Abbott} {et~al.}(2023{\natexlab{a}}){Abbott}, {Aguena}, {Alarcon},
  {Alves}, {Amon}, {Andrade-Oliveira}, {Annis}, {Avila}, {Bacon}, {Baxter},
  {Bechtol}, {Becker}, {Bernstein}, {Birrer}, {Blazek}, {Bocquet},
  {Brandao-Souza}, {Bridle}, {Brooks}, {Burke}, {Camacho}, {Campos}, {Carnero
  Rosell}, {Carrasco Kind}, {Carretero}, {Castander}, {Cawthon}, {Chang},
  {Chen}, {Chen}, {Choi}, {Conselice}, {Cordero}, {Costanzi}, {Crocce}, {da
  Costa}, {Pereira}, {Davis}, {Davis}, {DeRose}, {Desai}, {Di Valentino},
  {Diehl}, {Dodelson}, {Doel}, {Doux}, {Drlica-Wagner}, {Eckert}, {Eifler},
  {Elsner}, {Elvin-Poole}, {Everett}, {Fang}, {Farahi}, {Ferrero}, {Fert{\'e}},
  {Flaugher}, {Fosalba}, {Friedel}, {Friedrich}, {Frieman},
  {Garc{\'\i}a-Bellido}, {Gatti}, {Giani}, {Giannantonio}, {Giannini}, {Gruen},
  {Gruendl}, {Gschwend}, {Gutierrez}, {Hamaus}, {Harrison}, {Hartley},
  {Herner}, {Hinton}, {Hollowood}, {Honscheid}, {Huang}, {Huff}, {Huterer},
  {Jain}, {James}, {Jarvis}, {Jeffrey}, {Jeltema}, {Kovacs}, {Krause}, {Kuehn},
  {Kuropatkin}, {Lahav}, {Lee}, {Leget}, {Lemos}, {Leonard}, {Liddle}, {Lima},
  {Lin}, {MacCrann}, {Marshall}, {McCullough}, {Mena-Fern{\'a}ndez},
  {Menanteau}, {Miquel}, {Miranda}, {Mohr}, {Muir}, {Myles}, {Nadathur},
  {Navarro-Alsina}, {Nichol}, {Ogando}, {Omori}, {Palmese}, {Pandey}, {Park},
  {Paterno}, {Paz-Chinch{\'o}n}, {Percival}, {Pieres}, {Plazas Malag{\'o}n},
  {Porredon}, {Prat}, {Raveri}, {Rodriguez-Monroy}, {Rogozenski}, {Rollins},
  {Romer}, {Roodman}, {Rosenfeld}, {Ross}, {Rykoff}, {Samuroff}, {S{\'a}nchez},
  {Sanchez}, {Sanchez}, {Sanchez Cid}, {Scarpine}, {Scolnic}, {Secco},
  {Sevilla-Noarbe}, {Sheldon}, {Shin}, {Smith}, {Soares-Santos}, {Suchyta},
  {Tabbutt}, {Tarle}, {Thomas}, {To}, {Troja}, {Troxel}, {Tutusaus}, {Varga},
  {Vincenzi}, {Walker}, {Weaverdyck}, {Wechsler}, {Weller}, {Yanny}, {Yin},
  {Zhang}, {Zuntz}, \& {DES Collaboration}}]{Abbott2023c}
{Abbott}, T.~M.~C., {Aguena}, M., {Alarcon}, A., {et~al.} 2023{\natexlab{a}},
  \prd, 107, 083504

\bibitem[{{Abbott} {et~al.}(2023{\natexlab{b}}){Abbott}, {Aguena}, {Alarcon},
  {Alves}, {Amon}, {Andrade-Oliveira}, {Annis}, {Avila}, {Bacon}, {Baxter},
  {Bechtol}, {Becker}, {Bernstein}, {Birrer}, {Blazek}, {Bocquet},
  {Brandao-Souza}, {Bridle}, {Brooks}, {Burke}, {Camacho}, {Campos}, {Carnero
  Rosell}, {Carrasco Kind}, {Carretero}, {Castander}, {Cawthon}, {Chang},
  {Chen}, {Chen}, {Choi}, {Conselice}, {Cordero}, {Costanzi}, {Crocce}, {da
  Costa}, {Pereira}, {Davis}, {Davis}, {DeRose}, {Desai}, {Di Valentino},
  {Diehl}, {Dodelson}, {Doel}, {Doux}, {Drlica-Wagner}, {Eckert}, {Eifler},
  {Elsner}, {Elvin-Poole}, {Everett}, {Fang}, {Farahi}, {Ferrero}, {Fert{\'e}},
  {Flaugher}, {Fosalba}, {Friedel}, {Friedrich}, {Frieman},
  {Garc{\'\i}a-Bellido}, {Gatti}, {Giani}, {Giannantonio}, {Giannini}, {Gruen},
  {Gruendl}, {Gschwend}, {Gutierrez}, {Hamaus}, {Harrison}, {Hartley},
  {Herner}, {Hinton}, {Hollowood}, {Honscheid}, {Huang}, {Huff}, {Huterer},
  {Jain}, {James}, {Jarvis}, {Jeffrey}, {Jeltema}, {Kovacs}, {Krause}, {Kuehn},
  {Kuropatkin}, {Lahav}, {Lee}, {Leget}, {Lemos}, {Leonard}, {Liddle}, {Lima},
  {Lin}, {MacCrann}, {Marshall}, {McCullough}, {Mena-Fern{\'a}ndez},
  {Menanteau}, {Miquel}, {Miranda}, {Mohr}, {Muir}, {Myles}, {Nadathur},
  {Navarro-Alsina}, {Nichol}, {Ogando}, {Omori}, {Palmese}, {Pandey}, {Park},
  {Paterno}, {Paz-Chinch{\'o}n}, {Percival}, {Pieres}, {Plazas Malag{\'o}n},
  {Porredon}, {Prat}, {Raveri}, {Rodriguez-Monroy}, {Rogozenski}, {Rollins},
  {Romer}, {Roodman}, {Rosenfeld}, {Ross}, {Rykoff}, {Samuroff}, {S{\'a}nchez},
  {Sanchez}, {Sanchez}, {Sanchez Cid}, {Scarpine}, {Scolnic}, {Secco},
  {Sevilla-Noarbe}, {Sheldon}, {Shin}, {Smith}, {Soares-Santos}, {Suchyta},
  {Tabbutt}, {Tarle}, {Thomas}, {To}, {Troja}, {Troxel}, {Tutusaus}, {Varga},
  {Vincenzi}, {Walker}, {Weaverdyck}, {Wechsler}, {Weller}, {Yanny}, {Yin},
  {Zhang}, {Zuntz}, \& {DES Collaboration}}]{Abbott2023}
{Abbott}, T.~M.~C., {Aguena}, M., {Alarcon}, A., {et~al.} 2023{\natexlab{b}},
  \prd, 107, 083504

\bibitem[{{Aihara} {et~al.}(2018){Aihara}, {Armstrong}, {Bickerton}, {Bosch},
  {Coupon}, {Furusawa}, {Hayashi}, {Ikeda}, {Kamata}, {Karoji}, {Kawanomoto},
  {Koike}, {Komiyama}, {Lang}, {Lupton}, {Mineo}, {Miyatake}, {Miyazaki},
  {Morokuma}, {Obuchi}, {Oishi}, {Okura}, {Price}, {Takata}, {Tanaka},
  {Tanaka}, {Tanaka}, {Uchida}, {Uraguchi}, {Utsumi}, {Wang}, {Yamada},
  {Yamanoi}, {Yasuda}, {Arimoto}, {Chiba}, {Finet}, {Fujimori}, {Fujimoto},
  {Furusawa}, {Goto}, {Goulding}, {Gunn}, {Harikane}, {Hattori}, {Hayashi},
  {He{\l}miniak}, {Higuchi}, {Hikage}, {Ho}, {Hsieh}, {Huang}, {Huang},
  {Imanishi}, {Iwata}, {Jaelani}, {Jian}, {Kashikawa}, {Katayama}, {Kojima},
  {Konno}, {Koshida}, {Kusakabe}, {Leauthaud}, {Lee}, {Lin}, {Lin},
  {Mandelbaum}, {Matsuoka}, {Medezinski}, {Miyama}, {Momose}, {More}, {More},
  {Mukae}, {Murata}, {Murayama}, {Nagao}, {Nakata}, {Niida}, {Niikura},
  {Nishizawa}, {Oguri}, {Okabe}, {Ono}, {Onodera}, {Onoue}, {Ouchi}, {Pyo},
  {Shibuya}, {Shimasaku}, {Simet}, {Speagle}, {Spergel}, {Strauss}, {Sugahara},
  {Sugiyama}, {Suto}, {Suzuki}, {Tait}, {Takada}, {Terai}, {Toba}, {Turner},
  {Uchiyama}, {Umetsu}, {Urata}, {Usuda}, {Yeh}, \& {Yuma}}]{Aihara2018}
{Aihara}, H., {Armstrong}, R., {Bickerton}, S., {et~al.} 2018, \pasj, 70, S8

\bibitem[{{Aiola} {et~al.}(2020){Aiola}, {Calabrese}, {Maurin}, {Naess},
  {Schmitt}, {Abitbol}, {Addison}, {Ade}, {Alonso}, {Amiri}, {Amodeo},
  {Angile}, {Austermann}, {Baildon}, {Battaglia}, {Beall}, {Bean}, {Becker},
  {Bond}, {Bruno}, {Calafut}, {Campusano}, {Carrero}, {Chesmore}, {Cho},
  {Choi}, {Clark}, {Cothard}, {Crichton}, {Crowley}, {Darwish}, {Datta},
  {Denison}, {Devlin}, {Duell}, {Duff}, {Duivenvoorden}, {Dunkley},
  {D{\"u}nner}, {Essinger-Hileman}, {Fankhanel}, {Ferraro}, {Fox}, {Fuzia},
  {Gallardo}, {Gluscevic}, {Golec}, {Grace}, {Gralla}, {Guan}, {Hall},
  {Halpern}, {Han}, {Hargrave}, {Hasselfield}, {Helton}, {Henderson},
  {Hensley}, {Hill}, {Hilton}, {Hilton}, {Hincks}, {Hlo{\v{z}}ek}, {Ho},
  {Hubmayr}, {Huffenberger}, {Hughes}, {Infante}, {Irwin}, {Jackson}, {Klein},
  {Knowles}, {Koopman}, {Kosowsky}, {Lakey}, {Li}, {Li}, {Li}, {Lokken},
  {Louis}, {Lungu}, {MacInnis}, {Madhavacheril}, {Maldonado}, {Mallaby-Kay},
  {Marsden}, {McMahon}, {Menanteau}, {Moodley}, {Morton}, {Namikawa}, {Nati},
  {Newburgh}, {Nibarger}, {Nicola}, {Niemack}, {Nolta}, {Orlowski-Sherer},
  {Page}, {Pappas}, {Partridge}, {Phakathi}, {Pisano}, {Prince}, {Puddu}, {Qu},
  {Rivera}, {Robertson}, {Rojas}, {Salatino}, {Schaan}, {Schillaci}, {Sehgal},
  {Sherwin}, {Sierra}, {Sievers}, {Sifon}, {Sikhosana}, {Simon}, {Spergel},
  {Staggs}, {Stevens}, {Storer}, {Sunder}, {Switzer}, {Thorne}, {Thornton},
  {Trac}, {Treu}, {Tucker}, {Vale}, {Van Engelen}, {Van Lanen}, {Vavagiakis},
  {Wagoner}, {Wang}, {Ward}, {Wollack}, {Xu}, {Zago}, \& {Zhu}}]{Aiola2020}
{Aiola}, S., {Calabrese}, E., {Maurin}, L., {et~al.} 2020, \jcap, 2020, 047

\bibitem[{{Amon} {et~al.}(2022){Amon}, {Gruen}, {Troxel}, {MacCrann},
  {Dodelson}, {Choi}, {Doux}, {Secco}, {Samuroff}, {Krause}, {Cordero},
  {Myles}, {DeRose}, {Wechsler}, {Gatti}, {Navarro-Alsina}, {Bernstein},
  {Jain}, {Blazek}, {Alarcon}, {Fert{\'e}}, {Lemos}, {Raveri}, {Campos},
  {Prat}, {S{\'a}nchez}, {Jarvis}, {Alves}, {Andrade-Oliveira}, {Baxter},
  {Bechtol}, {Becker}, {Bridle}, {Camacho}, {Carnero Rosell}, {Carrasco Kind},
  {Cawthon}, {Chang}, {Chen}, {Chintalapati}, {Crocce}, {Davis}, {Diehl},
  {Drlica-Wagner}, {Eckert}, {Eifler}, {Elvin-Poole}, {Everett}, {Fang},
  {Fosalba}, {Friedrich}, {Gaztanaga}, {Giannini}, {Gruendl}, {Harrison},
  {Hartley}, {Herner}, {Huang}, {Huff}, {Huterer}, {Kuropatkin}, {Leget},
  {Liddle}, {McCullough}, {Muir}, {Pandey}, {Park}, {Porredon}, {Refregier},
  {Rollins}, {Roodman}, {Rosenfeld}, {Ross}, {Rykoff}, {Sanchez},
  {Sevilla-Noarbe}, {Sheldon}, {Shin}, {Troja}, {Tutusaus}, {Tutusaus},
  {Varga}, {Weaverdyck}, {Yanny}, {Yin}, {Zhang}, {Zuntz}, {Aguena}, {Allam},
  {Annis}, {Bacon}, {Bertin}, {Bhargava}, {Brooks}, {Buckley-Geer}, {Burke},
  {Carretero}, {Costanzi}, {da Costa}, {Pereira}, {De Vicente}, {Desai},
  {Dietrich}, {Doel}, {Ferrero}, {Flaugher}, {Frieman}, {Garc{\'\i}a-Bellido},
  {Gaztanaga}, {Gerdes}, {Giannantonio}, {Gschwend}, {Gutierrez}, {Hinton},
  {Hollowood}, {Honscheid}, {Hoyle}, {James}, {Kron}, {Kuehn}, {Lahav}, {Lima},
  {Lin}, {Maia}, {Marshall}, {Martini}, {Melchior}, {Menanteau}, {Miquel},
  {Mohr}, {Morgan}, {Ogando}, {Palmese}, {Paz-Chinch{\'o}n}, {Petravick},
  {Pieres}, {Romer}, {Sanchez}, {Scarpine}, {Schubnell}, {Serrano}, {Smith},
  {Soares-Santos}, {Tarle}, {Thomas}, {To}, {Weller}, \& {DES
  Collaboration}}]{Amon2022}
{Amon}, A., {Gruen}, D., {Troxel}, M.~A., {et~al.} 2022, \prd, 105, 023514

\bibitem[{{Artis} {et~al.}(2024){Artis}, {Ghirardini}, {Bulbul}, {Grandis},
  {Garrel}, {Clerc}, {Seppi}, {Comparat}, {Cataneo}, {Bahar}, {Balzer}, {Chiu},
  {Gruen}, {Kleinebreil}, {Kluge}, {Krippendorf}, {Li}, {Liu}, {Merloni},
  {Miyatake}, {Miyazaki}, {Nandra}, {Okabe}, {Pacaud}, {Predehl}, {Ramos-Ceja},
  {Reiprich}, {Sanders}, {Schrabback}, {Zelmer}, \& {Zhang}}]{Artis2024}
{Artis}, E., {Ghirardini}, V., {Bulbul}, E., {et~al.} 2024, arXiv e-prints,
  arXiv:2402.08459

\bibitem[{{Asencio} {et~al.}(2021){Asencio}, {Banik}, \&
  {Kroupa}}]{Ascencio2021}
{Asencio}, E., {Banik}, I., \& {Kroupa}, P. 2021, \mnras, 500, 5249

\bibitem[{{Astropy Collaboration} {et~al.}(2022){Astropy Collaboration},
  {Price-Whelan}, {Lim}, {Earl}, {Starkman}, {Bradley}, {Shupe}, {Patil},
  {Corrales}, {Brasseur}, {N{\"o}the}, {Donath}, {Tollerud}, {Morris},
  {Ginsburg}, {Vaher}, {Weaver}, {Tocknell}, {Jamieson}, {van Kerkwijk},
  {Robitaille}, {Merry}, {Bachetti}, {G{\"u}nther}, {Aldcroft},
  {Alvarado-Montes}, {Archibald}, {B{\'o}di}, {Bapat}, {Barentsen},
  {Baz{\'a}n}, {Biswas}, {Boquien}, {Burke}, {Cara}, {Cara}, {Conroy},
  {Conseil}, {Craig}, {Cross}, {Cruz}, {D'Eugenio}, {Dencheva}, {Devillepoix},
  {Dietrich}, {Eigenbrot}, {Erben}, {Ferreira}, {Foreman-Mackey}, {Fox},
  {Freij}, {Garg}, {Geda}, {Glattly}, {Gondhalekar}, {Gordon}, {Grant},
  {Greenfield}, {Groener}, {Guest}, {Gurovich}, {Handberg}, {Hart},
  {Hatfield-Dodds}, {Homeier}, {Hosseinzadeh}, {Jenness}, {Jones}, {Joseph},
  {Kalmbach}, {Karamehmetoglu}, {Ka{\l}uszy{\'n}ski}, {Kelley}, {Kern},
  {Kerzendorf}, {Koch}, {Kulumani}, {Lee}, {Ly}, {Ma}, {MacBride}, {Maljaars},
  {Muna}, {Murphy}, {Norman}, {O'Steen}, {Oman}, {Pacifici}, {Pascual},
  {Pascual-Granado}, {Patil}, {Perren}, {Pickering}, {Rastogi}, {Roulston},
  {Ryan}, {Rykoff}, {Sabater}, {Sakurikar}, {Salgado}, {Sanghi}, {Saunders},
  {Savchenko}, {Schwardt}, {Seifert-Eckert}, {Shih}, {Jain}, {Shukla}, {Sick},
  {Simpson}, {Singanamalla}, {Singer}, {Singhal}, {Sinha}, {Sip{\H{o}}cz},
  {Spitler}, {Stansby}, {Streicher}, {{\v{S}}umak}, {Swinbank}, {Taranu},
  {Tewary}, {Tremblay}, {de Val-Borro}, {Van Kooten}, {Vasovi{\'c}}, {Verma},
  {de Miranda Cardoso}, {Williams}, {Wilson}, {Winkel}, {Wood-Vasey}, {Xue},
  {Yoachim}, {Zhang}, {Zonca}, \& {Astropy Project Contributors}}]{Astropy2022}
{Astropy Collaboration}, {Price-Whelan}, A.~M., {Lim}, P.~L., {et~al.} 2022,
  \apj, 935, 167

\bibitem[{{Aymerich} {et~al.}(2024){Aymerich}, {Douspis}, {Pratt}, {Salvati},
  {Soubri{\'e}}, {Andrade-Santos}, {Forman}, {Jones}, {Aghanim}, {Kraft}, \&
  {van Weeren}}]{Aymerich2024}
{Aymerich}, G., {Douspis}, M., {Pratt}, G.~W., {et~al.} 2024, arXiv e-prints,
  arXiv:2402.04006

\bibitem[{{Basilakos} \& {Anagnostopoulos}(2020)}]{Basilakos2020}
{Basilakos}, S. \& {Anagnostopoulos}, F.~K. 2020, European Physical Journal C,
  80, 212

\bibitem[{{Batista}(2014)}]{Batista2014}
{Batista}, R.~C. 2014, \prd, 89, 123508

\bibitem[{Bayer {et~al.}(2015)Bayer, Osendorfer, Diot-Girard, Rueckstiess, \&
  Urban}]{Bayer2015}
Bayer, J., Osendorfer, C., Diot-Girard, S., Rueckstiess, T., \& Urban, S. 2015,
  TUM, Tech. Rep.

\bibitem[{{Beutler} {et~al.}(2014){Beutler}, {Saito}, {Seo}, {Brinkmann},
  {Dawson}, {Eisenstein}, {Font-Ribera}, {Ho}, {McBride}, {Montesano},
  {Percival}, {Ross}, {Ross}, {Samushia}, {Schlegel}, {S{\'a}nchez}, {Tinker},
  \& {Weaver}}]{Beutler2014}
{Beutler}, F., {Saito}, S., {Seo}, H.-J., {et~al.} 2014, \mnras, 443, 1065

\bibitem[{{Bianchini} {et~al.}(2020){Bianchini}, {Wu}, {Ade}, {Anderson},
  {Austermann}, {Avva}, {Beall}, {Bender}, {Benson}, {Bleem}, {Carlstrom},
  {Chang}, {Chaubal}, {Chiang}, {Citron}, {Moran}, {Crawford}, {Crites}, {de
  Haan}, {Dobbs}, {Everett}, {Gallicchio}, {George}, {Gilbert}, {Gupta},
  {Halverson}, {Harrington}, {Henning}, {Hilton}, {Holder}, {Holzapfel},
  {Hrubes}, {Huang}, {Hubmayr}, {Irwin}, {Knox}, {Lee}, {Li}, {Lowitz},
  {Manzotti}, {McMahon}, {Meyer}, {Millea}, {Mocanu}, {Montgomery}, {Nadolski},
  {Natoli}, {Nibarger}, {Noble}, {Novosad}, {Omori}, {Padin}, {Patil}, {Pryke},
  {Reichardt}, {Ruhl}, {Saliwanchik}, {Sayre}, {Schaffer}, {Sievers}, {Simard},
  {Smecher}, {Stark}, {Story}, {Tucker}, {Vanderlinde}, {Veach}, {Vieira},
  {Wang}, {Whitehorn}, \& {Yefremenko}}]{Bianchini2020}
{Bianchini}, F., {Wu}, W.~L.~K., {Ade}, P.~A.~R., {et~al.} 2020, \apj, 888, 119

\bibitem[{{Biffi} {et~al.}(2018){Biffi}, {Dolag}, \& {Merloni}}]{Biffi2018}
{Biffi}, V., {Dolag}, K., \& {Merloni}, A. 2018, \mnras, 481, 2213

\bibitem[{{Bocquet} {et~al.}(2024){Bocquet}, {Grandis}, {Bleem}, {Klein},
  {Mohr}, {Schrabback}, {Abbott}, {Ade}, {Aguena}, {Alarcon}, {Allam}, {Allen},
  {Alves}, {Amon}, {Anderson}, {Annis}, {Ansarinejad}, {Austermann}, {Avila},
  {Bacon}, {Bayliss}, {Beall}, {Bechtol}, {Becker}, {Bender}, {Benson},
  {Bernstein}, {Bhargava}, {Bianchini}, {Brodwin}, {Brooks}, {Bryant},
  {Campos}, {Canning}, {Carlstrom}, {Carnero Rosell}, {Carrasco Kind},
  {Carretero}, {Castander}, {Cawthon}, {Chang}, {Chang}, {Chaubal}, {Chen},
  {Chiang}, {Choi}, {Chou}, {Citron}, {Corbett Moran}, {Cordero}, {Costanzi},
  {Crawford}, {Crites}, {da Costa}, {Pereira}, {Davis}, {Davis}, {DeRose},
  {Desai}, {de Haan}, {Diehl}, {Dobbs}, {Dodelson}, {Doux}, {Drlica-Wagner},
  {Eckert}, {Elvin-Poole}, {Everett}, {Everett}, {Ferrero}, {Fert{\'e}},
  {Flores}, {Frieman}, {Gallicchio}, {Garc{\'\i}a-Bellido}, {Gatti}, {George},
  {Giannini}, {Gladders}, {Gruen}, {Gruendl}, {Gupta}, {Gutierrez},
  {Halverson}, {Harrison}, {Hartley}, {Herner}, {Hinton}, {Holder},
  {Hollowood}, {Holzapfel}, {Honscheid}, {Hrubes}, {Huang}, {Hubmayr}, {Huff},
  {Huterer}, {Irwin}, {James}, {Jarvis}, {Khullar}, {Kim}, {Knox}, {Kraft},
  {Krause}, {Kuehn}, {Kuropatkin}, {K{\'e}ruzor{\'e}}, {Lahav}, {Lee}, {Leget},
  {Li}, {Lin}, {Lowitz}, {MacCrann}, {Mahler}, {Mantz}, {Marshall},
  {McCullough}, {McDonald}, {McMahon}, {Mena-Fern{\'a}ndez}, {Menanteau},
  {Meyer}, {Miquel}, {Montgomery}, {Myles}, {Natoli}, {Navarro-Alsina},
  {Nibarger}, {Noble}, {Novosad}, {Ogando}, {Omori}, {Padin}, {Pandey},
  {Paschos}, {Patil}, {Pieres}, {Plazas Malag{\'o}n}, {Porredon}, {Prat},
  {Pryke}, {Raveri}, {Reichardt}, {Roberson}, {Rollins}, {Romero}, {Roodman},
  {Ruhl}, {Rykoff}, {Saliwanchik}, {Salvati}, {S{\'a}nchez}, {Sanchez},
  {Sanchez Cid}, {Saro}, {Schaffer}, {Secco}, {Sevilla-Noarbe}, {Sharon},
  {Sheldon}, {Shin}, {Sievers}, {Smecher}, {Smith}, {Somboonpanyakul},
  {Sommer}, {Stalder}, {Stark}, {Stephen}, {Strazzullo}, {Suchyta}, {Tarle},
  {To}, {Troxel}, {Tucker}, {Tutusaus}, {Varga}, {Veach}, {Vieira},
  {Vikhlinin}, {von der Linden}, {Wang}, {Weaverdyck}, {Weller}, {Whitehorn},
  {Wu}, {Yanny}, {Yefremenko}, {Yin}, {Young}, {Zebrowski}, {Zhang}, {Zohren},
  \& {Zuntz}}]{Bocquet2024}
{Bocquet}, S., {Grandis}, S., {Bleem}, L.~E., {et~al.} 2024, arXiv e-prints,
  arXiv:2401.02075

\bibitem[{{Bocquet} {et~al.}(2015){Bocquet}, {Saro}, {Mohr}, {Aird}, {Ashby},
  {Bautz}, {Bayliss}, {Bazin}, {Benson}, {Bleem}, {Brodwin}, {Carlstrom},
  {Chang}, {Chiu}, {Cho}, {Clocchiatti}, {Crawford}, {Crites}, {Desai}, {de
  Haan}, {Dietrich}, {Dobbs}, {Foley}, {Forman}, {Gangkofner}, {George},
  {Gladders}, {Gonzalez}, {Halverson}, {Hennig}, {Hlavacek-Larrondo}, {Holder},
  {Holzapfel}, {Hrubes}, {Jones}, {Keisler}, {Knox}, {Lee}, {Leitch}, {Liu},
  {Lueker}, {Luong-Van}, {Marrone}, {McDonald}, {McMahon}, {Meyer}, {Mocanu},
  {Murray}, {Padin}, {Pryke}, {Reichardt}, {Rest}, {Ruel}, {Ruhl},
  {Saliwanchik}, {Sayre}, {Schaffer}, {Shirokoff}, {Spieler}, {Stalder},
  {Stanford}, {Staniszewski}, {Stark}, {Story}, {Stubbs}, {Vanderlinde},
  {Vieira}, {Vikhlinin}, {Williamson}, {Zahn}, \& {Zenteno}}]{Bocquet2015}
{Bocquet}, S., {Saro}, A., {Mohr}, J.~J., {et~al.} 2015, \apj, 799, 214

\bibitem[{{Brout} {et~al.}(1978){Brout}, {Englert}, \& {Gunzig}}]{Brout1978}
{Brout}, R., {Englert}, F., \& {Gunzig}, E. 1978, Annals of Physics, 115, 78

\bibitem[{{Buchner}(2021)}]{Buchner2021}
{Buchner}, J. 2021, The Journal of Open Source Software, 6, 3001

\bibitem[{{Bulbul} {et~al.}(2024){Bulbul}, {Liu}, {Kluge}, {Zhang}, {Sanders},
  {Bahar}, {Ghirardini}, {Artis}, {Seppi}, {Garrel}, {Ramos-Ceja}, {Comparat},
  {Balzer}, {B{\"o}ckmann}, {Br{\"u}ggen}, {Clerc}, {Dennerl}, {Dolag},
  {Freyberg}, {Grandis}, {Gruen}, {Kleinebreil}, {Krippendorf}, {Lamer},
  {Merloni}, {Migkas}, {Nandra}, {Pacaud}, {Predehl}, {Reiprich}, {Schrabback},
  {Veronica}, {Weller}, \& {Zelmer}}]{Bulbul2023}
{Bulbul}, E., {Liu}, A., {Kluge}, M., {et~al.} 2024, \aap, 685, A106

\bibitem[{{Bullock}(2010)}]{Bullock2010}
{Bullock}, J.~S. 2010, arXiv e-prints, arXiv:1009.4505

\bibitem[{{Busillo} {et~al.}(2023){Busillo}, {Covone}, {Sereno}, {Ingoglia},
  {Radovich}, {Bardelli}, {Castignani}, {Giocoli}, {Lesci}, {Marulli},
  {Maturi}, {Moscardini}, {Puddu}, \& {Roncarelli}}]{Busillo2023}
{Busillo}, V., {Covone}, G., {Sereno}, M., {et~al.} 2023, \mnras, 524, 5050

\bibitem[{{Chisari} {et~al.}(2019){Chisari}, {Alonso}, {Krause}, {Leonard},
  {Bull}, {Neveu}, {Villarreal}, {Singh}, {McClintock}, {Ellison}, {Du},
  {Zuntz}, {Mead}, {Joudaki}, {Lorenz}, {Troester}, {Sanchez}, {Lanusse},
  {Ishak}, {Hlozek}, {Blazek}, {Campagne}, {Almoubayyed}, {Eifler}, {Kirby},
  {Kirkby}, {Plaszczynski}, {Slosar}, {Vrastil}, \& {Wagoner}}]{Chisari2019}
{Chisari}, N.~E., {Alonso}, D., {Krause}, E., {et~al.} 2019, {CCL: Core
  Cosmology Library}, Astrophysics Source Code Library, record ascl:1901.003

\bibitem[{{Chiu} {et~al.}(2023){Chiu}, {Klein}, {Mohr}, \&
  {Bocquet}}]{Chiu2023}
{Chiu}, I.~N., {Klein}, M., {Mohr}, J., \& {Bocquet}, S. 2023, \mnras, 522,
  1601

\bibitem[{{Clerc}(2023)}]{Clerc2023b}
{Clerc}, N. e.~a. 2023

\bibitem[{{Costanzi} {et~al.}(2019){Costanzi}, {Rozo}, {Simet}, {Zhang},
  {Evrard}, {Mantz}, {Rykoff}, {Jeltema}, {Gruen}, {Allen}, {McClintock},
  {Romer}, {von der Linden}, {Farahi}, {DeRose}, {Varga}, {Weller}, {Giles},
  {Hollowood}, {Bhargava}, {Bermeo-Hernandez}, {Chen}, {Abbott}, {Abdalla},
  {Avila}, {Bechtol}, {Brooks}, {Buckley-Geer}, {Burke}, {Rosell}, {Kind},
  {Carretero}, {Crocce}, {Cunha}, {da Costa}, {Davis}, {De Vicente}, {Diehl},
  {Dietrich}, {Doel}, {Eifler}, {Estrada}, {Flaugher}, {Fosalba}, {Frieman},
  {Garc{\'\i}a-Bellido}, {Gaztanaga}, {Gerdes}, {Giannantonio}, {Gruendl},
  {Gschwend}, {Gutierrez}, {Hartley}, {Honscheid}, {Hoyle}, {James}, {Krause},
  {Kuehn}, {Kuropatkin}, {Lima}, {Lin}, {Maia}, {March}, {Marshall}, {Martini},
  {Menanteau}, {Miller}, {Miquel}, {Mohr}, {Ogando}, {Plazas}, {Roodman},
  {Sanchez}, {Scarpine}, {Schindler}, {Schubnell}, {Serrano}, {Sevilla-Noarbe},
  {Sheldon}, {Smith}, {Soares-Santos}, {Sobreira}, {Suchyta}, {Swanson},
  {Tarle}, {Thomas}, \& {Wechsler}}]{Costanzi2019}
{Costanzi}, M., {Rozo}, E., {Simet}, M., {et~al.} 2019, \mnras, 488, 4779

\bibitem[{{Costanzi} {et~al.}(2021){Costanzi}, {Saro}, {Bocquet}, {Abbott},
  {Aguena}, {Allam}, {Amara}, {Annis}, {Avila}, {Bacon}, {Benson}, {Bhargava},
  {Brooks}, {Buckley-Geer}, {Burke}, {Carnero Rosell}, {Carrasco Kind},
  {Carretero}, {Choi}, {da Costa}, {Pereira}, {De Vicente}, {Desai}, {Diehl},
  {Dietrich}, {Doel}, {Eifler}, {Everett}, {Ferrero}, {Fert{\'e}}, {Flaugher},
  {Fosalba}, {Frieman}, {Garc{\'\i}a-Bellido}, {Gaztanaga}, {Gerdes},
  {Giannantonio}, {Giles}, {Grandis}, {Gruen}, {Gruendl}, {Gupta}, {Gutierrez},
  {Hartley}, {Hinton}, {Hollowood}, {Honscheid}, {James}, {Jeltema}, {Krause},
  {Kuehn}, {Kuropatkin}, {Lahav}, {Lima}, {MacCrann}, {Maia}, {Marshall},
  {Menanteau}, {Miquel}, {Mohr}, {Morgan}, {Myles}, {Ogando}, {Palmese},
  {Paz-Chinch{\'o}n}, {Plazas}, {Rapetti}, {Reichardt}, {Romer}, {Roodman},
  {Ruppin}, {Salvati}, {Samuroff}, {Sanchez}, {Scarpine}, {Serrano},
  {Sevilla-Noarbe}, {Singh}, {Smith}, {Soares-Santos}, {Stark}, {Suchyta},
  {Swanson}, {Tarle}, {Thomas}, {To}, {Tucker}, {Varga}, {Wechsler}, {Zhang},
  {DES}, \& {SPT Collaborations}}]{Costanzi2021}
{Costanzi}, M., {Saro}, A., {Bocquet}, S., {et~al.} 2021, \prd, 103, 043522

\bibitem[{{Costanzi} {et~al.}(2013){Costanzi}, {Villaescusa-Navarro}, {Viel},
  {Xia}, {Borgani}, {Castorina}, \& {Sefusatti}}]{Costanzi2013}
{Costanzi}, M., {Villaescusa-Navarro}, F., {Viel}, M., {et~al.} 2013, \jcap,
  2013, 012

\bibitem[{{Dalal} {et~al.}(2023){Dalal}, {Li}, {Nicola}, {Zuntz}, {Strauss},
  {Sugiyama}, {Zhang}, {Rau}, {Mandelbaum}, {Takada}, {More}, {Miyatake},
  {Kannawadi}, {Shirasaki}, {Taniguchi}, {Takahashi}, {Osato}, {Hamana},
  {Oguri}, {Nishizawa}, {Malag{\'o}n}, {Sunayama}, {Alonso}, {Slosar}, {Luo},
  {Armstrong}, {Bosch}, {Hsieh}, {Komiyama}, {Lupton}, {Lust}, {MacArthur},
  {Miyazaki}, {Murayama}, {Nishimichi}, {Okura}, {Price}, {Tait}, {Tanaka}, \&
  {Wang}}]{Dalal2023}
{Dalal}, R., {Li}, X., {Nicola}, A., {et~al.} 2023, \prd, 108, 123519

\bibitem[{{Dark Energy Survey and Kilo-Degree Survey Collaboration}
  {et~al.}(2023){Dark Energy Survey and Kilo-Degree Survey Collaboration},
  {Abbott}, {Aguena}, {Alarcon}, {Alves}, {Amon}, {Andrade-Oliveira}, {Asgari},
  {Avila}, {Bacon}, {Bechtol}, {Becker}, {Bernstein}, {Bertin}, {Bilicki},
  {Blazek}, {Bocquet}, {Brooks}, {Burger}, {Burke}, {Camacho}, {Campos},
  {Carnero Rosell}, {Carrasco Kind}, {Carretero}, {Castander}, {Cawthon},
  {Chang}, {Chen}, {Choi}, {Conselice}, {Cordero}, {Crocce}, {da Costa}, {da
  Silva Pereira}, {Dalal}, {Davis}, {de Jong}, {DeRose}, {Desai}, {Diehl},
  {Dodelson}, {Doel}, {Doux}, {Drlica-Wagner}, {Dvornik}, {Eckert}, {Eifler},
  {Elvin-Poole}, {Everett}, {Fang}, {Ferrero}, {Fert{\'e}}, {Flaugher},
  {Friedrich}, {Frieman}, {Garc{\'\i}a-Bellido}, {Gatti}, {Giannini}, {Giblin},
  {Gruen}, {Gruendl}, {Gutierrez}, {Harrison}, {Hartley}, {Herner}, {Heymans},
  {Hildebrandt}, {Hinton}, {Hoekstra}, {Hollowood}, {Honscheid}, {Huang},
  {Huff}, {Huterer}, {James}, {Jarvis}, {Jeffrey}, {Jeltema}, {Joachimi},
  {Joudaki}, {Kannawadi}, {Krause}, {Kuehn}, {Kuijken}, {Kuropatkin}, {Lahav},
  {Leget}, {Lemos}, {Li}, {Li}, {Liddle}, {Lima}, {Lin}, {Lin}, {MacCrann},
  {Mahony}, {Marshall}, {McCullough}, {Mena-Fern{\'a}ndez}, {Menanteau},
  {Miquel}, {Mohr}, {Muir}, {Myles}, {Napolitano}, {Navarro-Alsina}, {Ogando},
  {Palmese}, {Pandey}, {Park}, {Paterno}, {Peacock}, {Petravick}, {Pieres},
  {Plazas Malag{\'o}n}, {Porredon}, {Prat}, {Radovich}, {Raveri}, {Reischke},
  {Robertson}, {Rollins}, {Romer}, {Roodman}, {Rykoff}, {Samuroff},
  {S{\'a}nchez}, {Sanchez}, {Sanchez}, {Schneider}, {Secco}, {Sevilla-Noarbe},
  {Shan}, {Sheldon}, {Shin}, {Sif{\'o}n}, {Smith}, {Soares-Santos},
  {St{\"o}lzner}, {Suchyta}, {Swanson}, {Tarle}, {Thomas}, {To}, {Troxel},
  {Tr{\"o}ster}, {Tutusaus}, {van den Busch}, {Varga}, {Walker}, {Weaverdyck},
  {Wechsler}, {Weller}, {Wiseman}, {Wright}, {Yanny}, {Yin}, {Yoon}, {Zhang},
  \& {Zuntz}}]{Abbott2023b}
{Dark Energy Survey and Kilo-Degree Survey Collaboration}, {Abbott}, T.~M.~C.,
  {Aguena}, M., {et~al.} 2023, The Open Journal of Astrophysics, 6, 36

\bibitem[{{de Mattia} {et~al.}(2021){de Mattia}, {Ruhlmann-Kleider},
  {Raichoor}, {Ross}, {Tamone}, {Zhao}, {Alam}, {Avila}, {Burtin}, {Bautista},
  {Beutler}, {Brinkmann}, {Brownstein}, {Chapman}, {Chuang}, {Comparat}, {du
  Mas des Bourboux}, {Dawson}, {de la Macorra}, {Gil-Mar{\'\i}n},
  {Gonzalez-Perez}, {Gorgoni}, {Hou}, {Kong}, {Lin}, {Nadathur}, {Newman},
  {Mueller}, {Percival}, {Rezaie}, {Rossi}, {Schneider}, {Tiwari}, {Vivek},
  {Wang}, \& {Zhao}}]{deMattia2021}
{de Mattia}, A., {Ruhlmann-Kleider}, V., {Raichoor}, A., {et~al.} 2021, \mnras,
  501, 5616

\bibitem[{{DESI Collaboration} {et~al.}(2024){DESI Collaboration}, {Adame},
  {Aguilar}, {Ahlen}, {Alam}, {Alexander}, {Alvarez}, {Alves}, {Anand},
  {Andrade}, {Armengaud}, {Avila}, {Aviles}, {Awan}, {Bahr-Kalus}, {Bailey},
  {Baltay}, {Bault}, {Behera}, {BenZvi}, {Bera}, {Beutler}, {Bianchi}, {Blake},
  {Blum}, {Brieden}, {Brodzeller}, {Brooks}, {Buckley-Geer}, {Burtin},
  {Calderon}, {Canning}, {Carnero Rosell}, {Cereskaite}, {Cervantes-Cota},
  {Chabanier}, {Chaussidon}, {Chaves-Montero}, {Chen}, {Chen}, {Claybaugh},
  {Cole}, {Cuceu}, {Davis}, {Dawson}, {de la Macorra}, {de Mattia}, {Deiosso},
  {Dey}, {Dey}, {Ding}, {Doel}, {Edelstein}, {Eftekharzadeh}, {Eisenstein},
  {Elliott}, {Fagrelius}, {Fanning}, {Ferraro}, {Ereza}, {Findlay}, {Flaugher},
  {Font-Ribera}, {Forero-S{\'a}nchez}, {Forero-Romero}, {Frenk},
  {Garcia-Quintero}, {Gazta{\~n}aga}, {Gil-Mar{\'\i}n}, {Gontcho},
  {Gonzalez-Morales}, {Gonzalez-Perez}, {Gordon}, {Green}, {Gruen}, {Gsponer},
  {Gutierrez}, {Guy}, {Hadzhiyska}, {Hahn}, {Hanif}, {Herrera-Alcantar},
  {Honscheid}, {Howlett}, {Huterer}, {Ir{\v{s}}i{\v{c}}}, {Ishak}, {Juneau},
  {Kara{\c{c}}ayl{\i}}, {Kehoe}, {Kent}, {Kirkby}, {Kremin}, {Krolewski},
  {Lai}, {Lan}, {Landriau}, {Lang}, {Lasker}, {Le Goff}, {Le Guillou},
  {Leauthaud}, {Levi}, {Li}, {Linder}, {Lodha}, {Magneville}, {Manera},
  {Margala}, {Martini}, {Maus}, {McDonald}, {Medina-Varela}, {Meisner},
  {Mena-Fern{\'a}ndez}, {Miquel}, {Moon}, {Moore}, {Moustakas}, {Mudur},
  {Mueller}, {Mu{\~n}oz-Guti{\'e}rrez}, {Myers}, {Nadathur}, {Napolitano},
  {Neveux}, {Newman}, {Nguyen}, {Nie}, {Niz}, {Noriega}, {Padmanabhan},
  {Paillas}, {Palanque-Delabrouille}, {Pan}, {Penmetsa}, {Percival}, {Pieri},
  {Pinon}, {Poppett}, {Porredon}, {Prada}, {P{\'e}rez-Fern{\'a}ndez},
  {P{\'e}rez-R{\`a}fols}, {Rabinowitz}, {Raichoor}, {Ram{\'\i}rez-P{\'e}rez},
  {Ramirez-Solano}, {Ravoux}, {Rashkovetskyi}, {Rezaie}, {Rich}, {Rocher},
  {Rockosi}, {Roe}, {Rosado-Marin}, {Ross}, {Rossi}, {Ruggeri},
  {Ruhlmann-Kleider}, {Samushia}, {Sanchez}, {Saulder}, {Schlafly}, {Schlegel},
  {Schubnell}, {Seo}, {Shafieloo}, {Sharples}, {Silber}, {Slosar}, {Smith},
  {Sprayberry}, {Tan}, {Tarl{\'e}}, {Taylor}, {Trusov}, {Ure{\~n}a-L{\'o}pez},
  {Vaisakh}, {Valcin}, {Valdes}, {Vargas-Maga{\~n}a}, {Verde}, {Walther},
  {Wang}, {Wang}, {Weaver}, {Weaverdyck}, {Wechsler}, {Weinberg}, {White},
  {Yu}, {Yu}, {Yuan}, {Y{\`e}che}, {Zaborowski}, {Zarrouk}, {Zhang}, {Zhao},
  {Zhao}, {Zhou}, {Zhuang}, \& {Zou}}]{Desi2024}
{DESI Collaboration}, {Adame}, A.~G., {Aguilar}, J., {et~al.} 2024, arXiv
  e-prints, arXiv:2404.03002

\bibitem[{{Dey} {et~al.}(2019){Dey}, {Schlegel}, {Lang}, {Blum}, {Burleigh},
  {Fan}, {Findlay}, {Finkbeiner}, {Herrera}, {Juneau}, {Landriau}, {Levi},
  {McGreer}, {Meisner}, {Myers}, {Moustakas}, {Nugent}, {Patej}, {Schlafly},
  {Walker}, {Valdes}, {Weaver}, {Y{\`e}che}, {Zou}, {Zhou}, {Abareshi},
  {Abbott}, {Abolfathi}, {Aguilera}, {Alam}, {Allen}, {Alvarez}, {Annis},
  {Ansarinejad}, {Aubert}, {Beechert}, {Bell}, {BenZvi}, {Beutler}, {Bielby},
  {Bolton}, {Brice{\~n}o}, {Buckley-Geer}, {Butler}, {Calamida}, {Carlberg},
  {Carter}, {Casas}, {Castander}, {Choi}, {Comparat}, {Cukanovaite}, {Delubac},
  {DeVries}, {Dey}, {Dhungana}, {Dickinson}, {Ding}, {Donaldson}, {Duan},
  {Duckworth}, {Eftekharzadeh}, {Eisenstein}, {Etourneau}, {Fagrelius},
  {Farihi}, {Fitzpatrick}, {Font-Ribera}, {Fulmer}, {G{\"a}nsicke},
  {Gaztanaga}, {George}, {Gerdes}, {Gontcho}, {Gorgoni}, {Green}, {Guy},
  {Harmer}, {Hernandez}, {Honscheid}, {Huang}, {James}, {Jannuzi}, {Jiang},
  {Joyce}, {Karcher}, {Karkar}, {Kehoe}, {Kneib}, {Kueter-Young}, {Lan},
  {Lauer}, {Le Guillou}, {Le Van Suu}, {Lee}, {Lesser}, {Perreault Levasseur},
  {Li}, {Mann}, {Marshall}, {Mart{\'\i}nez-V{\'a}zquez}, {Martini}, {du Mas des
  Bourboux}, {McManus}, {Meier}, {M{\'e}nard}, {Metcalfe},
  {Mu{\~n}oz-Guti{\'e}rrez}, {Najita}, {Napier}, {Narayan}, {Newman}, {Nie},
  {Nord}, {Norman}, {Olsen}, {Paat}, {Palanque-Delabrouille}, {Peng},
  {Poppett}, {Poremba}, {Prakash}, {Rabinowitz}, {Raichoor}, {Rezaie},
  {Robertson}, {Roe}, {Ross}, {Ross}, {Rudnick}, {Safonova}, {Saha},
  {S{\'a}nchez}, {Savary}, {Schweiker}, {Scott}, {Seo}, {Shan}, {Silva},
  {Slepian}, {Soto}, {Sprayberry}, {Staten}, {Stillman}, {Stupak}, {Summers},
  {Sien Tie}, {Tirado}, {Vargas-Maga{\~n}a}, {Vivas}, {Wechsler}, {Williams},
  {Yang}, {Yang}, {Yapici}, {Zaritsky}, {Zenteno}, {Zhang}, {Zhang}, {Zhou}, \&
  {Zhou}}]{Dey2019}
{Dey}, A., {Schlegel}, D.~J., {Lang}, D., {et~al.} 2019, \aj, 157, 168

\bibitem[{{Di Valentino} {et~al.}(2021{\natexlab{a}}){Di Valentino},
  {Anchordoqui}, {Akarsu}, {Ali-Haimoud}, {Amendola}, {Arendse}, {Asgari},
  {Ballardini}, {Basilakos}, {Battistelli}, {Benetti}, {Birrer}, {Bouchet},
  {Bruni}, {Calabrese}, {Camarena}, {Capozziello}, {Chen}, {Chluba},
  {Chudaykin}, {Colg{\'a}in}, {Cyr-Racine}, {de Bernardis}, {de Cruz
  P{\'e}rez}, {Delabrouille}, {Dunkley}, {Escamilla-Rivera}, {Fert{\'e}},
  {Finelli}, {Freedman}, {Frusciante}, {Giusarma}, {G{\'o}mez-Valent},
  {Handley}, {Harrison}, {Hart}, {Heavens}, {Hildebrandt}, {Holz}, {Huterer},
  {Ivanov}, {Joudaki}, {Kamionkowski}, {Karwal}, {Knox}, {Kumar}, {Lamagna},
  {Lesgourgues}, {Lucca}, {Marra}, {Masi}, {Matarrese}, {Mazumdar},
  {Melchiorri}, {Mena}, {Mersini-Houghton}, {Miranda}, {Moreno-Pulido}, {Mota},
  {Muir}, {Mukherjee}, {Niedermann}, {Notari}, {Nunes}, {Pace},
  {Paliathanasis}, {Palmese}, {Pan}, {Paoletti}, {Pettorino}, {Piacentini},
  {Poulin}, {Raveri}, {Riess}, {Salzano}, {Saridakis}, {Sen}, {Shafieloo},
  {Shajib}, {Silk}, {Silvestri}, {Sloth}, {Smith}, {Sol{\`a} Peracaula}, {van
  de Bruck}, {Verde}, {Visinelli}, {Wandelt}, {Wang}, {Wang}, {Yadav}, \&
  {Yang}}]{DiValentino2021b}
{Di Valentino}, E., {Anchordoqui}, L.~A., {Akarsu}, {\"O}., {et~al.}
  2021{\natexlab{a}}, Astroparticle Physics, 131, 102604

\bibitem[{{Di Valentino} {et~al.}(2021{\natexlab{b}}){Di Valentino}, {Mena},
  {Pan}, {Visinelli}, {Yang}, {Melchiorri}, {Mota}, {Riess}, \&
  {Silk}}]{DiValentino2021}
{Di Valentino}, E., {Mena}, O., {Pan}, S., {et~al.} 2021{\natexlab{b}},
  Classical and Quantum Gravity, 38, 153001

\bibitem[{{Freedman} {et~al.}(2024){Freedman}, {Madore}, {Jang}, {Hoyt}, {Lee},
  \& {Owens}}]{Freedman2024}
{Freedman}, W.~L., {Madore}, B.~F., {Jang}, I.~S., {et~al.} 2024, arXiv
  e-prints, arXiv:2408.06153

\bibitem[{{Fumagalli} {et~al.}(2024){Fumagalli}, {Costanzi}, {Saro}, {Castro},
  \& {Borgani}}]{Fumagalli2024}
{Fumagalli}, A., {Costanzi}, M., {Saro}, A., {Castro}, T., \& {Borgani}, S.
  2024, \aap, 682, A148

\bibitem[{{Garc{\'\i}a-Garc{\'\i}a} {et~al.}(2021){Garc{\'\i}a-Garc{\'\i}a},
  {Ruiz-Zapatero}, {Alonso}, {Bellini}, {Ferreira}, {Mueller}, {Nicola}, \&
  {Ruiz-Lapuente}}]{GarciaGarcia2021}
{Garc{\'\i}a-Garc{\'\i}a}, C., {Ruiz-Zapatero}, J., {Alonso}, D., {et~al.}
  2021, \jcap, 2021, 030

\bibitem[{{Garrel} {et~al.}(2022){Garrel}, {Pierre}, {Valageas}, {Eckert},
  {Marulli}, {Veropalumbo}, {Pacaud}, {Clerc}, {Sereno}, {Umetsu},
  {Moscardini}, {Bhargava}, {Adami}, {Chiappetti}, {Gastaldello},
  {Koulouridis}, {Le Fevre}, \& {Plionis}}]{Garrel2022}
{Garrel}, C., {Pierre}, M., {Valageas}, P., {et~al.} 2022, \aap, 663, A3

\bibitem[{{Gatti} {et~al.}(2021){Gatti}, {Sheldon}, {Amon}, {Becker}, {Troxel},
  {Choi}, {Doux}, {MacCrann}, {Navarro-Alsina}, {Harrison}, {Gruen},
  {Bernstein}, {Jarvis}, {Secco}, {Fert{\'e}}, {Shin}, {McCullough}, {Rollins},
  {Chen}, {Chang}, {Pandey}, {Tutusaus}, {Prat}, {Elvin-Poole}, {Sanchez},
  {Plazas}, {Roodman}, {Zuntz}, {Abbott}, {Aguena}, {Allam}, {Annis}, {Avila},
  {Bacon}, {Bertin}, {Bhargava}, {Brooks}, {Burke}, {Carnero Rosell}, {Carrasco
  Kind}, {Carretero}, {Castander}, {Conselice}, {Costanzi}, {Crocce}, {da
  Costa}, {Davis}, {De Vicente}, {Desai}, {Diehl}, {Dietrich}, {Doel},
  {Drlica-Wagner}, {Eckert}, {Everett}, {Ferrero}, {Frieman},
  {Garc{\'\i}a-Bellido}, {Gerdes}, {Giannantonio}, {Gruendl}, {Gschwend},
  {Gutierrez}, {Hartley}, {Hinton}, {Hollowood}, {Honscheid}, {Hoyle}, {Huff},
  {Huterer}, {Jain}, {James}, {Jeltema}, {Krause}, {Kron}, {Kuropatkin},
  {Lima}, {Maia}, {Marshall}, {Miquel}, {Morgan}, {Myles}, {Palmese},
  {Paz-Chinch{\'o}n}, {Rykoff}, {Samuroff}, {Sanchez}, {Scarpine}, {Schubnell},
  {Serrano}, {Sevilla-Noarbe}, {Smith}, {Soares-Santos}, {Suchyta}, {Swanson},
  {Tarle}, {Thomas}, {To}, {Tucker}, {Varga}, {Wechsler}, {Weller}, {Wester},
  \& {Wilkinson}}]{Gatti2021}
{Gatti}, M., {Sheldon}, E., {Amon}, A., {et~al.} 2021, \mnras, 504, 4312

\bibitem[{{Ghirardini} {et~al.}(2024){Ghirardini}, {Bulbul}, {Artis}, {Clerc},
  {Garrel}, {Grandis}, {Kluge}, {Liu}, {Bahar}, {Balzer}, {Chiu}, {Comparat},
  {Gruen}, {Kleinebreil}, {Krippendorf}, {Merloni}, {Nandra}, {Okabe},
  {Pacaud}, {Predehl}, {Ramos-Ceja}, {Reiprich}, {Sanders}, {Schrabback},
  {Seppi}, {Zelmer}, {Zhang}, {Bornemann}, {Brunner}, {Burwitz}, {Coutinho},
  {Dennerl}, {Freyberg}, {Friedrich}, {Gaida}, {Gueguen}, {Haberl}, {Kink},
  {Lamer}, {Li}, {Liu}, {Maitra}, {Meidinger}, {Mueller}, {Miyatake},
  {Miyazaki}, {Robrade}, {Schwope}, \& {Stewart}}]{Ghirardini2023}
{Ghirardini}, V., {Bulbul}, E., {Artis}, E., {et~al.} 2024, arXiv e-prints,
  arXiv:2402.08458

\bibitem[{{Giblin} {et~al.}(2021){Giblin}, {Heymans}, {Asgari}, {Hildebrandt},
  {Hoekstra}, {Joachimi}, {Kannawadi}, {Kuijken}, {Lin}, {Miller},
  {Tr{\"o}ster}, {van den Busch}, {Wright}, {Bilicki}, {Blake}, {de Jong},
  {Dvornik}, {Erben}, {Getman}, {Napolitano}, {Schneider}, {Shan}, \&
  {Valentijn}}]{Giblin2021}
{Giblin}, B., {Heymans}, C., {Asgari}, M., {et~al.} 2021, \aap, 645, A105

\bibitem[{{Gil-Mar{\'\i}n} {et~al.}(2017){Gil-Mar{\'\i}n}, {Percival}, {Verde},
  {Brownstein}, {Chuang}, {Kitaura}, {Rodr{\'\i}guez-Torres}, \&
  {Olmstead}}]{Gil-Marin2017}
{Gil-Mar{\'\i}n}, H., {Percival}, W.~J., {Verde}, L., {et~al.} 2017, \mnras,
  465, 1757

\bibitem[{{GPy}(since 2012)}]{gpy2014}
{GPy}. since 2012, {GPy}: A Gaussian process framework in python,
  \url{http://github.com/SheffieldML/GPy}

\bibitem[{{Grandis} {et~al.}(2024{\natexlab{a}}){Grandis}, {Aric{\`o}},
  {Schneider}, \& {Linke}}]{Grandis2024}
{Grandis}, S., {Aric{\`o}}, G., {Schneider}, A., \& {Linke}, L.
  2024{\natexlab{a}}, \mnras, 528, 4379

\bibitem[{{Grandis} {et~al.}(2024{\natexlab{b}}){Grandis}, {Ghirardini},
  {Bocquet}, {Garrel}, {Mohr}, {Liu}, {Kluge}, {Kimmig}, {Reiprich}, {Alarcon},
  {Amon}, {Artis}, {Bahar}, {Balzer}, {Bechtol}, {Becker}, {Bernstein},
  {Bulbul}, {Campos}, {Carnero Rosell}, {Carrasco Kind}, {Cawthon}, {Chang},
  {Chen}, {Chiu}, {Choi}, {Clerc}, {Comparat}, {Cordero}, {Davis}, {Derose},
  {Diehl}, {Dodelson}, {Doux}, {Drlica-Wagner}, {Eckert}, {Elvin-Poole},
  {Everett}, {Ferte}, {Gatt}, {Giannini}, {Giles}, {Gruen}, {Gruendl},
  {Harrison}, {Hartley}, {Herner}, {Huf}, {Kleinebreil}, {Kuropatkin}, {Leget},
  {Maccrann}, {Mccullough}, {Merloni}, {Myles}, {Nandra}, {Navarro-Alsina},
  {Okabe}, {Pacaud}, {Pandey}, {Prat}, {Predehl}, {Ramos}, {Raveri}, {Rollins},
  {Roodman}, {Ross}, {Rykoff}, {Sanchez}, {Sanders}, {Schrabback}, {Secco},
  {Seppi}, {Sevilla-Noarbe}, {Sheldon}, {Shin}, {Troxel}, {Tutusaus}, {Varga},
  {Wu}, {Yanny}, {Yin}, {Zhang}, {Zhang}, {Alves}, {Bhargava}, {Brooks},
  {Burke}, {Carretero}, {Costanzi}, {da Costa}, {Pereira}, {De Vicente},
  {Desai}, {Doel}, {Ferrero}, {Flaugher}, {Friedel}, {Frieman},
  {Garc{\'\i}a-Bellido}, {Gutierrez}, {Hinton}, {Hollowood}, {Honscheid},
  {James}, {Jeffrey}, {Lahav}, {Lee}, {Marshall}, {Menanteau}, {Ogando},
  {Pieres}, {Plazas Malag{\'o}n}, {Romer}, {Sanchez}, {Schubnell}, {Smith},
  {Suchyta}, {Swanson}, {Tarle}, {Weaverdyck}, \& {Weller}}]{Grandis2023}
{Grandis}, S., {Ghirardini}, V., {Bocquet}, S., {et~al.} 2024{\natexlab{b}},
  arXiv e-prints, arXiv:2402.08455

\bibitem[{{Harnois-D{\'e}raps} {et~al.}(2021){Harnois-D{\'e}raps}, {Martinet},
  {Castro}, {Dolag}, {Giblin}, {Heymans}, {Hildebrandt}, \&
  {Xia}}]{Harnois2021}
{Harnois-D{\'e}raps}, J., {Martinet}, N., {Castro}, T., {et~al.} 2021, \mnras,
  506, 1623

\bibitem[{Harris {et~al.}(2020)Harris, Millman, van~der Walt, Gommers,
  Virtanen, Cournapeau, Wieser, Taylor, Berg, Smith, Kern, Picus, Hoyer, van
  Kerkwijk, Brett, Haldane, del R{\'{i}}o, Wiebe, Peterson,
  G{\'{e}}rard-Marchant, Sheppard, Reddy, Weckesser, Abbasi, Gohlke, \&
  Oliphant}]{Harris2020}
Harris, C.~R., Millman, K.~J., van~der Walt, S.~J., {et~al.} 2020, Nature, 585,
  357

\bibitem[{{Hildebrandt} {et~al.}(2021){Hildebrandt}, {van den Busch}, {Wright},
  {Blake}, {Joachimi}, {Kuijken}, {Tr{\"o}ster}, {Asgari}, {Bilicki}, {de
  Jong}, {Dvornik}, {Erben}, {Getman}, {Giblin}, {Heymans}, {Kannawadi}, {Lin},
  \& {Shan}}]{Hildebrandt2021}
{Hildebrandt}, H., {van den Busch}, J.~L., {Wright}, A.~H., {et~al.} 2021,
  \aap, 647, A124

\bibitem[{{Hinshaw} {et~al.}(2013){Hinshaw}, {Larson}, {Komatsu}, {Spergel},
  {Bennett}, {Dunkley}, {Nolta}, {Halpern}, {Hill}, {Odegard}, {Page}, {Smith},
  {Weiland}, {Gold}, {Jarosik}, {Kogut}, {Limon}, {Meyer}, {Tucker}, {Wollack},
  \& {Wright}}]{Hinshaw2013}
{Hinshaw}, G., {Larson}, D., {Komatsu}, E., {et~al.} 2013, \apjs, 208, 19

\bibitem[{{Huff} \& {Mandelbaum}(2017)}]{Huff2017}
{Huff}, E. \& {Mandelbaum}, R. 2017, arXiv e-prints, arXiv:1702.02600

\bibitem[{Hunter(2007)}]{Hunter2007matplotlib}
Hunter, J.~D. 2007, Computing in Science \& Engineering, 9, 90

\bibitem[{{Hwang} \& {Noh}(2006)}]{Hwang2006}
{Hwang}, J.-c. \& {Noh}, H. 2006, \mnras, 367, 1515

\bibitem[{{Jullo} {et~al.}(2019){Jullo}, {de la Torre}, {Cousinou},
  {Escoffier}, {Giocoli}, {Metcalf}, {Comparat}, {Shan}, {Makler}, {Kneib},
  {Prada}, {Yepes}, \& {Gottl{\"o}ber}}]{Jullo2019}
{Jullo}, E., {de la Torre}, S., {Cousinou}, M.~C., {et~al.} 2019, \aap, 627,
  A137

\bibitem[{{Kaiser}(1986)}]{Kaiser1986}
{Kaiser}, N. 1986, \mnras, 222, 323

\bibitem[{{Kanehisa} {et~al.}(2024){Kanehisa}, {Pawlowski}, {Heesters}, \&
  {M{\"u}ller}}]{Kanehisa2024}
{Kanehisa}, K.~J., {Pawlowski}, M.~S., {Heesters}, N., \& {M{\"u}ller}, O.
  2024, \aap, 686, A280

\bibitem[{{Kazantzidis} \& {Perivolaropoulos}(2021)}]{Kazantzidis2021}
{Kazantzidis}, L. \& {Perivolaropoulos}, L. 2021, in Modified Gravity and
  Cosmology; An Update by the CANTATA Network, ed. E.~N. {Saridakis},
  R.~{Lazkoz}, V.~{Salzano}, P.~V. {Moniz}, S.~{Capozziello}, J.~{Beltr{\'a}n
  Jim{\'e}nez}, M.~{De Laurentis}, \& G.~J. {Olmo}, 507--537

\bibitem[{{Kiakotou} {et~al.}(2008){Kiakotou}, {Elgar{\o}y}, \&
  {Lahav}}]{Kiakotou2008}
{Kiakotou}, A., {Elgar{\o}y}, {\O}., \& {Lahav}, O. 2008, \prd, 77, 063005

\bibitem[{{Kleinebreil} {et~al.}(2024){Kleinebreil}, {Grandis}, {Schrabback},
  {Ghirardini}, {Chiu}, {Liu}, {Kluge}, {Reiprich}, {Artis}, {Bahar}, {Balzer},
  {Bulbul}, {Clerc}, {Comparat}, {Garrel}, {Gruen}, {Li}, {Miyatake},
  {Miyazaki}, {Ramos-Ceja}, {Sanders}, {Seppi}, {Okabe}, \&
  {Zhang}}]{Kleinebreil2024}
{Kleinebreil}, F., {Grandis}, S., {Schrabback}, T., {et~al.} 2024, arXiv
  e-prints, arXiv:2402.08456

\bibitem[{{Kluge} {et~al.}(2024){Kluge}, {Comparat}, {Liu}, {Balzer}, {Bulbul},
  {Ider Chitham}, {Ghirardini}, {Garrel}, {Bahar}, {Artis}, {Bender}, {Clerc},
  {Dwelly}, {Fabricius}, {Grandis}, {Hern{\'a}ndez-Lang}, {Hill}, {Joshi},
  {Lamer}, {Merloni}, {Nandra}, {Pacaud}, {Predehl}, {Ramos-Ceja}, {Reiprich},
  {Salvato}, {Sanders}, {Schrabback}, {Seppi}, {Zelmer}, {Zenteno}, \&
  {Zhang}}]{Kluge2023}
{Kluge}, M., {Comparat}, J., {Liu}, A., {et~al.} 2024, arXiv e-prints,
  arXiv:2402.08453

\bibitem[{{Kuijken} {et~al.}(2019){Kuijken}, {Heymans}, {Dvornik},
  {Hildebrandt}, {de Jong}, {Wright}, {Erben}, {Bilicki}, {Giblin}, {Shan},
  {Getman}, {Grado}, {Hoekstra}, {Miller}, {Napolitano}, {Paolilo}, {Radovich},
  {Schneider}, {Sutherland}, {Tewes}, {Tortora}, {Valentijn}, \& {Verdoes
  Kleijn}}]{Kuijken2019}
{Kuijken}, K., {Heymans}, C., {Dvornik}, A., {et~al.} 2019, \aap, 625, A2

\bibitem[{{Lesci} {et~al.}(2022){Lesci}, {Nanni}, {Marulli}, {Moscardini},
  {Veropalumbo}, {Maturi}, {Sereno}, {Radovich}, {Bellagamba}, {Roncarelli},
  {Bardelli}, {Castignani}, {Covone}, {Giocoli}, {Ingoglia}, \&
  {Puddu}}]{Lesci2022}
{Lesci}, G.~F., {Nanni}, L., {Marulli}, F., {et~al.} 2022, \aap, 665, A100

\bibitem[{{Lewis} \& {Challinor}(2011)}]{Lewis2011CAMB}
{Lewis}, A. \& {Challinor}, A. 2011, {CAMB: Code for Anisotropies in the
  Microwave Background}, Astrophysics Source Code Library, record ascl:1102.026

\bibitem[{{Li} {et~al.}(2022){Li}, {Miyatake}, {Luo}, {More}, {Oguri},
  {Hamana}, {Mandelbaum}, {Shirasaki}, {Takada}, {Armstrong}, {Kannawadi},
  {Takita}, {Miyazaki}, {Nishizawa}, {Plazas Malagon}, {Strauss}, {Tanaka}, \&
  {Yoshida}}]{Li2022}
{Li}, X., {Miyatake}, H., {Luo}, W., {et~al.} 2022, \pasj, 74, 421

\bibitem[{{Li} {et~al.}(2023){Li}, {Zhang}, {Sugiyama}, {Dalal}, {Rau},
  {Mandelbaum}, {Takada}, {More}, {Strauss}, {Miyatake}, {Shirasaki}, {Hamana},
  {Oguri}, {Luo}, {Nishizawa}, {Takahashi}, {Nicola}, {Osato}, {Kannawadi},
  {Sunayama}, {Armstrong}, {Komiyama}, {Lupton}, {Lust}, {Miyazaki},
  {Murayama}, {Nishimichi}, {Okura}, {Price}, {Tait}, {Tanaka}, \&
  {Wang}}]{Li2023}
{Li}, X., {Zhang}, T., {Sugiyama}, S., {et~al.} 2023, arXiv e-prints,
  arXiv:2304.00702

\bibitem[{{Lin} {et~al.}(2024){Lin}, {Jain}, {Raveri}, {Baxter}, {Chang},
  {Gatti}, {Lee}, \& {Muir}}]{Lin2024}
{Lin}, M.-X., {Jain}, B., {Raveri}, M., {et~al.} 2024, \prd, 109, 063523

\bibitem[{{Linder} \& {Cahn}(2007)}]{Linder2007}
{Linder}, E.~V. \& {Cahn}, R.~N. 2007, Astroparticle Physics, 28, 481

\bibitem[{{Liu} {et~al.}(2023){Liu}, {Yuan}, {Pan}, {Zhang}, {Wang}, \&
  {Fan}}]{Liu2023}
{Liu}, X., {Yuan}, S., {Pan}, C., {et~al.} 2023, \mnras, 519, 594

\bibitem[{{Madhavacheril} {et~al.}(2024){Madhavacheril}, {Qu}, {Sherwin},
  {MacCrann}, {Li}, {Abril-Cabezas}, {Ade}, {Aiola}, {Alford}, {Amiri},
  {Amodeo}, {An}, {Atkins}, {Austermann}, {Battaglia}, {Battistelli}, {Beall},
  {Bean}, {Beringue}, {Bhandarkar}, {Biermann}, {Bolliet}, {Bond}, {Cai},
  {Calabrese}, {Calafut}, {Capalbo}, {Carrero}, {Challinor}, {Chesmore}, {Cho},
  {Choi}, {Clark}, {C{\'o}rdova Rosado}, {Cothard}, {Coughlin}, {Coulton},
  {Crowley}, {Dalal}, {Darwish}, {Devlin}, {Dicker}, {Doze}, {Duell}, {Duff},
  {Duivenvoorden}, {Dunkley}, {D{\"u}nner}, {Fanfani}, {Fankhanel}, {Farren},
  {Ferraro}, {Freundt}, {Fuzia}, {Gallardo}, {Garrido}, {Givans}, {Gluscevic},
  {Golec}, {Guan}, {Hall}, {Halpern}, {Han}, {Harrison}, {Hasselfield},
  {Healy}, {Henderson}, {Hensley}, {Herv{\'\i}as-Caimapo}, {Hill}, {Hilton},
  {Hilton}, {Hincks}, {Hlo{\v{z}}ek}, {Ho}, {Huber}, {Hubmayr}, {Huffenberger},
  {Hughes}, {Irwin}, {Isopi}, {Jense}, {Keller}, {Kim}, {Knowles}, {Koopman},
  {Kosowsky}, {Kramer}, {Kusiak}, {La Posta}, {Lague}, {Lakey}, {Lee}, {Li},
  {Limon}, {Lokken}, {Louis}, {Lungu}, {MacInnis}, {Maldonado}, {Maldonado},
  {Mallaby-Kay}, {Marques}, {McMahon}, {Mehta}, {Menanteau}, {Moodley},
  {Morris}, {Mroczkowski}, {Naess}, {Namikawa}, {Nati}, {Newburgh}, {Nicola},
  {Niemack}, {Nolta}, {Orlowski-Scherer}, {Page}, {Pandey}, {Partridge},
  {Prince}, {Puddu}, {Radiconi}, {Robertson}, {Rojas}, {Sakuma}, {Salatino},
  {Schaan}, {Schmitt}, {Sehgal}, {Shaikh}, {Sierra}, {Sievers}, {Sif{\'o}n},
  {Simon}, {Sonka}, {Spergel}, {Staggs}, {Storer}, {Switzer}, {Tampier},
  {Thornton}, {Trac}, {Treu}, {Tucker}, {Ullom}, {Vale}, {Van Engelen}, {Van
  Lanen}, {van Marrewijk}, {Vargas}, {Vavagiakis}, {Wagoner}, {Wang}, {Wenzl},
  {Wollack}, {Xu}, {Zago}, \& {Zheng}}]{Madhavacheril2024}
{Madhavacheril}, M.~S., {Qu}, F.~J., {Sherwin}, B.~D., {et~al.} 2024, \apj,
  962, 113

\bibitem[{{Mantz} {et~al.}(2015){Mantz}, {von der Linden}, {Allen},
  {Applegate}, {Kelly}, {Morris}, {Rapetti}, {Schmidt}, {Adhikari}, {Allen},
  {Burchat}, {Burke}, {Cataneo}, {Donovan}, {Ebeling}, {Shand era}, \&
  {Wright}}]{Mantz2015}
{Mantz}, A.~B., {von der Linden}, A., {Allen}, S.~W., {et~al.} 2015, \mnras,
  446, 2205

\bibitem[{{Marques} {et~al.}(2024){Marques}, {Liu}, {Shirasaki}, {Thiele},
  {Grand{\'o}n}, {Huffenberger}, {Cheng}, {Harnois-D{\'e}raps}, {Osato}, \&
  {Coulton}}]{Marques2024}
{Marques}, G.~A., {Liu}, J., {Shirasaki}, M., {et~al.} 2024, \mnras, 528, 4513

\bibitem[{{Martinet} {et~al.}(2018){Martinet}, {Schneider}, {Hildebrandt},
  {Shan}, {Asgari}, {Dietrich}, {Harnois-D{\'e}raps}, {Erben}, {Grado},
  {Heymans}, {Hoekstra}, {Klaes}, {Kuijken}, {Merten}, \&
  {Nakajima}}]{Martinet2018}
{Martinet}, N., {Schneider}, P., {Hildebrandt}, H., {et~al.} 2018, \mnras, 474,
  712

\bibitem[{{Merloni} {et~al.}(2024){Merloni}, {Lamer}, {Liu}, {Ramos-Ceja},
  {Brunner}, {Bulbul}, {Dennerl}, {Doroshenko}, {Freyberg}, {Friedrich},
  {Gatuzz}, {Georgakakis}, {Haberl}, {Igo}, {Kreykenbohm}, {Liu}, {Maitra},
  {Malyali}, {Mayer}, {Nandra}, {Predehl}, {Robrade}, {Salvato}, {Sanders},
  {Stewart}, {Tub{\'\i}n-Arenas}, {Weber}, {Wilms}, {Arcodia}, {Artis},
  {Aschersleben}, {Avakyan}, {Aydar}, {Bahar}, {Balzer}, {Becker}, {Berger},
  {Boller}, {Bornemann}, {Br{\"u}ggen}, {Brusa}, {Buchner}, {Burwitz},
  {Camilloni}, {Clerc}, {Comparat}, {Coutinho}, {Czesla}, {Dannhauer},
  {Dauner}, {Dauser}, {Dietl}, {Dolag}, {Dwelly}, {Egg}, {Ehl}, {Freund},
  {Friedrich}, {Gaida}, {Garrel}, {Ghirardini}, {Gokus}, {Gr{\"u}nwald},
  {Grandis}, {Grotova}, {Gruen}, {Gueguen}, {H{\"a}mmerich}, {Hamaus},
  {Hasinger}, {Haubner}, {Homan}, {Ider Chitham}, {Joseph}, {Joyce},
  {K{\"o}nig}, {Kaltenbrunner}, {Khokhriakova}, {Kink}, {Kirsch}, {Kluge},
  {Knies}, {Krippendorf}, {Krumpe}, {Kurpas}, {Li}, {Liu}, {Locatelli},
  {Lorenz}, {M{\"u}ller}, {Magaudda}, {Mannes}, {McCall}, {Meidinger},
  {Michailidis}, {Migkas}, {Mu{\~n}oz-Giraldo}, {Musiimenta}, {Nguyen-Dang},
  {Ni}, {Olechowska}, {Ota}, {Pacaud}, {Pasini}, {Perinati}, {Pires},
  {Pommranz}, {Ponti}, {Poppenhaeger}, {P{\"u}hlhofer}, {Rau}, {Reh},
  {Reiprich}, {Roster}, {Saeedi}, {Santangelo}, {Sasaki}, {Schmitt},
  {Schneider}, {Schrabback}, {Schuster}, {Schwope}, {Seppi}, {Serim},
  {Shreeram}, {Sokolova-Lapa}, {Starck}, {Stelzer}, {Stierhof}, {Suleimanov},
  {Tenzer}, {Traulsen}, {Tr{\"u}mper}, {Tsuge}, {Urrutia}, {Veronica},
  {Waddell}, {Willer}, {Wolf}, {Yeung}, {Zainab}, {Zangrandi}, {Zhang},
  {Zhang}, \& {Zheng}}]{Merloni2023}
{Merloni}, A., {Lamer}, G., {Liu}, T., {et~al.} 2024, \aap, 682, A34

\bibitem[{{Migkas} {et~al.}(2020){Migkas}, {Schellenberger}, {Reiprich},
  {Pacaud}, {Ramos-Ceja}, \& {Lovisari}}]{Migkas2020}
{Migkas}, K., {Schellenberger}, G., {Reiprich}, T.~H., {et~al.} 2020, \aap,
  636, A15

\bibitem[{{Moresco} \& {Marulli}(2017)}]{Moresco2017}
{Moresco}, M. \& {Marulli}, F. 2017, \mnras, 471, L82

\bibitem[{{Nesseris} \& {Perivolaropoulos}(2008)}]{Nesseris2008}
{Nesseris}, S. \& {Perivolaropoulos}, L. 2008, \prd, 77, 023504

\bibitem[{{Nguyen} {et~al.}(2023){Nguyen}, {Huterer}, \& {Wen}}]{Nguyen2023}
{Nguyen}, N.-M., {Huterer}, D., \& {Wen}, Y. 2023, \prl, 131, 111001

\bibitem[{{Palanque-Delabrouille} {et~al.}(2020){Palanque-Delabrouille},
  {Y{\`e}che}, {Sch{\"o}neberg}, {Lesgourgues}, {Walther}, {Chabanier}, \&
  {Armengaud}}]{PalanqueDelabrouille2020}
{Palanque-Delabrouille}, N., {Y{\`e}che}, C., {Sch{\"o}neberg}, N., {et~al.}
  2020, \jcap, 2020, 038

\bibitem[{{Peebles}(2022)}]{Peebles2022}
{Peebles}, P.~J.~E. 2022, Annals of Physics, 447, 169159

\bibitem[{{Perivolaropoulos} \& {Skara}(2022)}]{Perivolaropoulos2022}
{Perivolaropoulos}, L. \& {Skara}, F. 2022, \nar, 95, 101659

\bibitem[{{Planck Collaboration} {et~al.}(2014){Planck Collaboration}, {Ade},
  {Aghanim}, {Armitage-Caplan}, {Arnaud}, {Ashdown}, {Atrio-Barandela},
  {Aumont}, {Baccigalupi}, {Banday}, {Barreiro}, {Bartlett}, {Battaner},
  {Benabed}, {Beno{\^\i}t}, {Benoit-L{\'e}vy}, {Bernard}, {Bersanelli},
  {Bielewicz}, {Bobin}, {Bock}, {Bonaldi}, {Bond}, {Borrill}, {Bouchet},
  {Bridges}, {Bucher}, {Burigana}, {Butler}, {Calabrese}, {Cappellini},
  {Cardoso}, {Catalano}, {Challinor}, {Chamballu}, {Chary}, {Chen}, {Chiang},
  {Chiang}, {Christensen}, {Church}, {Clements}, {Colombi}, {Colombo},
  {Couchot}, {Coulais}, {Crill}, {Curto}, {Cuttaia}, {Danese}, {Davies},
  {Davis}, {de Bernardis}, {de Rosa}, {de Zotti}, {Delabrouille}, {Delouis},
  {D{\'e}sert}, {Dickinson}, {Diego}, {Dolag}, {Dole}, {Donzelli}, {Dor{\'e}},
  {Douspis}, {Dunkley}, {Dupac}, {Efstathiou}, {Elsner}, {En{\ss}lin},
  {Eriksen}, {Finelli}, {Forni}, {Frailis}, {Fraisse}, {Franceschi}, {Gaier},
  {Galeotta}, {Galli}, {Ganga}, {Giard}, {Giardino}, {Giraud-H{\'e}raud},
  {Gjerl{\o}w}, {Gonz{\'a}lez-Nuevo}, {G{\'o}rski}, {Gratton}, {Gregorio},
  {Gruppuso}, {Gudmundsson}, {Haissinski}, {Hamann}, {Hansen}, {Hanson},
  {Harrison}, {Henrot-Versill{\'e}}, {Hern{\'a}ndez-Monteagudo}, {Herranz},
  {Hildebrandt}, {Hivon}, {Hobson}, {Holmes}, {Hornstrup}, {Hou}, {Hovest},
  {Huffenberger}, {Jaffe}, {Jaffe}, {Jewell}, {Jones}, {Juvela},
  {Keih{\"a}nen}, {Keskitalo}, {Kisner}, {Kneissl}, {Knoche}, {Knox}, {Kunz},
  {Kurki-Suonio}, {Lagache}, {L{\"a}hteenm{\"a}ki}, {Lamarre}, {Lasenby},
  {Lattanzi}, {Laureijs}, {Lawrence}, {Leach}, {Leahy}, {Leonardi},
  {Le{\'o}n-Tavares}, {Lesgourgues}, {Lewis}, {Liguori}, {Lilje},
  {Linden-V{\o}rnle}, {L{\'o}pez-Caniego}, {Lubin}, {Mac{\'\i}as-P{\'e}rez},
  {Maffei}, {Maino}, {Mandolesi}, {Maris}, {Marshall}, {Martin},
  {Mart{\'\i}nez-Gonz{\'a}lez}, {Masi}, {Massardi}, {Matarrese}, {Matthai},
  {Mazzotta}, {Meinhold}, {Melchiorri}, {Melin}, {Mendes}, {Menegoni},
  {Mennella}, {Migliaccio}, {Millea}, {Mitra}, {Miville-Desch{\^e}nes},
  {Moneti}, {Montier}, {Morgante}, {Mortlock}, {Moss}, {Munshi}, {Murphy},
  {Naselsky}, {Nati}, {Natoli}, {Netterfield}, {N{\o}rgaard-Nielsen},
  {Noviello}, {Novikov}, {Novikov}, {O'Dwyer}, {Osborne}, {Oxborrow}, {Paci},
  {Pagano}, {Pajot}, {Paladini}, {Paoletti}, {Partridge}, {Pasian},
  {Patanchon}, {Pearson}, {Pearson}, {Peiris}, {Perdereau}, {Perotto},
  {Perrotta}, {Pettorino}, {Piacentini}, {Piat}, {Pierpaoli}, {Pietrobon},
  {Plaszczynski}, {Platania}, {Pointecouteau}, {Polenta}, {Ponthieu}, {Popa},
  {Poutanen}, {Pratt}, {Pr{\'e}zeau}, {Prunet}, {Puget}, {Rachen}, {Reach},
  {Rebolo}, {Reinecke}, {Remazeilles}, {Renault}, {Ricciardi}, {Riller},
  {Ristorcelli}, {Rocha}, {Rosset}, {Roudier}, {Rowan-Robinson},
  {Rubi{\~n}o-Mart{\'\i}n}, {Rusholme}, {Sandri}, {Santos}, {Savelainen},
  {Savini}, {Scott}, {Seiffert}, {Shellard}, {Spencer}, {Starck}, {Stolyarov},
  {Stompor}, {Sudiwala}, {Sunyaev}, {Sureau}, {Sutton}, {Suur-Uski}, {Sygnet},
  {Tauber}, {Tavagnacco}, {Terenzi}, {Toffolatti}, {Tomasi}, {Tristram},
  {Tucci}, {Tuovinen}, {T{\"u}rler}, {Umana}, {Valenziano}, {Valiviita}, {Van
  Tent}, {Vielva}, {Villa}, {Vittorio}, {Wade}, {Wandelt}, {Wehus}, {White},
  {White}, {Wilkinson}, {Yvon}, {Zacchei}, \&
  {Zonca}}]{PlanckCollaboration2014}
{Planck Collaboration}, {Ade}, P.~A.~R., {Aghanim}, N., {et~al.} 2014, \aap,
  571, A16

\bibitem[{{Planck Collaboration} {et~al.}(2016){Planck Collaboration}, {Ade},
  {Aghanim}, {Arnaud}, {Ashdown}, {Aumont}, {Baccigalupi}, {Banday},
  {Barreiro}, {Bartlett}, {Bartolo}, {Battaner}, {Battye}, {Benabed},
  {Beno{\^\i}t}, {Benoit-L{\'e}vy}, {Bernard}, {Bersanelli}, {Bielewicz},
  {Bock}, {Bonaldi}, {Bonavera}, {Bond}, {Borrill}, {Bouchet}, {Boulanger},
  {Bucher}, {Burigana}, {Butler}, {Calabrese}, {Cardoso}, {Catalano},
  {Challinor}, {Chamballu}, {Chary}, {Chiang}, {Chluba}, {Christensen},
  {Church}, {Clements}, {Colombi}, {Colombo}, {Combet}, {Coulais}, {Crill},
  {Curto}, {Cuttaia}, {Danese}, {Davies}, {Davis}, {de Bernardis}, {de Rosa},
  {de Zotti}, {Delabrouille}, {D{\'e}sert}, {Di Valentino}, {Dickinson},
  {Diego}, {Dolag}, {Dole}, {Donzelli}, {Dor{\'e}}, {Douspis}, {Ducout},
  {Dunkley}, {Dupac}, {Efstathiou}, {Elsner}, {En{\ss}lin}, {Eriksen},
  {Farhang}, {Fergusson}, {Finelli}, {Forni}, {Frailis}, {Fraisse},
  {Franceschi}, {Frejsel}, {Galeotta}, {Galli}, {Ganga}, {Gauthier}, {Gerbino},
  {Ghosh}, {Giard}, {Giraud-H{\'e}raud}, {Giusarma}, {Gjerl{\o}w},
  {Gonz{\'a}lez-Nuevo}, {G{\'o}rski}, {Gratton}, {Gregorio}, {Gruppuso},
  {Gudmundsson}, {Hamann}, {Hansen}, {Hanson}, {Harrison}, {Helou},
  {Henrot-Versill{\'e}}, {Hern{\'a}ndez-Monteagudo}, {Herranz}, {Hildebrandt},
  {Hivon}, {Hobson}, {Holmes}, {Hornstrup}, {Hovest}, {Huang}, {Huffenberger},
  {Hurier}, {Jaffe}, {Jaffe}, {Jones}, {Juvela}, {Keih{\"a}nen}, {Keskitalo},
  {Kisner}, {Kneissl}, {Knoche}, {Knox}, {Kunz}, {Kurki-Suonio}, {Lagache},
  {L{\"a}hteenm{\"a}ki}, {Lamarre}, {Lasenby}, {Lattanzi}, {Lawrence}, {Leahy},
  {Leonardi}, {Lesgourgues}, {Levrier}, {Lewis}, {Liguori}, {Lilje},
  {Linden-V{\o}rnle}, {L{\'o}pez-Caniego}, {Lubin}, {Mac{\'\i}as-P{\'e}rez},
  {Maggio}, {Maino}, {Mandolesi}, {Mangilli}, {Marchini}, {Maris}, {Martin},
  {Martinelli}, {Mart{\'\i}nez-Gonz{\'a}lez}, {Masi}, {Matarrese}, {McGehee},
  {Meinhold}, {Melchiorri}, {Melin}, {Mendes}, {Mennella}, {Migliaccio},
  {Millea}, {Mitra}, {Miville-Desch{\^e}nes}, {Moneti}, {Montier}, {Morgante},
  {Mortlock}, {Moss}, {Munshi}, {Murphy}, {Naselsky}, {Nati}, {Natoli},
  {Netterfield}, {N{\o}rgaard-Nielsen}, {Noviello}, {Novikov}, {Novikov},
  {Oxborrow}, {Paci}, {Pagano}, {Pajot}, {Paladini}, {Paoletti}, {Partridge},
  {Pasian}, {Patanchon}, {Pearson}, {Perdereau}, {Perotto}, {Perrotta},
  {Pettorino}, {Piacentini}, {Piat}, {Pierpaoli}, {Pietrobon}, {Plaszczynski},
  {Pointecouteau}, {Polenta}, {Popa}, {Pratt}, {Pr{\'e}zeau}, {Prunet},
  {Puget}, {Rachen}, {Reach}, {Rebolo}, {Reinecke}, {Remazeilles}, {Renault},
  {Renzi}, {Ristorcelli}, {Rocha}, {Rosset}, {Rossetti}, {Roudier},
  {Rouill{\'e} d'Orfeuil}, {Rowan-Robinson}, {Rubi{\~n}o-Mart{\'\i}n},
  {Rusholme}, {Said}, {Salvatelli}, {Salvati}, {Sandri}, {Santos},
  {Savelainen}, {Savini}, {Scott}, {Seiffert}, {Serra}, {Shellard}, {Spencer},
  {Spinelli}, {Stolyarov}, {Stompor}, {Sudiwala}, {Sunyaev}, {Sutton},
  {Suur-Uski}, {Sygnet}, {Tauber}, {Terenzi}, {Toffolatti}, {Tomasi},
  {Tristram}, {Trombetti}, {Tucci}, {Tuovinen}, {T{\"u}rler}, {Umana},
  {Valenziano}, {Valiviita}, {Van Tent}, {Vielva}, {Villa}, {Wade}, {Wandelt},
  {Wehus}, {White}, {White}, {Wilkinson}, {Yvon}, {Zacchei}, \&
  {Zonca}}]{Planckcollaboration2016b}
{Planck Collaboration}, {Ade}, P.~A.~R., {Aghanim}, N., {et~al.} 2016, \aap,
  594, A13

\bibitem[{{Planck Collaboration} {et~al.}(2020){Planck Collaboration},
  {Aghanim}, {Akrami}, {Ashdown}, {Aumont}, {Baccigalupi}, {Ballardini},
  {Banday}, {Barreiro}, {Bartolo}, {Basak}, {Battye}, {Benabed}, {Bernard},
  {Bersanelli}, {Bielewicz}, {Bock}, {Bond}, {Borrill}, {Bouchet}, {Boulanger},
  {Bucher}, {Burigana}, {Butler}, {Calabrese}, {Cardoso}, {Carron},
  {Challinor}, {Chiang}, {Chluba}, {Colombo}, {Combet}, {Contreras}, {Crill},
  {Cuttaia}, {de Bernardis}, {de Zotti}, {Delabrouille}, {Delouis}, {Di
  Valentino}, {Diego}, {Dor{\'e}}, {Douspis}, {Ducout}, {Dupac}, {Dusini},
  {Efstathiou}, {Elsner}, {En{\ss}lin}, {Eriksen}, {Fantaye}, {Farhang},
  {Fergusson}, {Fernandez-Cobos}, {Finelli}, {Forastieri}, {Frailis},
  {Fraisse}, {Franceschi}, {Frolov}, {Galeotta}, {Galli}, {Ganga},
  {G{\'e}nova-Santos}, {Gerbino}, {Ghosh}, {Gonz{\'a}lez-Nuevo}, {G{\'o}rski},
  {Gratton}, {Gruppuso}, {Gudmundsson}, {Hamann}, {Handley}, {Hansen},
  {Herranz}, {Hildebrandt}, {Hivon}, {Huang}, {Jaffe}, {Jones}, {Karakci},
  {Keih{\"a}nen}, {Keskitalo}, {Kiiveri}, {Kim}, {Kisner}, {Knox},
  {Krachmalnicoff}, {Kunz}, {Kurki-Suonio}, {Lagache}, {Lamarre}, {Lasenby},
  {Lattanzi}, {Lawrence}, {Le Jeune}, {Lemos}, {Lesgourgues}, {Levrier},
  {Lewis}, {Liguori}, {Lilje}, {Lilley}, {Lindholm}, {L{\'o}pez-Caniego},
  {Lubin}, {Ma}, {Mac{\'\i}as-P{\'e}rez}, {Maggio}, {Maino}, {Mandolesi},
  {Mangilli}, {Marcos-Caballero}, {Maris}, {Martin}, {Martinelli},
  {Mart{\'\i}nez-Gonz{\'a}lez}, {Matarrese}, {Mauri}, {McEwen}, {Meinhold},
  {Melchiorri}, {Mennella}, {Migliaccio}, {Millea}, {Mitra},
  {Miville-Desch{\^e}nes}, {Molinari}, {Montier}, {Morgante}, {Moss}, {Natoli},
  {N{\o}rgaard-Nielsen}, {Pagano}, {Paoletti}, {Partridge}, {Patanchon},
  {Peiris}, {Perrotta}, {Pettorino}, {Piacentini}, {Polastri}, {Polenta},
  {Puget}, {Rachen}, {Reinecke}, {Remazeilles}, {Renzi}, {Rocha}, {Rosset},
  {Roudier}, {Rubi{\~n}o-Mart{\'\i}n}, {Ruiz-Granados}, {Salvati}, {Sandri},
  {Savelainen}, {Scott}, {Shellard}, {Sirignano}, {Sirri}, {Spencer},
  {Sunyaev}, {Suur-Uski}, {Tauber}, {Tavagnacco}, {Tenti}, {Toffolatti},
  {Tomasi}, {Trombetti}, {Valenziano}, {Valiviita}, {Van Tent}, {Vibert},
  {Vielva}, {Villa}, {Vittorio}, {Wandelt}, {Wehus}, {White}, {White},
  {Zacchei}, \& {Zonca}}]{PlanckCollaboration2020}
{Planck Collaboration}, {Aghanim}, N., {Akrami}, Y., {et~al.} 2020, \aap, 641,
  A6

\bibitem[{{Polarski} \& {Gannouji}(2008)}]{Polarski2008}
{Polarski}, D. \& {Gannouji}, R. 2008, Physics Letters B, 660, 439

\bibitem[{{Polarski} {et~al.}(2016){Polarski}, {Starobinsky}, \&
  {Giacomini}}]{Polarski2016}
{Polarski}, D., {Starobinsky}, A.~A., \& {Giacomini}, H. 2016, \jcap, 2016, 037

\bibitem[{{Preston} {et~al.}(2023){Preston}, {Amon}, \&
  {Efstathiou}}]{Preston2023}
{Preston}, C., {Amon}, A., \& {Efstathiou}, G. 2023, \mnras, 525, 5554

\bibitem[{{Rapetti} {et~al.}(2009){Rapetti}, {Allen}, {Mantz}, \&
  {Ebeling}}]{Rapetti2009}
{Rapetti}, D., {Allen}, S.~W., {Mantz}, A., \& {Ebeling}, H. 2009, \mnras, 400,
  699

\bibitem[{{Rapetti} {et~al.}(2013){Rapetti}, {Blake}, {Allen}, {Mantz},
  {Parkinson}, \& {Beutler}}]{Rapetti2013}
{Rapetti}, D., {Blake}, C., {Allen}, S.~W., {et~al.} 2013, \mnras, 432, 973

\bibitem[{{Riess} {et~al.}(1998){Riess}, {Filippenko}, {Challis},
  {Clocchiatti}, {Diercks}, {Garnavich}, {Gilliland}, {Hogan}, {Jha},
  {Kirshner}, {Leibundgut}, {Phillips}, {Reiss}, {Schmidt}, {Schommer},
  {Smith}, {Spyromilio}, {Stubbs}, {Suntzeff}, \& {Tonry}}]{Riess1998}
{Riess}, A.~G., {Filippenko}, A.~V., {Challis}, P., {et~al.} 1998, \aj, 116,
  1009

\bibitem[{{Rubin} \& {Ford}(1970)}]{Rubin1970}
{Rubin}, V.~C. \& {Ford}, W.~Kent, J. 1970, \apj, 159, 379

\bibitem[{{Salvati} {et~al.}(2018){Salvati}, {Douspis}, \&
  {Aghanim}}]{Salvati2018}
{Salvati}, L., {Douspis}, M., \& {Aghanim}, N. 2018, \aap, 614, A13

\bibitem[{{Salvati} {et~al.}(2022){Salvati}, {Saro}, {Bocquet}, {Costanzi},
  {Ansarinejad}, {Benson}, {Bleem}, {Calzadilla}, {Carlstrom}, {Chang},
  {Chown}, {Crites}, {de Haan}, {Dobbs}, {Everett}, {Floyd}, {Grandis},
  {George}, {Halverson}, {Holder}, {Holzapfel}, {Hrubes}, {Lee}, {Luong-Van},
  {McDonald}, {McMahon}, {Meyer}, {Millea}, {Mocanu}, {Mohr}, {Natoli},
  {Omori}, {Padin}, {Pryke}, {Reichardt}, {Ruhl}, {Ruppin}, {Schaffer},
  {Schrabback}, {Shirokoff}, {Staniszewski}, {Stark}, {Vieira}, \&
  {Williamson}}]{Salvati2022}
{Salvati}, L., {Saro}, A., {Bocquet}, S., {et~al.} 2022, \apj, 934, 129

\bibitem[{{Samushia} {et~al.}(2013){Samushia}, {Reid}, {White}, {Percival},
  {Cuesta}, {Lombriser}, {Manera}, {Nichol}, {Schneider}, {Bizyaev},
  {Brewington}, {Malanushenko}, {Malanushenko}, {Oravetz}, {Pan}, {Simmons},
  {Shelden}, {Snedden}, {Tinker}, {Weaver}, {York}, \& {Zhao}}]{Samushia2013}
{Samushia}, L., {Reid}, B.~A., {White}, M., {et~al.} 2013, \mnras, 429, 1514

\bibitem[{{Sheldon} \& {Huff}(2017)}]{Sheldon2017}
{Sheldon}, E.~S. \& {Huff}, E.~M. 2017, \apj, 841, 24

\bibitem[{{Sunayama} {et~al.}(2023){Sunayama}, {Miyatake}, {Sugiyama}, {More},
  {Li}, {Dalal}, {Rau}, {Shi}, {Chiu}, {Shirasaki}, {Zhang}, \&
  {Nishizawa}}]{Sunayama2023}
{Sunayama}, T., {Miyatake}, H., {Sugiyama}, S., {et~al.} 2023, arXiv e-prints,
  arXiv:2309.13025

\bibitem[{{Tinker} {et~al.}(2008){Tinker}, {Kravtsov}, {Klypin}, {Abazajian},
  {Warren}, {Yepes}, {Gottl{\"o}ber}, \& {Holz}}]{Tinker2008}
{Tinker}, J., {Kravtsov}, A.~V., {Klypin}, A., {et~al.} 2008, \apj, 688, 709

\bibitem[{{Truemper}(1993)}]{Truemper1993}
{Truemper}, J. 1993, Science, 260, 1769

\bibitem[{Virtanen {et~al.}(2020)Virtanen, Gommers, Oliphant, Haberland, Reddy,
  Cournapeau, Burovski, Peterson, Weckesser, Bright, {van der Walt}, Brett,
  Wilson, Millman, Mayorov, Nelson, Jones, Kern, Larson, Carey, Polat, Feng,
  Moore, {VanderPlas}, Laxalde, Perktold, Cimrman, Henriksen, Quintero, Harris,
  Archibald, Ribeiro, Pedregosa, {van Mulbregt}, \& {SciPy 1.0
  Contributors}}]{Virtanen2020SciPy}
Virtanen, P., Gommers, R., Oliphant, T.~E., {et~al.} 2020, Nature Methods, 17,
  261

\bibitem[{{von der Linden} {et~al.}(2014){von der Linden}, {Allen},
  {Applegate}, {Kelly}, {Allen}, {Ebeling}, {Burchat}, {Burke}, {Donovan},
  {Morris}, {Blandford}, {Erben}, \& {Mantz}}]{vonderLinden2014}
{von der Linden}, A., {Allen}, M.~T., {Applegate}, D.~E., {et~al.} 2014,
  \mnras, 439, 2

\bibitem[{{Wang} \& {Steinhardt}(1998)}]{Wang1998}
{Wang}, L. \& {Steinhardt}, P.~J. 1998, \apj, 508, 483

\bibitem[{{Wen} {et~al.}(2023){Wen}, {Nguyen}, \& {Huterer}}]{Wen2023}
{Wen}, Y., {Nguyen}, N.-M., \& {Huterer}, D. 2023, \jcap, 2023, 028

\bibitem[{{White} {et~al.}(2022){White}, {Zhou}, {DeRose}, {Ferraro}, {Chen},
  {Kokron}, {Bailey}, {Brooks}, {Garc{\'\i}a-Bellido}, {Guy}, {Honscheid},
  {Kehoe}, {Kremin}, {Levi}, {Palanque-Delabrouille}, {Poppett}, {Schlegel}, \&
  {Tarle}}]{White2022}
{White}, M., {Zhou}, R., {DeRose}, J., {et~al.} 2022, \jcap, 2022, 007

\bibitem[{{Wright} {et~al.}(2020){Wright}, {Hildebrandt}, {van den Busch}, \&
  {Heymans}}]{Wright2020}
{Wright}, A.~H., {Hildebrandt}, H., {van den Busch}, J.~L., \& {Heymans}, C.
  2020, \aap, 637, A100

\bibitem[{{Zarrouk} {et~al.}(2018){Zarrouk}, {Burtin}, {Gil-Mar{\'\i}n},
  {Ross}, {Tojeiro}, {P{\^a}ris}, {Dawson}, {Myers}, {Percival}, {Chuang},
  {Zhao}, {Bautista}, {Comparat}, {Gonz{\'a}lez-P{\'e}rez}, {Habib},
  {Heitmann}, {Hou}, {Laurent}, {Le Goff}, {Prada}, {Rodr{\'\i}guez-Torres},
  {Rossi}, {Ruggeri}, {S{\'a}nchez}, {Schneider}, {Tinker}, {Wang},
  {Y{\`e}che}, {Baumgarten}, {Brownstein}, {de la Torre}, {du Mas des
  Bourboux}, {Kneib}, {Mariappan}, {Palanque-Delabrouille}, {Peacock},
  {Petitjean}, {Seo}, \& {Zhao}}]{Zarrouk2018}
{Zarrouk}, P., {Burtin}, E., {Gil-Mar{\'\i}n}, H., {et~al.} 2018, \mnras, 477,
  1639

\bibitem[{{Zubeldia} \& {Challinor}(2019)}]{Zubeldia2019}
{Zubeldia}, {\'I}. \& {Challinor}, A. 2019, \mnras, 489, 401

\bibitem[{{Z{\"u}rcher} {et~al.}(2022){Z{\"u}rcher}, {Fluri}, {Sgier},
  {Kacprzak}, {Gatti}, {Doux}, {Whiteway}, {R{\'e}fr{\'e}gier}, {Chang},
  {Jeffrey}, {Jain}, {Lemos}, {Bacon}, {Alarcon}, {Amon}, {Bechtol}, {Becker},
  {Bernstein}, {Campos}, {Chen}, {Choi}, {Davis}, {Derose}, {Dodelson},
  {Elsner}, {Elvin-Poole}, {Everett}, {Ferte}, {Gruen}, {Harrison}, {Huterer},
  {Jarvis}, {Leget}, {Maccrann}, {Mccullough}, {Muir}, {Myles}, {Navarro
  Alsina}, {Pandey}, {Prat}, {Raveri}, {Rollins}, {Roodman}, {Sanchez},
  {Secco}, {Sheldon}, {Shin}, {Troxel}, {Tutusaus}, {Yin}, {Aguena}, {Allam},
  {Andrade-Oliveira}, {Annis}, {Bertin}, {Brooks}, {Burke}, {Carnero Rosell},
  {Carrasco Kind}, {Carretero}, {Castander}, {Cawthon}, {Conselice},
  {Costanzi}, {da Costa}, {da Silva Pereira}, {Davis}, {De Vicente}, {Desai},
  {Diehl}, {Dietrich}, {Doel}, {Eckert}, {Evrard}, {Ferrero}, {Flaugher},
  {Fosalba}, {Friedel}, {Frieman}, {Garcia-Bellido}, {Gaztanaga}, {Gerdes},
  {Giannantonio}, {Gruendl}, {Gschwend}, {Gutierrez}, {Hinton}, {Hollowood},
  {Honscheid}, {Hoyle}, {James}, {Kuehn}, {Kuropatkin}, {Lahav}, {Lidman},
  {Lima}, {Maia}, {Marshall}, {Melchior}, {Menanteau}, {Miquel}, {Morgan},
  {Palmese}, {Paz-Chinchon}, {Pieres}, {Plazas Malag{\'o}n}, {Reil}, {Rodriguez
  Monroy}, {Romer}, {Sanchez}, {Scarpine}, {Schubnell}, {Serrano}, {Sevilla},
  {Smith}, {Suchyta}, {Tarle}, {Thomas}, {To}, {Varga}, {Weller}, {Wilkinson},
  \& {DES Collaboration}}]{Zurcher2022}
{Z{\"u}rcher}, D., {Fluri}, J., {Sgier}, R., {et~al.} 2022, \mnras, 511, 2075

\end{thebibliography}
\begin{appendix}
\section{$\gamma-\Lambda\mathrm{CDM}$ posteriors}
\label{app:glcdm} 
In appendix \ref{app:glcdm}, we show the full cosmological corner plot in the case of the $\gamma \Lambda$CDM model and the correlation of the cosmological parameters with the scaling relation parameters. The constraints presented are obtained from cluster count only. We only show the main parameters of interest.
\begin{figure}[!h]
\noindent\parbox{\textwidth}{
\includegraphics[scale=0.43]{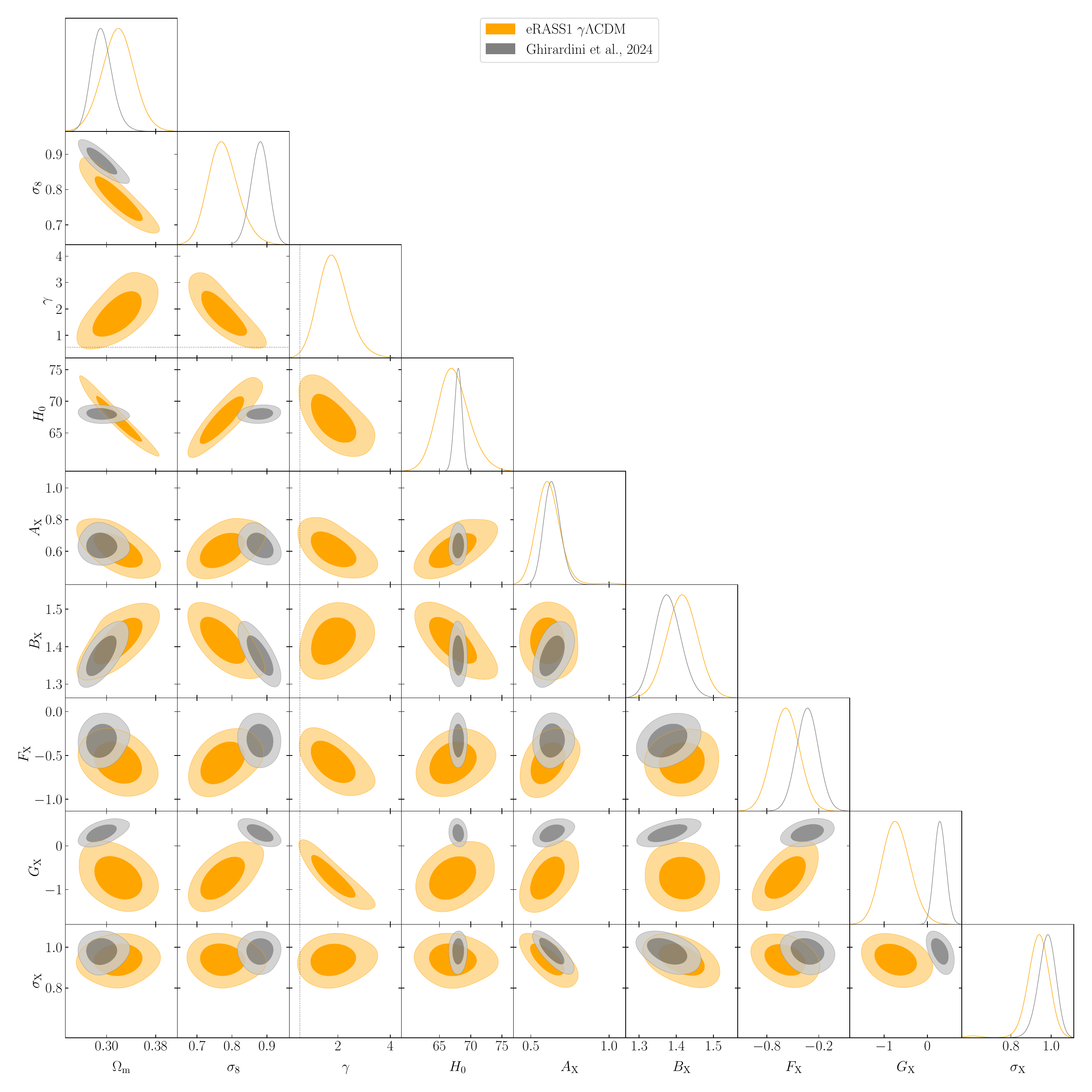}
}
\noindent\parbox{\textwidth}{\caption{The full cosmological corner plot of the parameters obtained in the $\gamma\Lambda\mathrm{CDM}$ case is shown in orange color. For comparison purposes, we also show the controls of the $\Lambda\mathrm{CDM}$ model from \citet{Ghirardini2023} in gray contours.}
}
\label{fig:corner_glcdm}
\end{figure} 
\newpage\phantom{skippage}
\newpage\phantom{skippage}

\section{$\gamma-w\mathrm{CDM}$ posteriors}
\label{app:wcdm}
In appendix \ref{app:wcdm}, we show the full cosmological corner plot in the case of the $\gamma - w$CDM model and the correlation of the cosmological parameters with the scaling relation parameters.
\begin{figure}[!h]
\noindent\parbox{\textwidth}{
\includegraphics[scale=0.38]{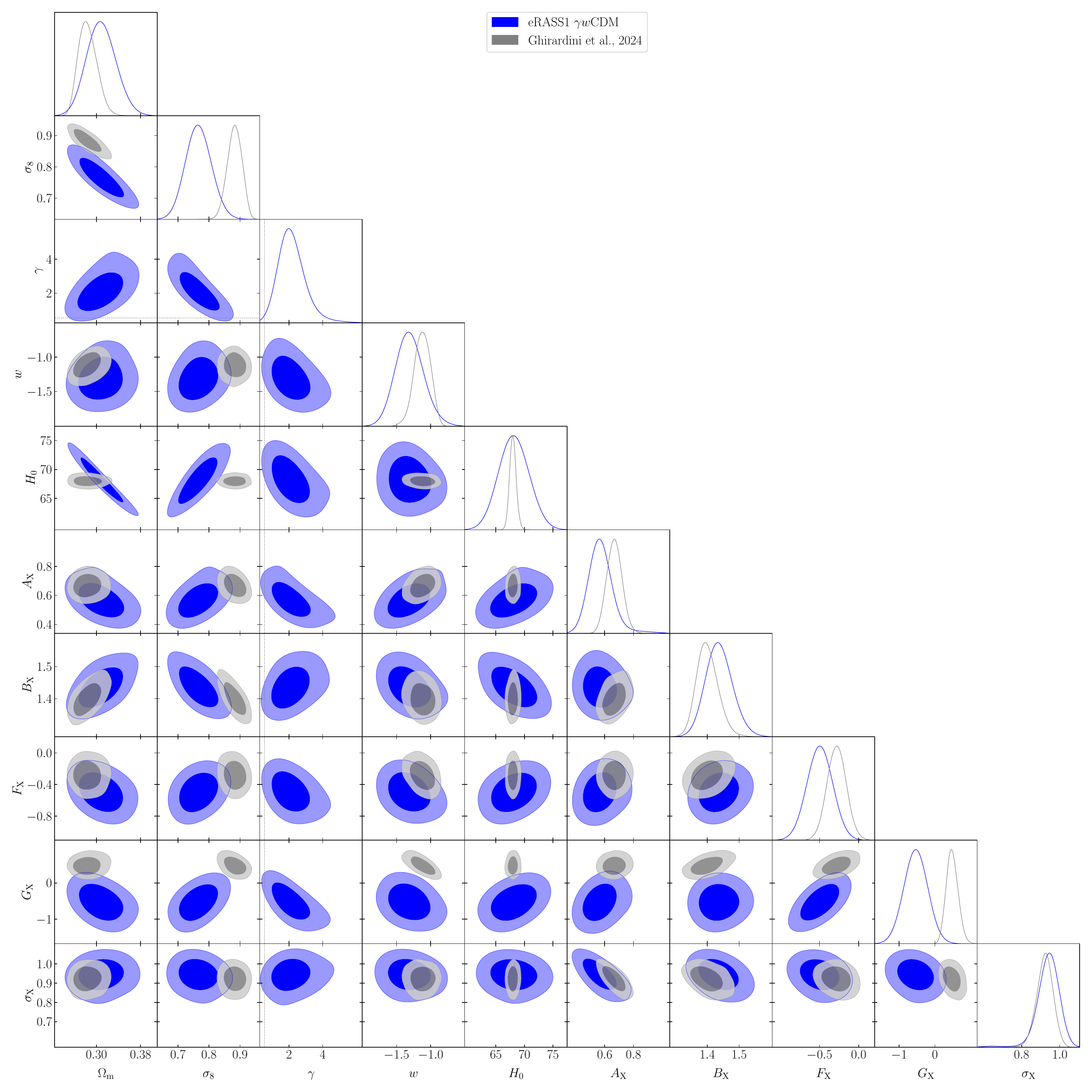}
}
\noindent\parbox{\textwidth}{\caption{Full cosmological corner plot obtained for the $\gamma-w\mathrm{CDM}$ model. We do not show the parameters that are not constrained by clusters.}
}
\label{fig:corner_wcdm}
\end{figure} 
\newpage\phantom{skippage}
\newpage\phantom{skippage}
\section{Scaling relations}
\label{app:scaling_relations}
In this appendix, we show the consistency of the scaling relations parameters fitted by different models. We first show the consistency in the case of $\gamma-\Lambda$CDM and $\gamma-w$CDM compared to the results of the $\Lambda$CDM analysis presented in \cite{Ghirardini2023}, and the weak lensing constraints obtained by \cite{Grandis2023}. Then, we compare the result of the scaling relations in different redshift bins. We note that in the case of the variation of the cosmic linear growth index, the value of the parameter $G_\mathrm{X}$ is lowered. This is due to the reduction of structure formation, which impacts the redshift dependence of the x-ray observable to mass scaling relation.
\begin{figure}[!h]
\noindent\parbox{\textwidth}{
\includegraphics[scale=0.6]{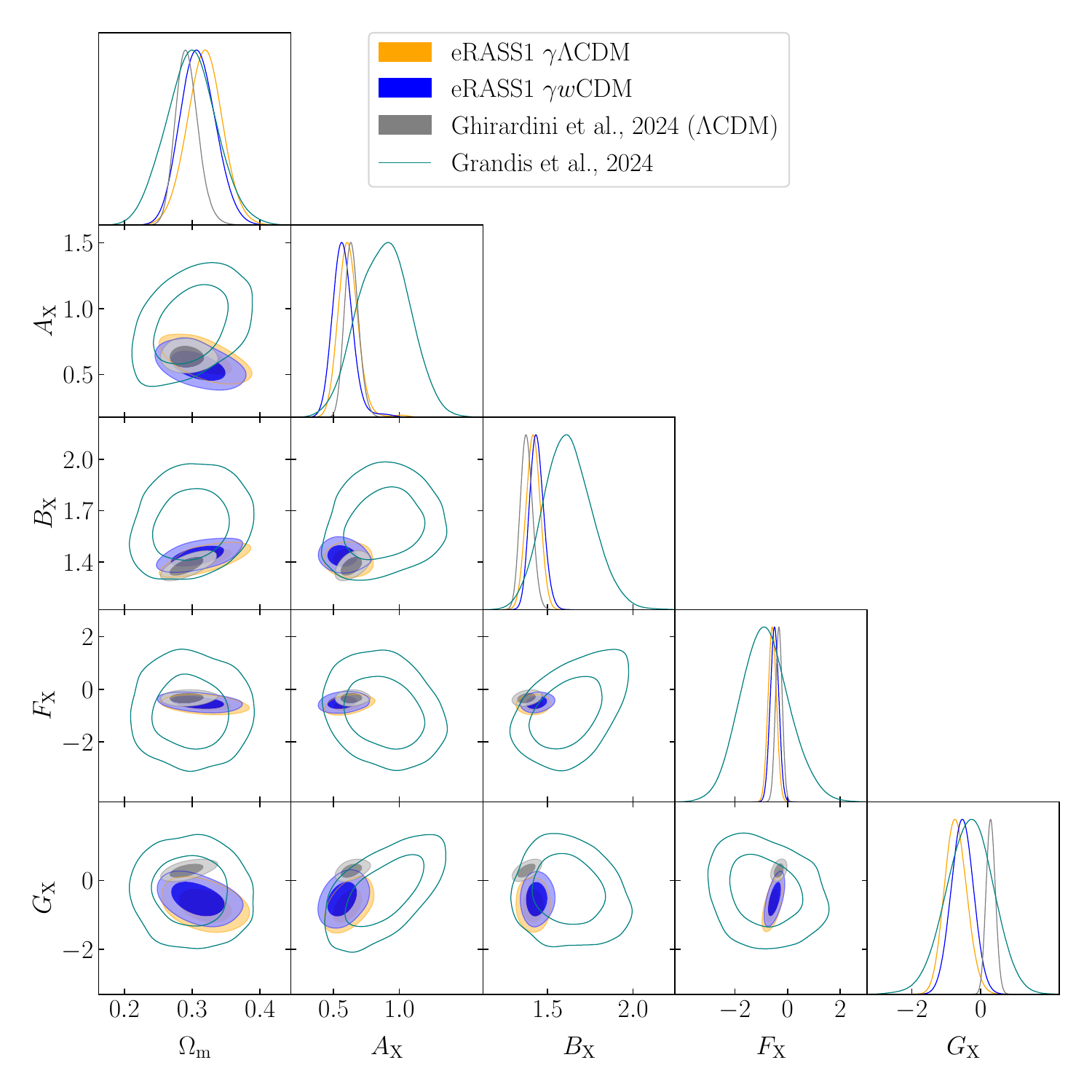}
}
\noindent\parbox{\textwidth}{
\caption{
Posterior distribution of parameters of the X-ray scaling relation presented in Section~\ref{sec:s_inference}. The $\Lambda\mathrm{CDM}$ contours are shown in gray, the $\gamma-\Lambda\mathrm{CDM}$ ones in orange, and the $\gamma-w\mathrm{CDM}$ in blue. The parameters of the scaling relation are mostly consistent against the changes in the modeling, except for the $G_\mathrm{X}$, quantifying the redshift dependence of the count rate to mass ($C_\mathrm{R}-M$) and redshift relation. }
}
\label{fig:corner_scaling}
\end{figure} 
\newpage\phantom{skippage}
\newpage\phantom{skippage}
\begin{figure}[!h]
\noindent\parbox{\textwidth}{
\includegraphics[width=\textwidth,height=21cm]{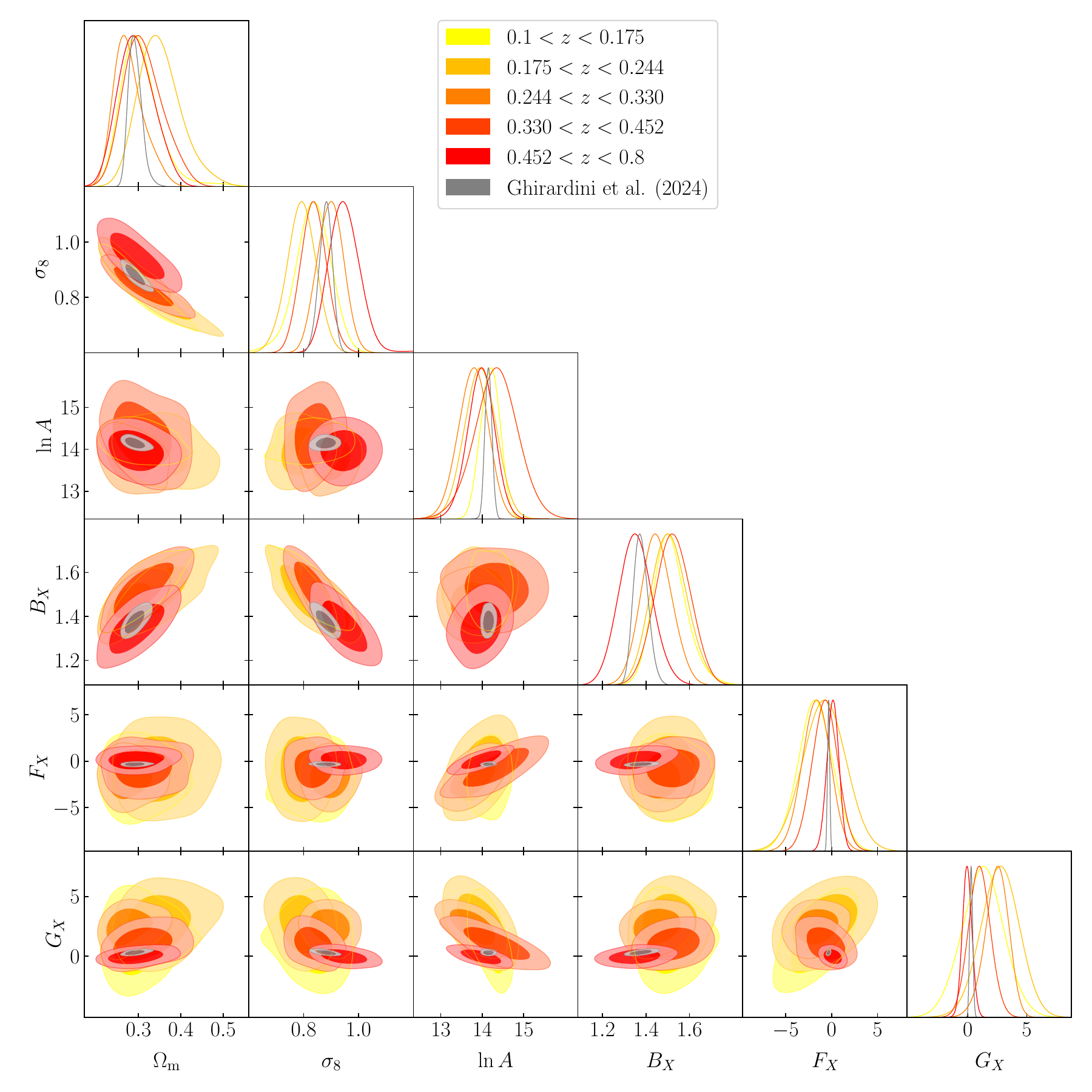}
}
\noindent\parbox{\textwidth}{\caption{Scaling relation and cosmological parameters obtained from the different redshift bins. The color scheme corresponds to the one described in Figure~\ref{fig:sample_bin}. Despite all the bins having the same number of clusters, the constraints are slightly smaller as we increase the redshift, as higher redshift bins are larger, thus providing more redshift leverage to constrain the scaling relation. All posteriors are in good agreement. Note that the pivot redshift is changed in every bin (see Section~\ref{sec:direct_strucutre_growth}). As a consequence, the normalization of the scaling relation $A_\mathrm{X}$ is different in every bin. We rescale it and plot $\ln A = \ln A_\mathrm{X} + F_\mathrm{X}\ln(1+z_{\mathrm{p},i})\ln(2) +2\ln(D_\mathrm{L}(z_{\mathrm{p},i}))-2\ln(E_\mathrm{X}(z_{\mathrm{p},i}))-G_\mathrm{X}\ln(1+z_{\mathrm{p},i})$, where $z_{\mathrm{p},i}$ represent the pivot redshift in each redshift bin. }
}
\label{fig:corner_scaling_bins}
\end{figure} 
\newpage\phantom{skippage}
\newpage\phantom{skippage}
\section{Redshift dependant systematic effect}
\label{app:red_dep}

As shown in Figure~\ref{fig:bin_om_s8}, our highest redshift bin $(0.452<z<0.8)$ yields a best-fit $\sigma_8$ values that are higher than the one obtained in the other redshift bins. Although we stay in statistical agreement, this motivates us to investigate this process in detail. This is also seen in the $\Sh$ measurement in Figure~\ref{fig:s8_z}. Finally, it can also be seen in the redshift dependence of the scaling relation parameters corner plot shown in appendix \ref{app:scaling_relations}. This might be the hint for an unknown systematic effect in the selection of the high redshift cluster population. By construction, this does not affect our model-agnostic binned results. To check the robustness of our results on model-based growth of structures against this putative effect, we rerun the whole pipeline, excluding the highest redshift bin. The results of the fit of the cosmic linear growth index are shown in the figure below. The obtained subcatalog contains a sample of 4196 clusters (80\% of the full cosmology sample).

In the absence of the high-$z$ clusters, we find a good agreement between both cluster catalogs, as well as with other LSS surveys that tested for the cosmic linear growth index. However, the tension with the standard GR value ($3.7\sigma$ in the standard $\gamma\Lambda$CDM case) is reduced to $2.4 \sigma$ if we ignore the high redshift clusters. The systematic effect will be investigated in the next eRASS data releases, but our main results are conserved.
\begin{figure}[h!]
    \centering
    \includegraphics[scale=0.6]{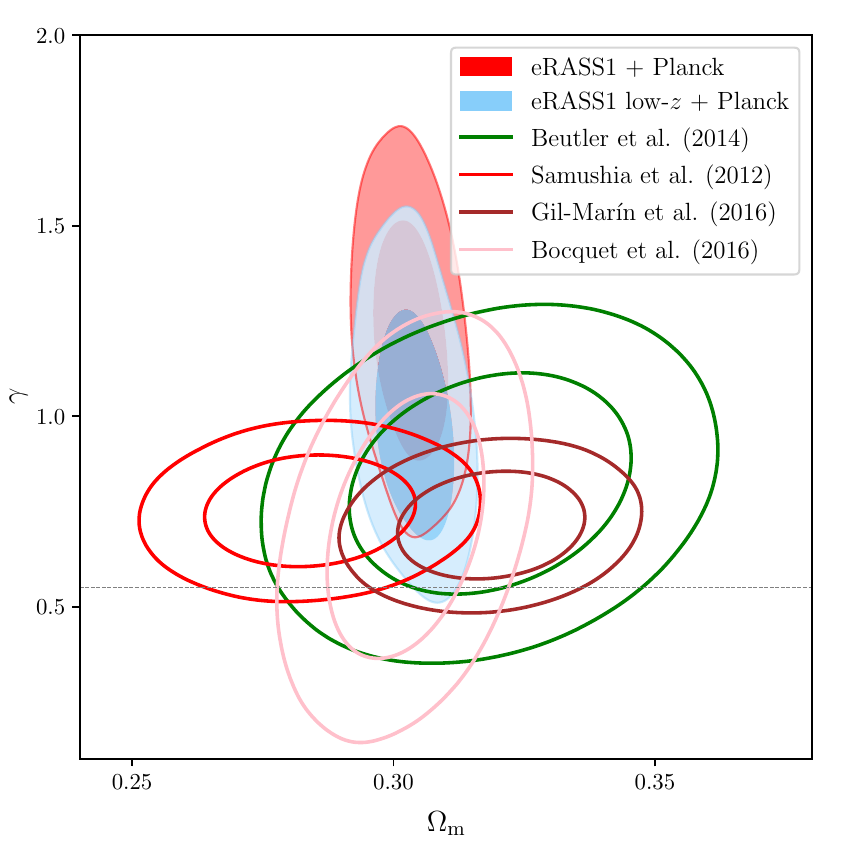}
    \caption{Impact of the high redshift clusters on the $\gamma\Lambda \rm CDM$ analysis. The constraints obtained while removing the sample are shown in light blue}
    \label{fig:low_z_impact}
\end{figure}

\section{Sound horizon priors}
\label{app:sound_horizon}

 Through this paper, we consider as priors the sound horizon scale at recombination. as this constitutes one of the most robust cosmological measurements to date:
\begin{equation}
    \label{eq:theta_star}
    \theta^* = \frac{r^*}{D_\mathrm{A}(z^*)},
\end{equation}
where $r^*$ is the maximum distance sound waves can travel, $z^*$ is the recombination redshift, and $D_\mathrm{A}$ is the angular diameter distance. The priors on $\theta^*$ and $z^*$ are taken from the latest cosmic microwave background measurements (\citetalias{PlanckCollaboration2020}). Assuming that the recombination physics is well understood, the sound horizon is one of the most robust cosmological measurements to date. As it is an angular distance measurement, it is mostly sensitive to a degeneracy between $H_\mathrm{0}$ and $\Omega_\mathrm{m}$, illustrated in this appendix.
\begin{figure}
    \centering
    \includegraphics[scale=0.6]{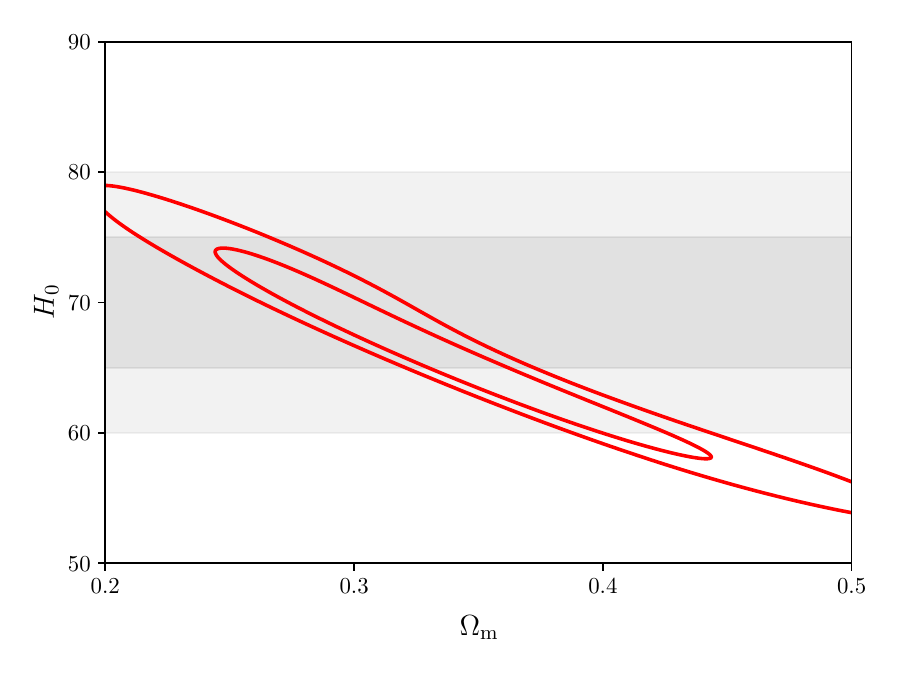}
    \caption{Cosmological constraints obtained from the sound horizon scale alone, measured in the CMB data from \citetalias{PlanckCollaboration2020} CMB in red. Those constraints are combined with a conservative $H_\mathrm{0}\sim \mathcal{N}(70, 5^2)$.}
    \label{fig:sound_horizon_prior}
\end{figure}
This prior is combined with a conservative $H_\mathrm{0}\sim\mathcal{N}(70,5^2)$, and mostly provides a measurement of the matter density parameter. This provides an anchor for our measurements since our constraints are significantly altered in the three cases we are exploring.
The $\Omega_\mathrm{m}$ prior that we obtain is $\Omega_\mathrm{m}\sim \mathcal{N}(0.31, 0.04)$.

\end{appendix}
\end{document}